\documentclass[a4paper]{article}
\usepackage[english]{babel}
\usepackage[utf8x]{inputenc}
\usepackage[T1]{fontenc}

\usepackage[a4paper,top=3cm,bottom=2cm,left=2.5cm,right=2.5cm,marginparwidth=1.75cm]{geometry}
\usepackage{multicol}
\usepackage{parskip}
\usepackage[section]{placeins}

\usepackage{amsmath}
\usepackage{graphicx}
\usepackage[colorinlistoftodos]{todonotes}
\usepackage[colorlinks=true, allcolors=blue]{hyperref}
\usepackage{float}
\usepackage{hyperref}
\usepackage{caption}%
\usepackage[symbol]{footmisc}

\newcommand\nc\newcommand
\nc\be{\begin{equation}}
\nc\ee{\end{equation}}
\nc\mss{\mathrm{MSS}}
\nc\demi{\frac{1}{2}}
\newcommand\vect[1]{\boldmath{#1}}
\everymath{\displaystyle}

\def\abs#1{\vert #1\vert}

 \title{New features of the sea-surface slope distribution revealed by IASI observations: Directional properties, influence of the wave height, large wave tilts}
  
 \author{Charles-Antoine Gu\'erin\footnote{Mediterranean Institute of Oceanography (MIO), Univ Toulon, Aix-Marseille Univ, CNRS, IRD, Toulon, France}, Virginie Capelle\footnote{Laboratoire de M\'et\'eorologie Dynamique/IPSL, CNRS, Ecole Polytechnique, Institut polytechnique de Paris, Sorbonne Universit\'e, Ecole Normale Sup\'erieure,  PSL research university, Palaiseau, France} and Jean-Michel Hartmann$^2$}
\begin{document}

\makeatletter
\let\@fnsymbol\@arabic
\makeatother
\maketitle

\setlength{\parskip}{0.1in}
\setlength{\parindent}{15pt}

\begin{abstract}
 By analysis of the reflected solar contribution to mid-infrared radiance spectra collected by the IASI instrument, we investigate the influences of both the wind and significant wave height on the wave-slope probability distribution function (PDF). We show that, at equal wind speed, smaller wave heights enhance the directional (upwind-downwind and upwind-crosswind) asymmetries of the probabilities. We also point out that the sea-surface mean square slopes slightly decrease as the wave height increases (by about $5\%$ per meter for winds slower than 6 m/s), an original result for which we propose possible causes (short-scale damping by long waves and/or effect of compounding statistical processes). We also investigate the so far practically unstudied  case of steep slopes and find large deviations from the quasi Gaussian behavior of the central part of the PDF, with a quite universal exponential decay of the probabilities with increasing tilt. We here observe larger probabilities in the downwind direction, but similar values in the up- and cross-wind ones. The physical mechanisms responsible for these trends are discussed. Finally, we propose refined parametrizations of  the mean square slope versus wind-speed, which departs from the Cox and Munk linear relationships at moderate wind speed (5-8 m/s).
\end{abstract}

%\begin{multicols}{2}
%% \usepackage{amssymb}
%% %% The amsthm package provides extended theorem environments
%% %% \usepackage{amsthm}
%% \usepackage{amsmath}
%% %% The lineno packages adds line numbers. Start line numbering with
%% %% \begin{linenumbers}, end it with \end{linenumbers}. Or switch it on
%% %% for the whole article with \linenumbers.
%% %% \usepackage{lineno}
%% \usepackage{color}
%% \usepackage[T1]{fontenc}




\section{Introduction}\label{intro}
Accurately describing the probability distribution function (PDF) of the wave-facets slopes on the ocean surface is essential for many geosciences studies, as demonstrated by the thousands of studies which have used and cited the parameterizations proposed by  \cite{Cox54}. Many investigations have been carried out after this pioneering work, as discussed in \cite{Guerin_RSE23,revue_vagues}, based on observations from boats, plateforms, airplanes and satellites, collected using various  techniques. With respect to field campaigns, space-borne sensings generally have the great advantage of providing huge amounts of data, practically world wide and over long time periods, thus probing a large variety of sea-surface states. These include active measurements using LIDARs \cite{hu_atmosChem08,Lenain_JPO19} and radars \cite{jackson_JGR92,Vandemark_JPO04,Nouguier_GRSL16}, as well as passive ones using the flux of solar photons reflected by the sea surface, with examples given in \cite{Breon, Guerin_RSE23} and references cited therein. Note that the latter often involve large variations of the zenith and azimuth angles of both the observation line-of-sight (thanks to the swath) and sun position  (thanks to its changes with season and latitude). This enables to probe a large portion of the wave-slope PDF and to quantify the up- versus cross-wind differences as well as the deviations from a purely Gaussian shape. Recall, for completeness, that, due to their significant wavelengths, radar soundings may not be sensitive to sea-surface roughnesses at small scales whose contributions are filtered, a limitation that does not affect  observations in the optical domain.
Following \cite{Cox54}, most studies probing the wave-slope PDF and providing detailed information on its shape, see \cite{Breon, Lenain_JPO19, Guerin_RSE23} and those cited therein,  assume  that wind is the only factor driving the distribution. However, a few other works have also considered the influences of the wave height, of breaking waves, of the swell, and of the degree of atmospheric stability, with examples respectively given by \cite{Nouguier_GRSL16,Voronovich,Hwang_JGR88,shaw}. A further discussion on this issue is made in Appendix A.5. 

In the present paper, we extend the study of \cite{Guerin_RSE23} by now not only considering the influence of the wind $\overrightarrow{U}$, but also that of the significant wave height $H_s$, on the wave-slope PDF and the mean square slopes (MSSs). As  reviewed in this previous study, the effect of the former has been largely investigated. In contrast, the influence of $H_s$, despite its consequences for wind-speed determination \cite{Lefevre,Li_IntJournRemSen13}, has received much less attention, with, to the best of our knowledge, no study using  the optical range (which makes the present work the first of its kind). Indeed, the few investigations are all based on data collected using microwave radar backscatter, which can only measure filtered MSSs \cite{Hwang_JGR88,Hauser_JGR08,Li_IntJournRemSen13,Nouguier_GRSL16}. Note that the last reference shows that the total MSS significantly increases with increasing $H_s$ for winds slower than about 5 m/s, and that the influence of $H_s$ progressively vanishes as $U$ increases. However, the effect of the swell, which can be at the origin of increased $H_s$ values, is still unclear as discussed in \cite{Hwang_JGR88,hwang_JGR08,hwang_JGR09comment,hauser_JGR09reply}.

The present investigation uses the wave-slope probabilities  retrieved \cite{Guerin_RSE23}, as detailed in \cite{Capelledaytime}, from radiances collected by the  satellite-based Infrared Atmospheric Sounder Interferometer. The former are  analyzed within bins of both $\overrightarrow{U}$ and $H_s$, providing the influences of these atmospheric and oceanic variables on the MSSs and on the parameters involved in the representations of the associated PDFs. The remainder of this paper is organized as follows. After a description  of the input data used (Sec. \ref{Dataused}), the equations proposed by \cite{Cox54} and our approach for the representation of the PDF are presented in Sec.3, our main findings concerning the latter being discussed in Sec. 4. Secton 5 then presents the results obtained for the MSSs-shape, which then enable the determinations of the MSSs together with those of the skewness and kurtosis coefficients (Sec. 6). Finally, the cases of large wave tilts and strong winds are the subjects of Sec. 7, while Sec. 8 presents a summary and discussion of the results of the present study. 

\section{Input data used}\label{Dataused}

\subsection{Probabilities, winds and wave heights}
The wave-slope probabilities  used in the present work were obtained \cite{Guerin_RSE23} from about 150 million daytime observations, under clear sky conditions, of  sea surfaces by the Infrared Atmospheric Sounder Interferometer (IASI) onboard the Metop platforms \cite{IASI}. For each of them a probability $p(s_x,s_y)$ was deduced \cite{Capelledaytime} from the contribution of reflected solar photons around 3.8 $\mu$m, where $s_x$ and $s_y$ are the slopes along the North-South and East-West directions determined  from the sun position and instrument line of sight. $s_x$ and $s_y$ were then converted  to  up- and cross-wind slopes, $s_u$ and $s_c$, by using the wind-direction information described below. 

Wind velocities  $\vec{U}$ (at 10 m), collocated in time and space with each IASI-observed spot, were taken from the European Centre for Medium-range Weather Forcast (ECMWF) ERA5 reanalysis dataset \cite{ERA5}, which provides \cite{ERA5data} world-wide values on hourly and $0.25^\circ\times0.25^\circ$ spatial-grid bases. For the speed $U=||\vec{U}||$, it was shown \cite{Guerin_RSE23}, through comparisons with ASCAT \cite{ASCAT} measurements, that the ERA5 data have an accuracy better than 1 m/s with typical RMSE values after correction for the mean bias of about 1 m/s. For the wind direction, the mean error is within $\pm 4^\circ$ (with a change of sign when switching from the North to the South Hemisphere) with a RMSE (after correction for the bias) which decreases with increasing $U$, from about $20^\circ$ for 3 m/s down to $6^\circ$ for 15 m/s. These findings and numbers are consistent with those of \cite{Belmonte} and \cite{ERA5wind}. A discussion of the implications of errors in the ERA5 wind information on the IASI-retrieved probabilities is made in Appendices A.4.1,2.

For the significant wave height, $H_s$, we also used values provided by ERA5. Their accuracy was extensively assessed \cite{Wang04032022} through more than 16 million comparisons with in situ values collected by buoys. The results show an overall excellent correlation (R=0.961) between the ERA5 and buoys $H_s$ values, with a very small bias -0.058 m and a RMSE of 0.325 m. The error increases with  $H_s$, with an underestimation under extreme wave conditions, but the accuracy of ERA5 is found very satisfactory for the most frequent sea states (0.5 m < $H_s$ < 4 m) with a bias and RMSE ranging from -0.236 m to 0.045 m and from 0.212 m to 0.435 m, respectively. Other studies based on more limited  datasets, e.g. \cite{Shi,Wang04032022,ANUSREE,Liu_Atmos22},  lead to similar conclusions and numbers. A discussion of the influence  of errors on $H_s$ on the IASI-retrieved probabilities Appendix A.4.3. Note that the above given numbers for $U$ and $H_s$ in ERA5 are comparable with those of an extensive study for the Mediteranean \cite{Schuckmann20082021} with found, for $U$, an overall bias of -0.26 m/s and a mean absolute error of 1.08 m/s, and, for $H_s$, a bias of -0.12 m with a mean absolute error of 0.26 m.

\begin{figure}[H]\centering
 \includegraphics[scale=0.35,angle=-0]{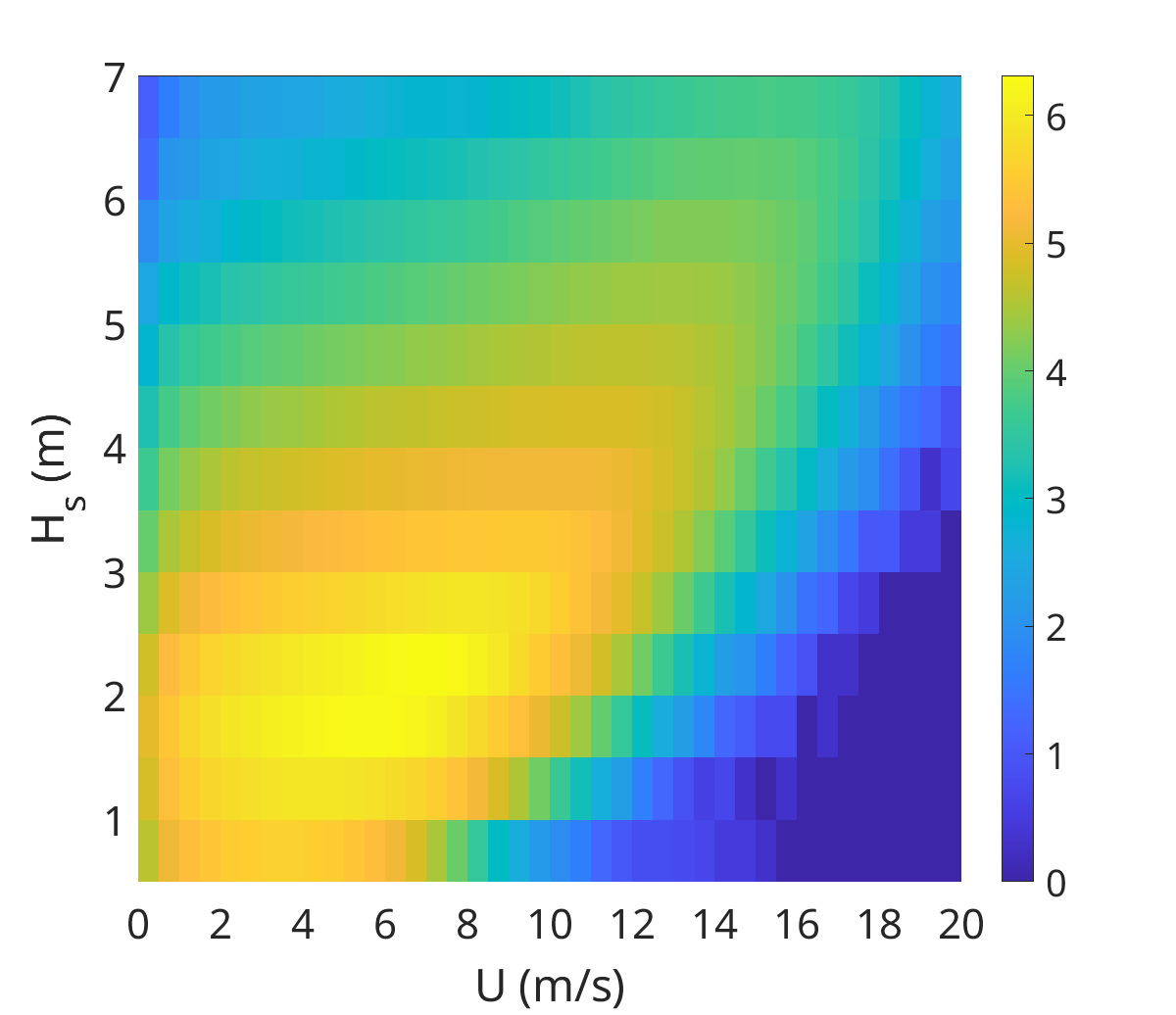}
 \caption{Number of IASI-retrieved probabilities (given in decimal logarithm) for each $(U,H_s)$ bin, using bin widths of $1$ m/s and $0.5$ m, respectively.} 
   \label{fig:statIASI}
\end{figure}

\begin{figure}[H]\centering
 \includegraphics[scale=0.35,angle=-0]{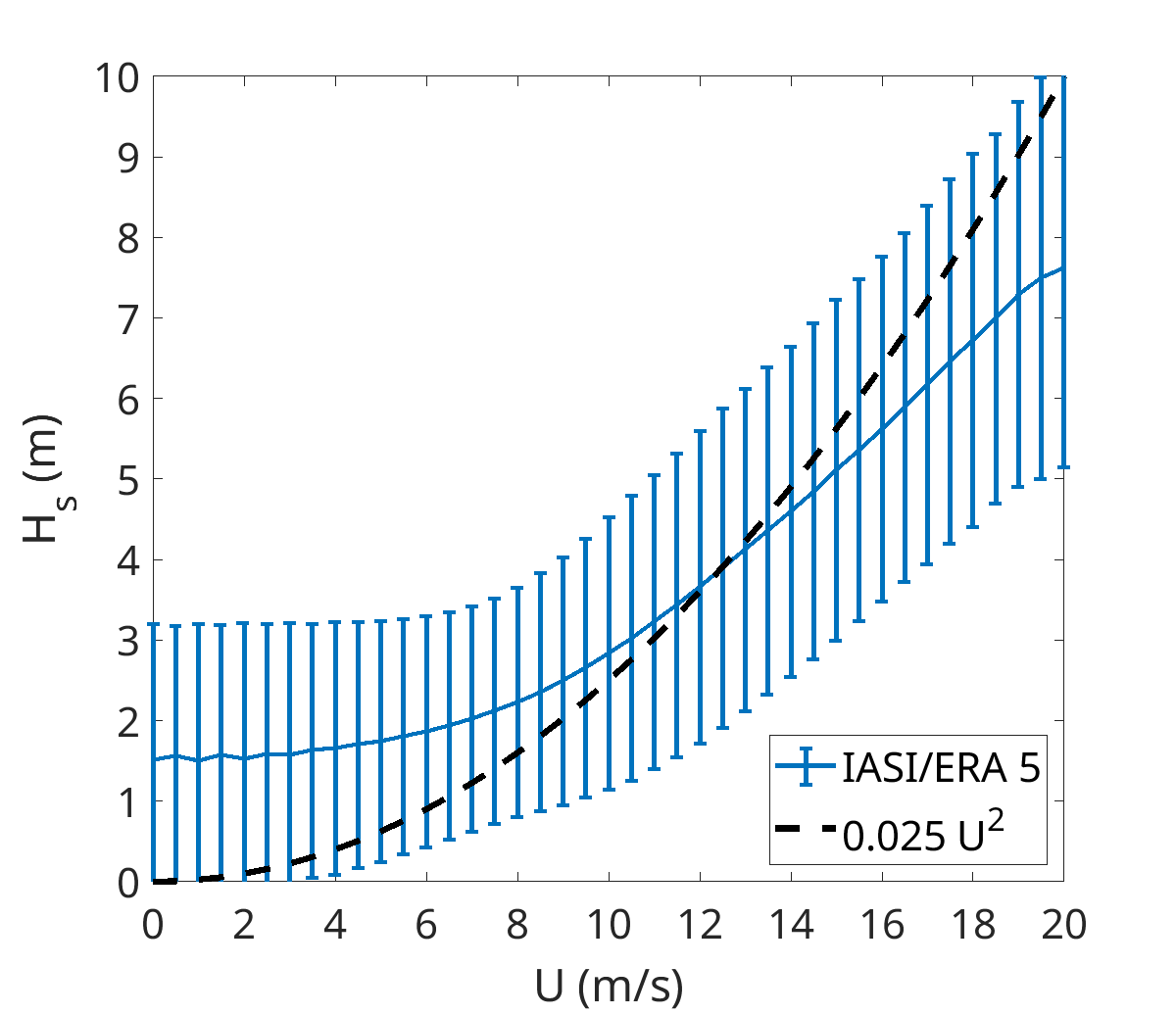}
 \caption{Mean $H_s$ associated to the IASI observations according to ERA 5, as a function of wind speed. The bars indicate $\pm 2$ standard deviations about the mean. The black dashed line gives the expected $H_s$ for fully developed seas, according to: $H_s=0.025\ U^2$ (see text).}
   \label{fig:statHs}
\end{figure}

Figure \ref{fig:statIASI} displays the numbers of individual IASI-retrieved probabilities after  sorting the data in both $U$ and $H_s$ using centered bins of widths $1$ m/s and $0.5$ m, respectively, by steps of $0.5$ m/s and $0.5$ m. Between $10^5$ and $10^6$ observations are available for small to moderate $U$ (2-8 m/s) and $H_s$ (1-3 m), the statistics being much sparser for larger values.  In addition, Fig. \ref{fig:statHs} shows the mean $H_s$ associated to the IASI observations for each wind speed. The error bars indicate $\pm 2$ standard deviations about the mean, showing that the effective span of variation of $H_s$ is of about 3 m at small wind speed. The  dashed  black line in this plot indicates the dependence of $H_s$ on wind speed for fully developed seas according to \cite{pierson}, $H_s=0.214/U_{19.5}^2$, which was here converted \cite{holthuisen_book10} with the 10 m wind speed $U$ as $H_s=0.025 U_{10}^2$, in good agreement with the result of \cite{Hs-vs-U}. As can be seen, the $H_s$ values associated with the IASI observations exceed this law for small $U$ values, showing that most of them are for the open ocean, as could be statistically expected. 

%/home/guerin/Dropbox/COXMUNK_HARTMANN/IASI_HS_2025/PRG_HS/PRG_FIGURES_PAPIER/plot_statIASI.m 
%/home/guerin/Dropbox/COXMUNK_HARTMANN/IASI_HS_2025/PRG_HS/stat_vent_Hs.m

%/home/guerin/Dropbox/COXMUNK_HARTMANN/IASI_HS_2025/PRG_HS/PRG_FIGURES_PAPIER/plot_statIASI.m 

\subsection{Missing points, scatter, and the tail in the IASI-retrieved probabilities}

\begin{figure}[H]\centering
  \includegraphics[scale=0.4,angle=-0]{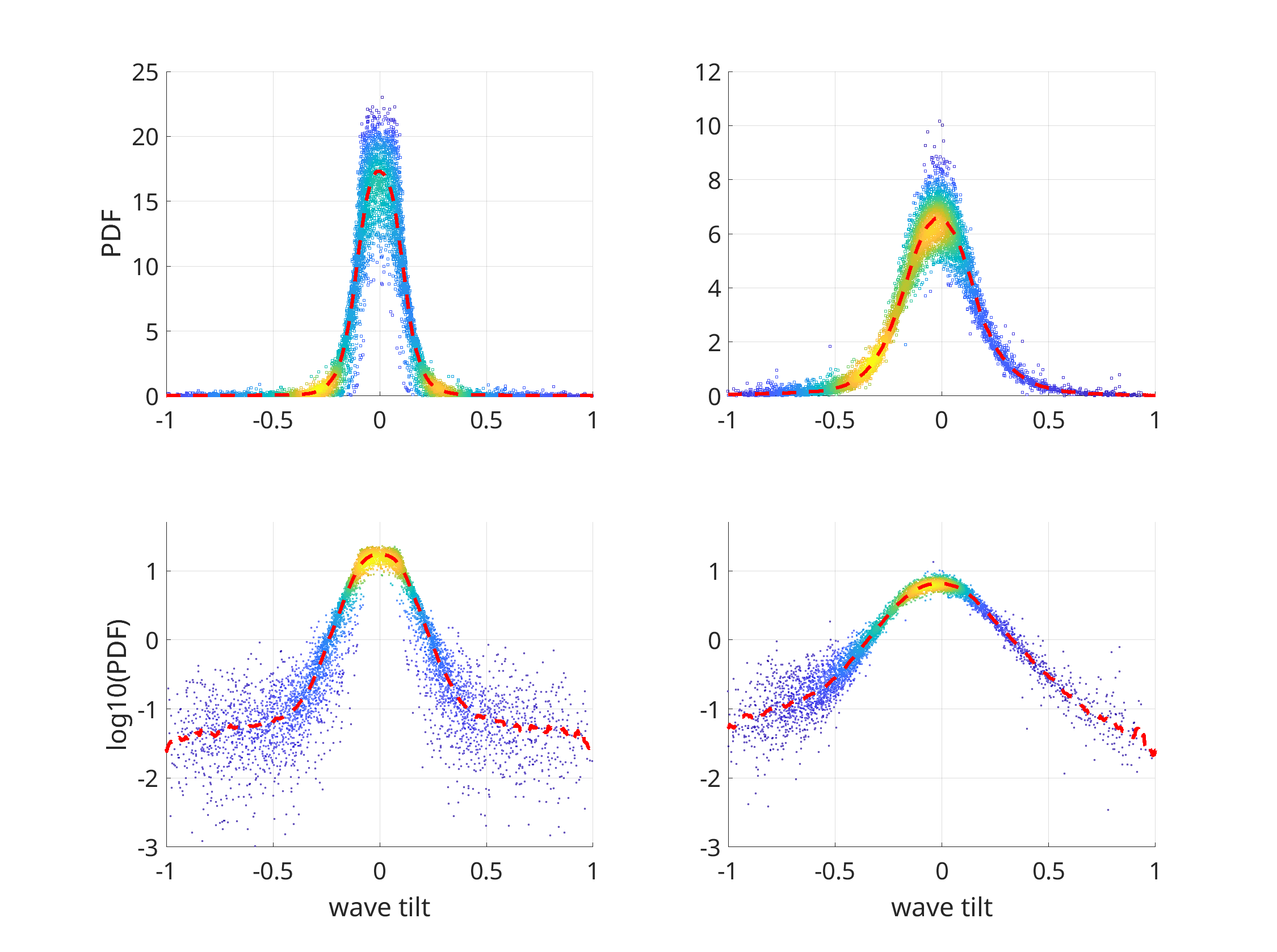}
          \caption{Cut of the wave slope PDF in the principal plane along the wind direction ($s_c=0,\ s=s_u$) versus the wave tilt $s$  for  $U$=2$\pm 0.5$ m/s, $H_s=1\pm 0.25$ m (left) and  $U$=9$\pm 0.5 $ m/s, $H_s=2\pm 0.25$ m (right). Hot colors indicate high densities of points, in arbitrary units. \label{fig:examplespdf_vent2_9}}
\end{figure}

Figure \ref{fig:examplespdf_vent2_9} shows the IASI-retrieved probabilities restricted to the principal plane along the wind direction for small ($U=2$ m/s, $H_s=1$ m) and large ($U=9$ m/s, $H_s=2$ m) values of $U$ and $H_s$. Several preliminary remarks can be made.
%\end{multicols}

%/home/guerin/Dropbox/COXMUNK_HARTMANN/IASI_HS_2025/PRG_HS/PRG_FIGURES_PAPIER/plot_scatterkde_ventHs.m

%\begin{multicols}{2}

The first is that there is a large inhomogeneity of the density of points, due to the limited variations (with season and latitude) of the sun position at the IASI passing time (9:30 AM LT) and to the restricted ($\pm 30^{\circ}$) deviations from nadir viewing of the observations retained in the present work \cite{Capelledaytime}. This irregular density of points hinders a direct statistical analysis through the fitting of a parametric distribution, as over-represented wave slopes could result in biased estimates.

Another issue is the thresholding of the PDF for probabilities larger than about 22, as seen in the upper left panel of Fig.  \ref{fig:examplespdf_vent2_9}. This results from the fact that some IASI spectra are not kept because the collected radiance is greater than a given limit. Measured values then exceed the on-board coding tables capacity and cause an overflow leading to the rejection of the spectrum \cite{IASIoverflow}. It is the case  when the contribution of the solar photons reflected by the sea surface is too large, which happens for small wind speeds  and wave slopes, resulting in a clipping of the large probabilities. This led  \cite{Guerin_RSE23} to only analyze the IASI data associated with $U \ge$ 3 m/s, a limitation released in the present study thanks to a new analysis method.

Figure \ref{fig:examplespdf_vent2_9} also shows that the IASI-retrieved probabilities  show a significant scatter. Several sources can contribute to the latter, including errors on the input data and model  used \cite{Capelledaytime} for the retrieval of the probabilities, the eventual contribution of scattering of solar photons by aerosols white caps and foam, improper values of  $U$ and $H_s$ due to errors in the ERA5 data and/or inhomogeneities within the IASI-observed pixel, and the fact that oceanic and/or atmospheric characteristics other than $U$ and $H_s$ participate to the driving of the wave slopes. These issues are further discussed in the Appendix, where it is shown that  the main source of scatter near the maximum of the distribution is wind-speed variability or errors on the assumed $U$, while aerosols and breaking are (likely) responsible for the scatter in the tail.

\section{Representation of the PDF of sea surface slopes}

\subsection{Classical representation of the PDF}
We here briefly recall the analysis procedure used by \cite{Cox54,Cox54b,Cox56} , later on applied in other investigations, e.g.   \cite{Breon, Guerin_RSE23}. The up- and cross-wind MSSs, $m_u$ and $m_c$, are, for a given wind speed, defined as the second moments of the PDF, i.e.:
\begin{equation}
  \begin{split}
    m_u&=\int_{-\infty}^{+\infty} s_u^2p(s_u,s_c)ds_uds_c,\\
    m_c&=\int_{-\infty}^{+\infty} s_c^2p(s_u,s_c)ds_uds_c ,
    \end{split}
\label{defsigma}
\end{equation}
the total MSS being $m_t=m_u+m_c$. The main findings of the above-cited studies are that: i) $m_u$, $m_c$ and $m_t$ increase quasi-linearly with the wind speed ; ii) $p(s_u,s_c)$ is symmetrical with respect to the crosswind direction but negatively skewed with respect to the upwind one and iii) it is close to a bidimensional normal distribution, with $m_u>m_c$ and a null mean slope, but, iv) with respect to Gaussians, it is slightly more peaky around the origin ($s_u=s_c=0$), and decreases more slowly at large slopes. These  deviations were  described by a Gram-Charlier (GC) series expansion, i.e.:
\begin{equation}
\begin{split}
p(s_u,s_c)&=(2\pi\sqrt{m_um_c})^{-1}e^{-\demi u^2}e^{-\demi c^2}\times \left(1-\demi C_{12}u(c^2-1)-\frac{1}{6}C_{30}(u^3-3u)\right.\\
&\left. +\frac{1}{24}C_{40}(u^4-6u^2+3)+\frac{1}{4}C_{22}(u^2-1)(c^2-1)+\frac{1}{24}C_{04}(c^4-6c^2+3) \right) , \end{split}
\label{eq:GC}
\end{equation}
where $u=s_u/\sqrt{m_u}$ and $c=s_c/\sqrt{m_c}$. $C_{12}$ and $C_{30}$ are the skewness coefficients along the wind direction, which account for the distribution asymmetry, while the kurtosis coefficients, $C_{40}$,  $C_{22}$ and $C_{04}$, describe the peakedness. 
For consistency with many studies in radar and optical remote sensing, we chose, as in \cite{Guerin_RSE23},  the azimuth reference axis $x$ as the upwind direction, which was the $y$ axis in  \cite{Cox54,Breon}. This simply implies that our $C_{ji}$ values corresponds to those of the $C_{ij}$'s in the above-cited studies. Finally, we recall the broadly used linear regressions proposed by \cite{Cox54}, later on confirmed by \cite{Breon}:
\begin{equation}
\begin{split}
  m_u&=3.16\ 10^{-3}\ U_{12.5}\\
  m_c&=3.\ 10^{-3} +1.92\ 10^{-3} U_{12.5},
\end{split}
\label{loiCM}
\end{equation}
where the wind speed at 12.5 m, $U_{12.5}$, can be converted to the more standard value $U$ at 10 m using the approximation $U\simeq0.98U_{12.5}$ \cite{Capelledaytime}.

\subsection{Our approach to represent the PDF}

 The significant magnitude of the probabilities   for large slopes and their large scatter, shown in Fig. \ref{fig:examplespdf_vent2_9}, cannot be properly described due to the difficulty to represent and separate the various physical processes potentially involved (steep breaking waves, white caps and foam, aerosols) and the natural variability of measurements (sub-pixel variability of wind speed, inaccuracy in collocated wind speed, etc). We therefore restrict our analysis to the central part of the distribution, which does not suffer from these limitations. Then, as \cite{Cox54}, we assume that the observed wave-slope PDFs can be fitted with  near-Gaussian models, which raises the usual issues that: i) the experimental distribution is truncated and must in some way be extrapolated to large wave slopes; ii) the PDFs estimated from IASI observations are obtained by averaging individual probabilities for non-equiprobable values of the wave slope (as seen in Fig.  \ref{fig:examplespdf_vent2_9} and discussed above). As a result, the retrieved PDFs do not satisfy the proper normalization condition of a probability density. In the following, we propose an improved method to circumvent this issue, based on a versatile description of the 2D distribution in terms of azimuthal harmonics and ``Gaussian times polynomial'' functions. ﻿We begin the analysis by an azimuthal harmonics approach similar to that used in \cite{Guerin_RSE23} but now applied within bins of both wind speed and significant wave height. The data were thus sorted both by $U$, using a $0.5$ m/s step and centered bins of width $\pm 0.5$ m/s, and by $H_s$, using a $0.5$ m step with centered bins of width $\pm 0.25$ m. For every slope vector $\vect{s}=(s_u,s_c)$, we considered each observed probability, $p_{obs}(s_u,s_c,U,H_s)$, as a random variable fluctuating around a mean value $p(s_u,s_c,U,H_s)$.  The latter was estimated by averaging the $p_{obs}(s_u,s_c,U,H_s)$'s inside bins of slope $s$ in polar coordinates $[s=(s_u^2+s_c^2)^{1/2}, \theta=\tan^{-1}(s_c/s_u)]$. For this, we sorted the data into a fixed number (400) of $s$ bins ensuring a minimal number of observations ($\sim 5\ 10^2$), which restricted the analysis to the ($U,H_s$) bins for which at least $2\ 10^5$ IASI-retrieved probabilities are available. Note that, due to the varying density of points within each wave-slope bin, the slopes $s$ were sampled with an irregular step $\Delta s$. Within each wave-slope interval, the probabilities  within regular azimuthal bins, of width $\Delta\theta=10^{\circ}$  by step of 1$^{\circ}$, were regrouped and, in order to minimize the impact of outliers, we defined the distribution  $p(s,\theta,U,H_s)$ as the median rather than the mean of the observed probabilities inside each bin. For each  $(U,H_s)$ pair, the resulting PDF is modeled using  a second order (tests show that this is sufficient) expansion, i.e. :
﻿
\begin{equation}
  p(s,\theta)=a_0(s)+a_1(s)\cos\theta+a_2(s)\cos(2\theta) .
  \label{eq:azimfit}
  \end{equation}
﻿
The isotropic coefficient $a_0(s)$, which is the largest and represents the omnidirectional probability, is obtained from:
\be
a_0(s)=\frac{1}{2\pi}\int_{-\pi}^{\pi}  p(s,\theta)d\theta ,
\ee
and constrained by the PDF normalization condition:
\be
\label{eq:normalization}
\int_0^\infty 2\pi s\ a_0(s)ds=1 .
\ee

The first ($a_1$) and second ($a_2$) harmonics amplitudes,  respectively quantify the upwind-downwind and upwind-crosswind asymmetries, since:
\be 2a_1(s)= p(s,0)-p(s,\pi),
\ee
and
\be
2a_2(s)=\demi [p(s,0)+p(s,\pi)]-p(s,\pi/2) .
\ee
﻿\section{First findings concerning the wave-slope PDF characteristics}\label{subsec:firstfindings}
\begin{figure}[H]\centering
\includegraphics[scale=0.5,angle=-0]{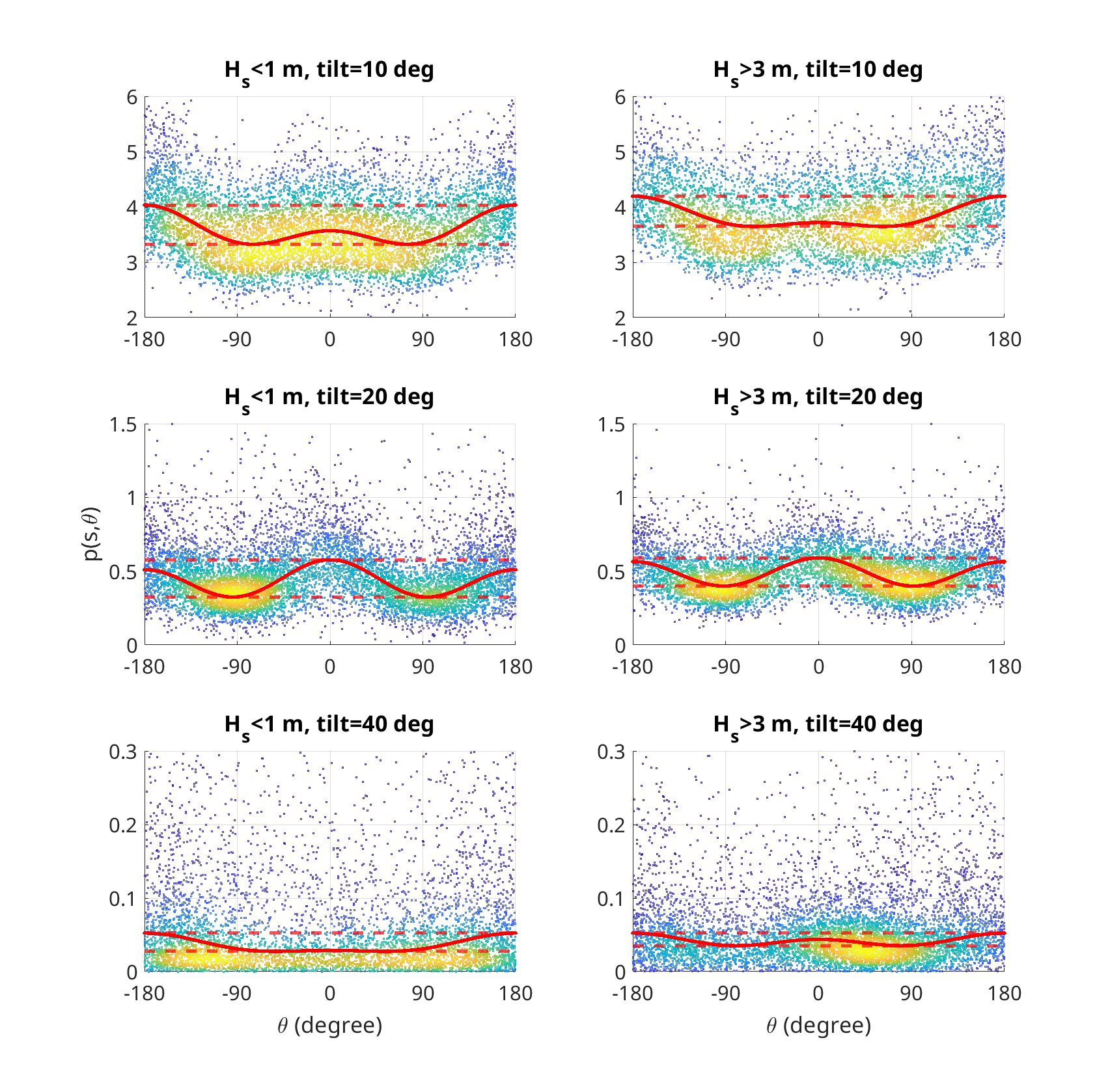}
\caption{\label{fig:examplefita0_azimuth}Observed probabilities $p(s,\theta)$ as a function of the azimuth angle $\theta$ with respect to the upwind direction for various wave tilt angles (whose tangent is equal to the slope $s$), for $U$=7 m/s together with small (<1 m, left column) and large (>3 m, right column) values of $H_s$.}
\end{figure}

Figure \ref{fig:examplefita0_azimuth} shows typical results, for $U$=7 m/s together with small (<1 m) and large (>3 m) values of $H_s$ and various wave-tilt angles=tan$^{-1}(s)$.  Several characteristic features of the wave-slope distribution can be pointed out in this example, as discussed below.
%\end{multicols}

\begin{figure}[H]\centering
\includegraphics[scale=0.35,angle=-0]{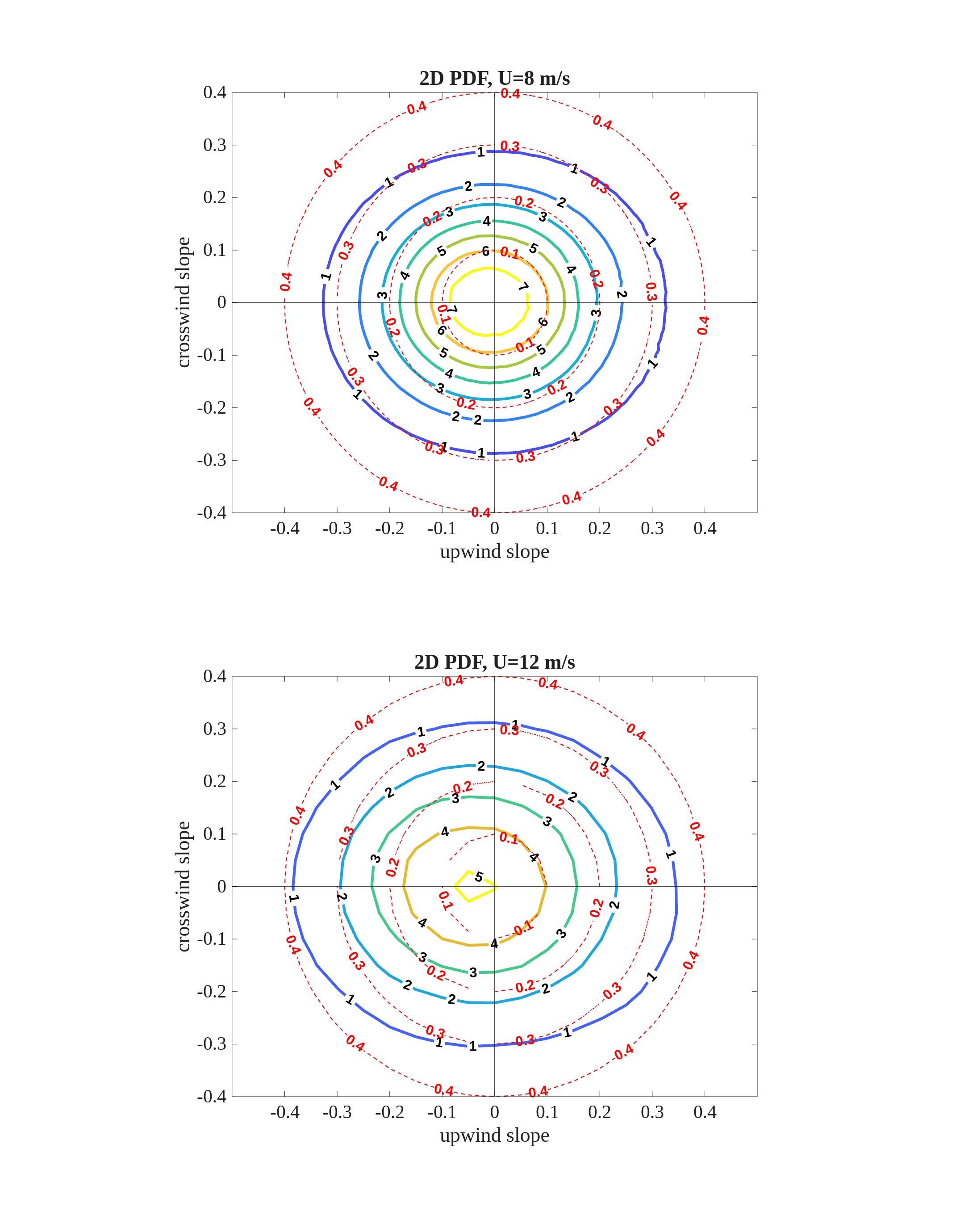}
\caption{Iso-value contours of the 2D wave slope PDF for wind speed 8 and 12 m/s (full colored lines for probabilities  of 1, 2, 3 etc,  indicated on the lines). The red dashed circles indicate the 0.1, 0.2, 0.3 and 0.4 iso-values of the wave tilts. \label{fig:hist2D}}
\end{figure}

\begin{figure}[H]\centering
  \includegraphics[scale=0.4,angle=-0]{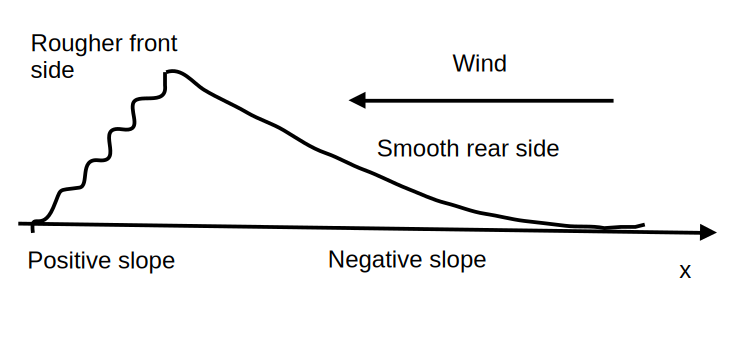}
\caption{Under the action of wind, the leeward side of the wave is roughened by parasitic capillary waves generated by microbreaking, while the windward side is smoother and broader with a gentle slope.\label{fig:frontback}}
\end{figure}

%/home/guerin/Dropbox/COXMUNK_HARTMANN/IASI_HS_2025/PRG_HS/PRG_FIGURES_PAPIER/plot_fita0_azimuth.m
%\begin{multicols}{2}
The most probable slope at a given wave tilt, i.e. the position of the PDF maximum, is generally found in the downwind direction ($\theta=\pm 180^{\circ}$). This results from to the well-known fact \cite{Cox54} that the PDF in the upwind plane is skewed along the wind direction and shifted toward negative slopes, in such a way that the most probable slope is slightly negative. This is confirmed by the fact that the most probable tilt angle decreases monotonically from 0 to about -3$^{\circ}$ when $U$ increases from 1 to 12 m/s, a result not shown here but provided by the last column of Table 1 of \cite{Guerin_RSE23}. The local maximum in the upwind direction  ($\theta=0^{\circ}$) vanishes for steep slopes, as obvious for the $40^{\circ}$ wave tilt, where the PDF is almost constant and minimal for $H_s$<1 m and $\theta$ between $\pm 80^{\circ}$. This behavior can be better pointed out by looking at the contour lines of the wave-slope PDF, obtained this time through the averaging of the IASI-retrieved probabilities  in cartesian slope bins, shown in Fig. \ref{fig:hist2D}. As seen, the iso-values (full colored lines) have elliptical shapes and are centered about negative upwind slopes. Looking at the distances between these curves and the  dashed red circles representing the iso-values of the wave-slope $s$, it is seen that, beyond some wave tilt, the PDF decreases monotonically from the downwind direction as the azimuthal angle is varied, the effect being more pronounced at high wind speed. This trend, which to our best knowledge was never observed in the optical domain, is opposite to what was found in microwave radar remote sensing, where the maximum is always located in the upwind direction and the downwind maximum eventually vanishes at grazing angles  \cite{masuko_JGR86,mouche_JGR07,guerraou_GRS16,guerraou_igarss18}. This is usually explained by an enhanced small-scale roughness on the front side of waves due to parasitic capillary ripples generated by micro-breaking \cite{LH63,zhang_JFM95,caulliez_JGR12,caulliez_JGR13} or by horse-shoe patterns formation through gravity waves resonant interactions \cite{shrira_JFM96}. In the optical domain, where the solar-photons redirection mechanism is specular reflection by surface facets rather than Bragg scattering by small ripples, the dominant part of the solar signal comes from the smoother and larger area on the rear side of steep waves (see Fig. \ref{fig:frontback}), which have negative slopes given our axis convention (the x-axis is pointing upwind).

\begin{figure}[H]\centering
	  \includegraphics[scale=0.3,angle=-0]{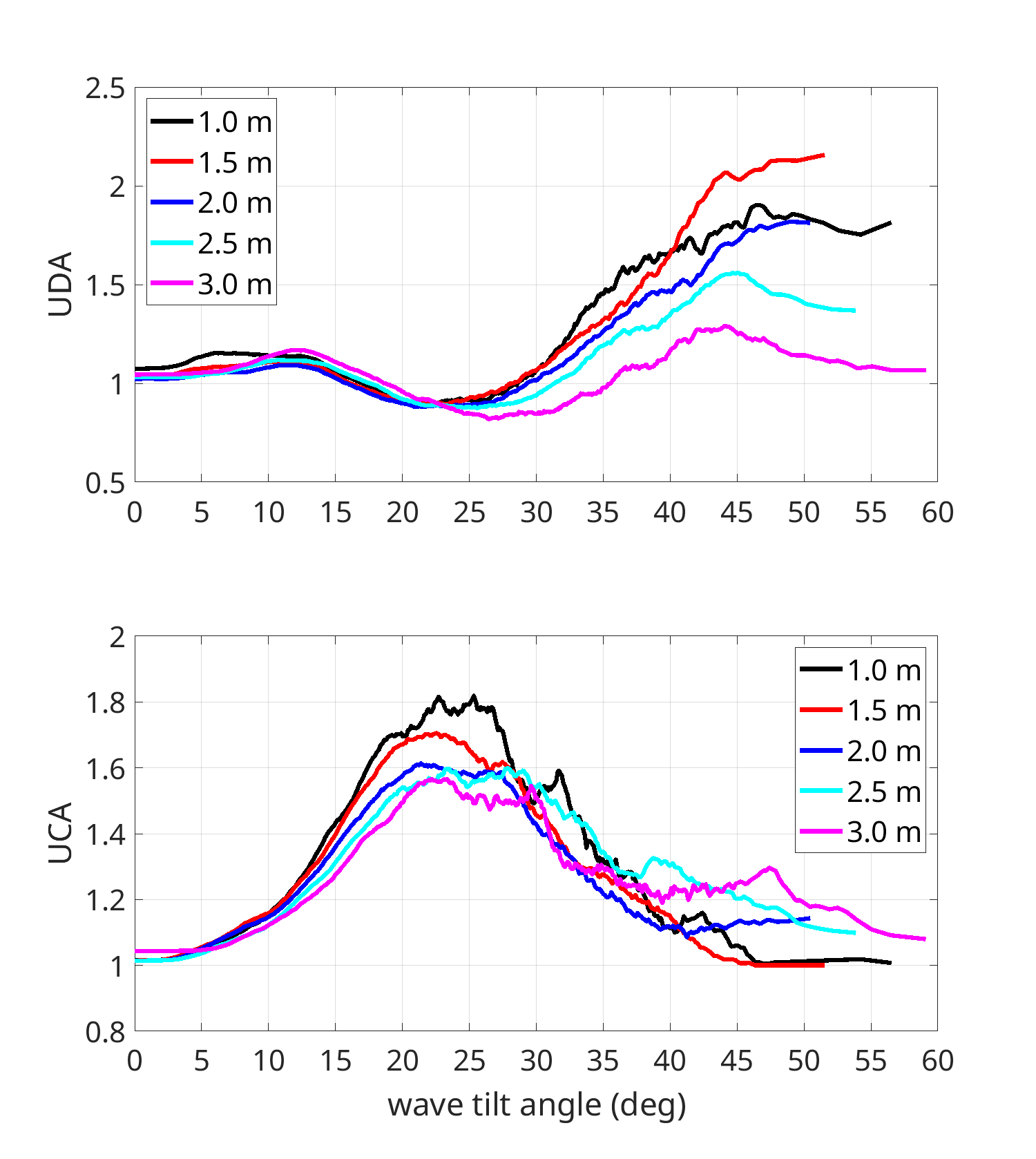}
          \caption{Variation of the upwind-downwind (top panel) and upwind-crosswind (bottom pane) asymmetry coefficients with the wave slope (converted here into a tilt angle in degree) for a wind speed of 7 m/s and various values of $H_s$. The asymmetries are enhanced by smaller $H_s$.}
          \label{fig:UDA_UCA} 
\end{figure}

\begin{figure}[H]\centering
	  \includegraphics[scale=0.3,angle=-0]{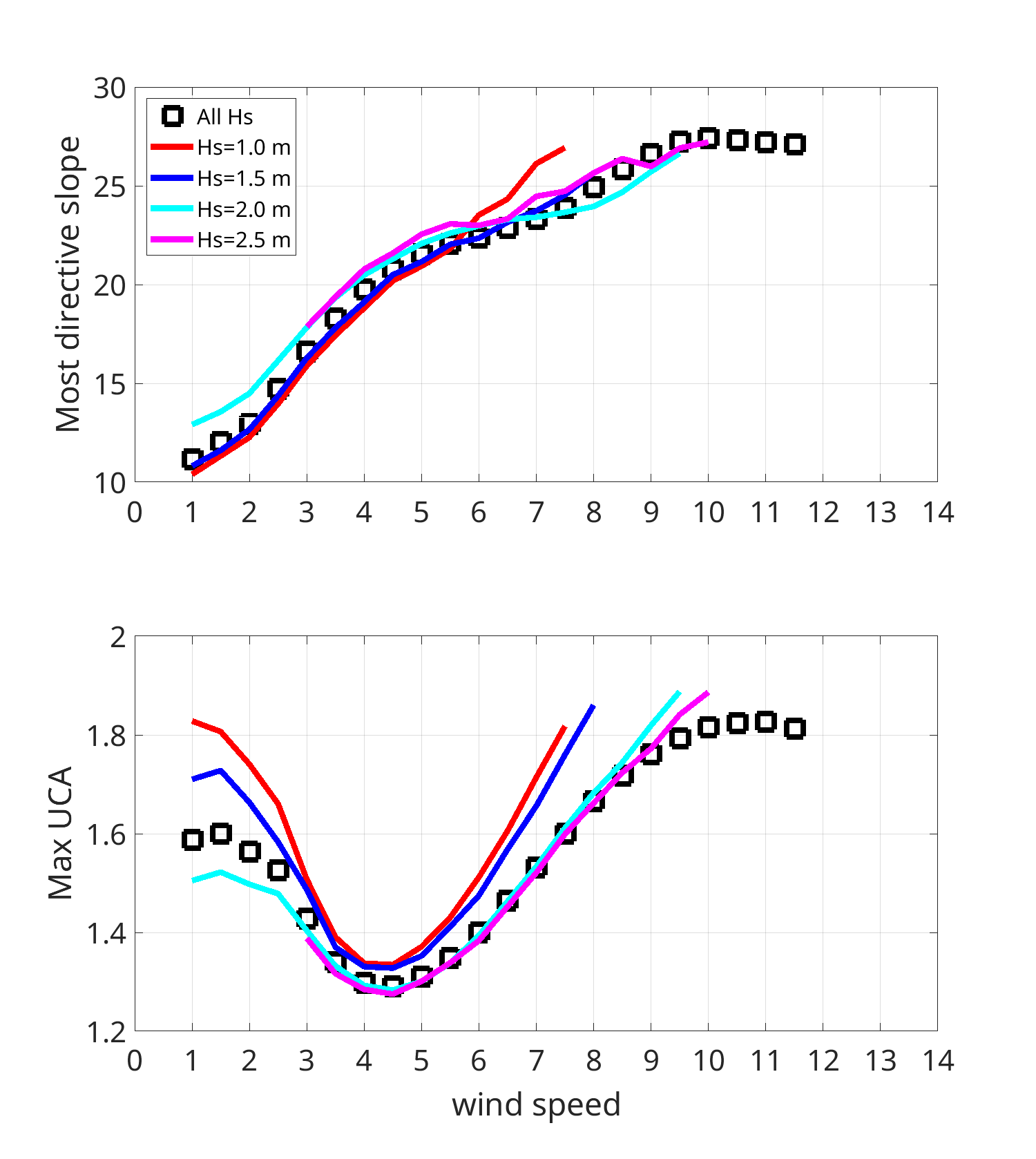}
\caption{Variation of the most directive slope, converted into a tilt angle, (top panel) and the maximal UCA (bottom panel) with wind speed for various $H_s$. \label{fig:directiveslope}}
\end{figure}

The PDF is symmetric in the crosswind direction, with identical probabilities for $\theta=\pm 90^{\circ}$, as also shown by Fig. \ref{fig:hist2D} where values for $\pm$ a given crosswind slope are the same. This absence of crosswind asymmetry, first observed by \cite{Cox54}, is the obvious result of balanced wind action around the wind axis. While the effect of $U$ on the asymmetry is well-known, e.g. \cite{Guerin_RSE23}, the effect of wave height is less obvious but, on this issued, Fig. \ref{fig:examplefita0_azimuth} provides an original result. Indeed, the amplitude of the azimuthal variations, which corresponds to the vertical distance between the dashed red lines, slightly decreases with increasing $H_s$, all the more when the wave tilt is small. This seems to indicate that, for a given wind speed, ocean surfaces with small significant wave heights (young seas) are more directive (in terms of wave slopes) than those with large-amplitude waves. To verify this statement, we quantified the upwind-crosswind asymmetry (UCA) through:
\be
UCA(s)=\frac{p(s,0)+p(s,\pi)}{p(s,\pi/2)+p(s,-\pi/2)}=\frac{a_0+a_2}{a_0-a_2},
\ee
and the upwind-downwind asymmetry (UDA) through:
\be
UDA(s)=\frac{p(s,\pi)}{p(s,0)}=\frac{a_0+a_2-a_1}{a_0+a_2+a_1}.
\ee
The variations of UCA and UDA with the wave slope $s$, which for convenience we convert into a tilt angle $\tan^{-1}(s)$ in degree, are shown in Fig. \ref{fig:UDA_UCA} for $U$=7 m/s and various $H_s$ values. They confirm that seas with a smaller $H_s$ (at equal wind speed) are more directive (higher UCA) and more asymmetric (higher UDA). We then define the most directive slope (or tilt angle) as the value for which the UCA reaches its maximum. These two quantities are shown as functions of $U$ and $H_s$ in Fig. \ref{fig:directiveslope}. The most directive tilt angle evolves from about 10$^{\circ}$ to 30$^{\circ}$, with a corresponding UCA between 1.3 and 2, as the wind speed increases from 1 m/s to 12 m/s. The influence  of $H_s$ on the wave directivity is here obvious, which is another original result of the present study, as the most directive slope angle increases with $H_s$ (with a span of variation of 3-4$^{\circ}$) while the UCA decreases (with a span of variation of about 0.2). A minimum of directivity is obtained at 4 m/s where the maximal UCA is about 1.25 for the larger $H_s$. This tends to show that, for a given wind speed, young or short-fetched seas, which have smaller wave heights, have more directive wave slopes than developed ones. Since the largest part of the MSSs comes from the contribution of small and intermediate waves \cite{hwang_JGR05}, this amounts to an enhanced directivity at those scales. It is also interesting to note that the maximal UCA is not monotonic with wind speed and reaches a minimum at 4 m/s. At this step, we have found no obvious physical mechanism explaining this observation.

%/home/guerin/Dropbox/COXMUNK_HARTMANN/IASI_HS_2025/PRG_HS/PRG_FIGURES_PAPIER/plot_hist2D.m

%/home/guerin/Dropbox/COXMUNK_HARTMANN/IASI_HS_2025/PRG_HS/PRG_FIGURES_PAPIER/frontbackasym.fig Xfig

%/home/guerin/Dropbox/COXMUNK_HARTMANN/IASI_HS_2025/PRG_HS/PRG_FIGURES_PAPIER/plot_UDA.m
﻿
﻿

%/home/guerin/Dropbox/COXMUNK_HARTMANN/IASI_HS_2025/PRG_HS/PRG_FIGURES_PAPIER/plot_mosprobableslope.m

\section{The MSS-shape}\label{sec:mssshape}
\begin{figure}[H]\centering
     \includegraphics[scale=0.4,angle=-0]{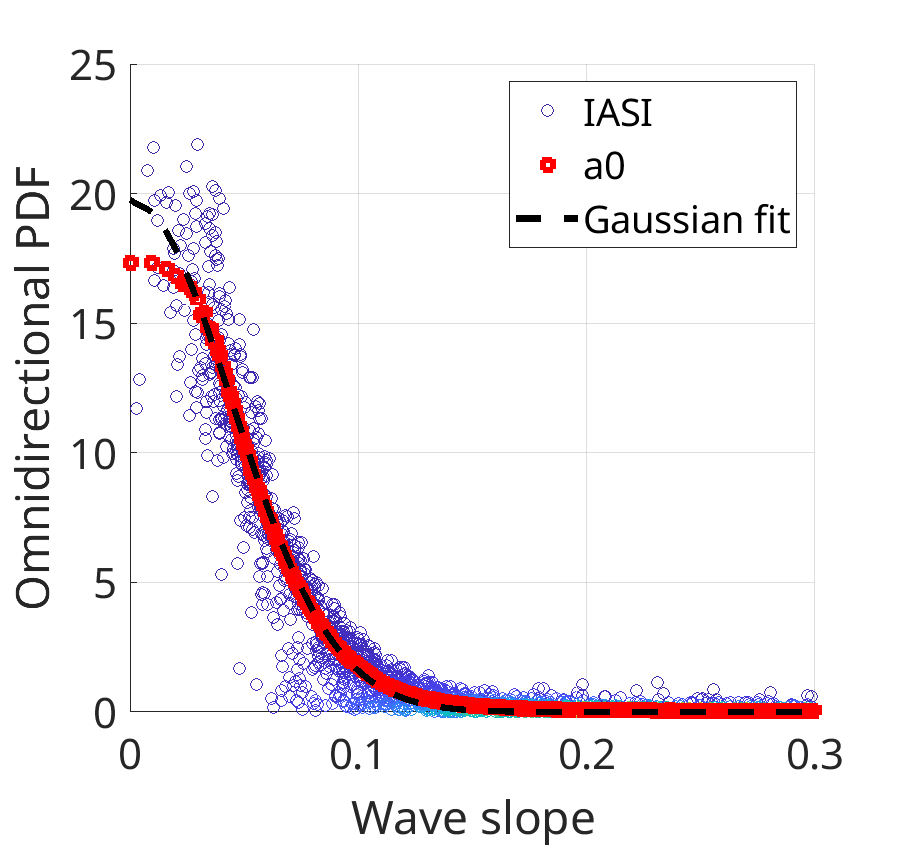}
\caption{Gaussian fitting of the omnidirectional PDF and correction of the saturation region for $U$=2 m/s and $H_s$=1 m. \label{fig:figmssshape_fit2parametres}}
\end{figure}

As discussed above, the parameterization of the wave-slope statistics from fits of the IASI-retrieved probabilities raises several technical issues which hinder the calculation of the MSSs from a direct integration using Eq. (\ref{defsigma}). However, the so-called MSS-shape can be obtained more easily, and its estimation is robust to both an inaccurate calibration of the probabilities and noise in the tail. It also plays an important role in microwave radar remote sensing as the ``radar-filtered'' MSS is in fact obtained by fitting a Gaussian shape to the near-nadir angular variations of the radar cross-section \cite{Hauser_JGR08,Nouguier_GRSL16,Guerin_GRS17}. The MSS-shape is defined as the MSS of the Gaussian distribution that best fits the PDF around $s$=0. The possible upwind-downwind asymmetry is thus ignored, and the analysis only considers its symmetric part:
\be\label{eq:p_sym}
p_{s}(s,\theta)=a_0(s)+a_2(s)\cos(2\theta).
\ee
Based on the azimuthal coefficients, we define the omnidirectional ($m_{0s}$), upwind ($m_{us}$), crosswind ($m_{cs}$) and total  ($m_{ts}$) MSS-shape as:
\be\label{eq:defmssshape}
\begin{split}
  a_0(s)&\simeq A_0 e^{-\demi s/m_{0s}},\\
   a_0(s)+a_2(s)&\simeq A_u e^{-\demi s_u^2/m_{us}} \\
   a_0(s)-a_2(s)&\simeq A_c e^{-\demi s_c^2/m_{cs}},\\
   m_{ts}&=m_{us}+m_{cs}
   \end{split}
  \ee
  for some constants $A_0$, $A_u$, and $A_c$ that give the values of the distributions at the origin. Due to the aformentioned issue of the clipping of the high wave slope probabilities around $s$=0 (see Fig. \ref{fig:examplespdf_vent2_9}), to the insufficient number of observations at very small wave tilts and to the  scatter of these observations, the peak of the distribution cannot be used for reliable fits of the parameters in Eq. (\ref{eq:defmssshape}). We therefore used another technique that does not involve the smallest slopes, based on the fact that, for $\nu>0$,
\be
g(A,\nu,s)=A s^\nu e^{-\demi s^2/m}
\ee
reaches its maximum for $s_{max}(\nu)=\sqrt{\nu m}$, where it takes the value $g_{max}(A,\nu)=A (\nu m)^{\nu/2} e^{-\nu/2}$,
i.e.:
\be
\log[g_{max}(A,\nu)]=\demi\nu\log(m)+\log(A)-\demi\nu(1-\log\nu) .
\ee
Hence, by finding $g_{max}$ for a range of values of $\nu$ and performing a linear fit of log($g_{max}$) using the above equation and floating $\log(A)$ and $\log(m)$, one can evaluate both $m$ and $A$. In practice, using $100$ values of $\nu$ in the range $0-1.5$  provides a very stable estimate, which is not impacted by the lack of some probability values for small slopes,  as shown by Fig. \ref{fig:figmssshape_fit2parametres}. Indeed, for the considered slow wind, the missing of high probability values ($\ge$22, see Fig. \ref{fig:examplespdf_vent2_9}) close to $s$=0 results in a bias of the obtained coefficient $a_0$, while estimating the MSS-shape with the aforementioned method properly restores the peak of the optimal Gaussian distribution.
 
\begin{figure}[H]\centering
\includegraphics[scale=0.28,angle=-0]{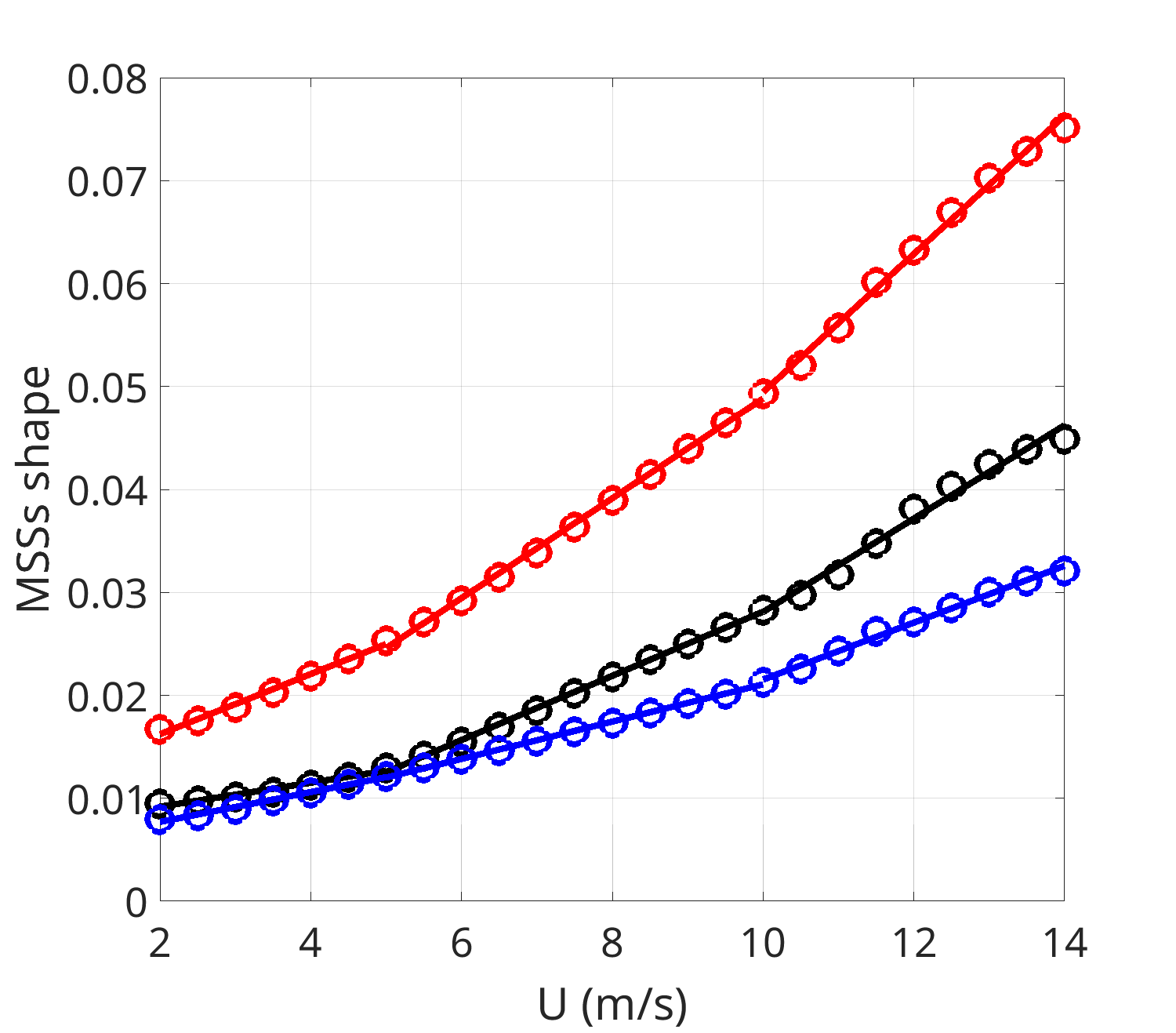}
\caption{MSSs-shape ($m_{ts}$ in red, $m_{us}$ in black, and $m_{cs}$ in blue) as a functions of wind speed only (circles), thus for "all $H_s$", and the corresponding linear fits using Eqs. (\ref{eq:fitmssshape}) (solid lines). \label{fig:fitmssshape}}
\end{figure}

%/home/guerin/Dropbox/COXMUNK_HARTMANN/IASI_HS_2025/PRG_HS/PRG_FIGURES_PAPIER/plot_example_calcul_mss_shape.m
This technique was applied to the IASI-retrieved probabilities, first binned in $U$ only, leading to the results in Fig. \ref{fig:fitmssshape}. The slightly non-linear evolutions of the directional and total MSSs-shape  with wind speed between 2 and 14 m/s can be well fitted with a 3 parts piecewise linear function:

\be\label{eq:fitmssshape}
\begin{split}
  m_{us}&=\alpha_{u}U+\beta_{u}\\
  m_{cs}&=\alpha_{c}U+\beta_{c}\\
   m_{ts}&=\alpha_{t}U+\beta_{t}
\end{split}
\ee

based on the coefficients given in Table \ref{table:mssshape} for each of the 3 wind speed ranges.

%/home/guerin/Dropbox/COXMUNK_HARTMANN/IASI_HS_2025/PRG_HS/PRG_FIGURES_PAPIER/plot_fitmss.m

%  \begin{table}
\begin{center}
  \begin{tabular}{| l | l | l | l | l |}
    \hline
$U_{10}$& 2-5 m/s & 5-10 m/s & 10-14 m/s \\
    \hline
    $1000\alpha_u$ &1.17 & 3.12 & 4.54 \\
    $1000\beta_u$ &6.85 &-3.05 &-17.30 \\
    \hline
      $1000\alpha_c$ &1.45 & 1.81 & 2.76 \\
      $1000\beta_c$ &4.8 & 2.96 & -6.10 \\
      \hline
        $1000\alpha_t$ &2.92 & 4.85 & 6.72 \\
    $1000\beta_t$ &10.40 &0.34 & -17.70 \\
\hline

  \end{tabular}
  \captionof{table}{Values of the coefficients of the partial linear laws in Eq. (\ref{eq:fitmssshape}) describing the evolution of the MSSs-shape with wind speed. $\beta$  is given in units of s/m.\label{table:mssshape}}
 %\end{table}
\end{center}

\begin{figure}[H]\centering
\includegraphics[scale=0.25,angle=-0]{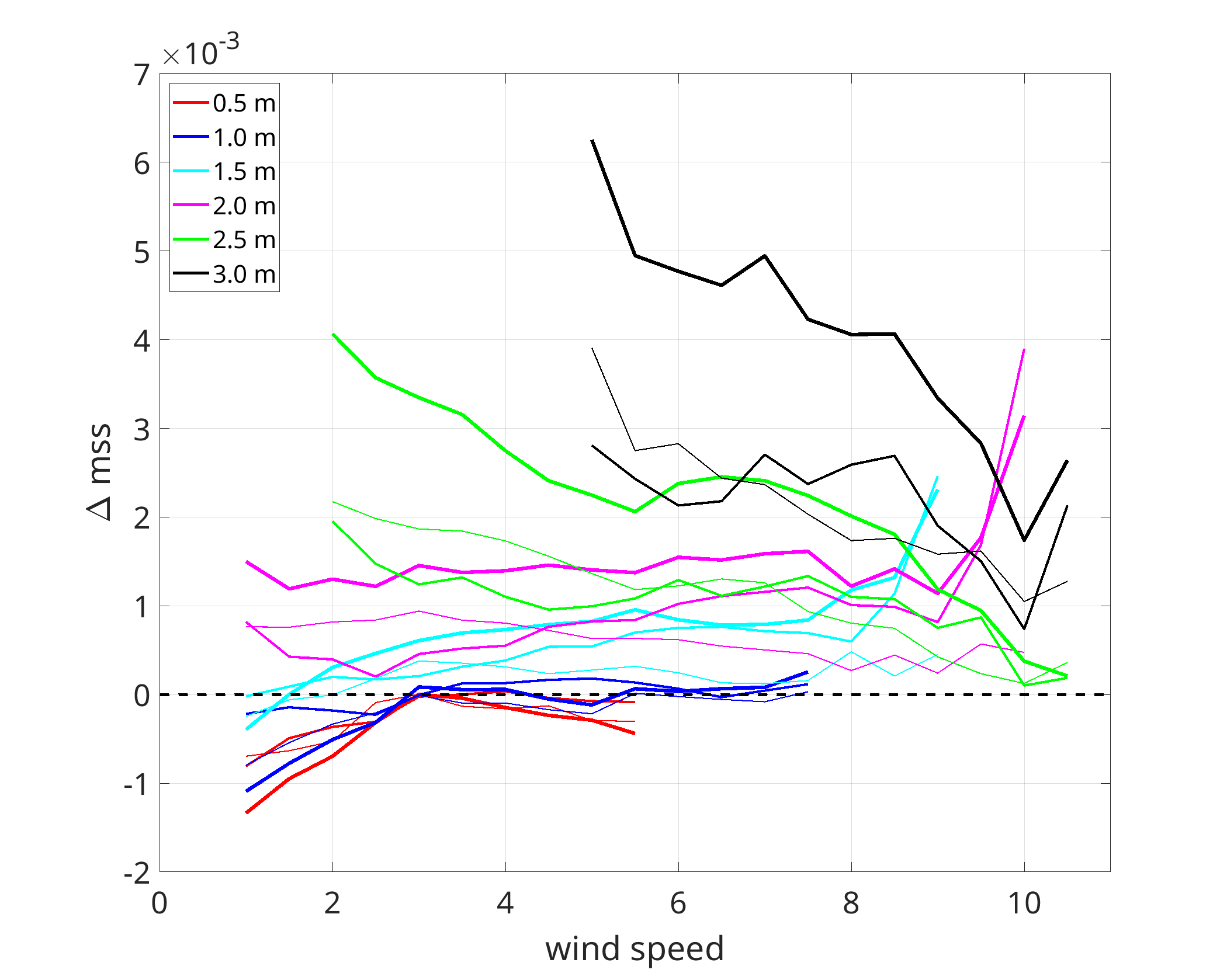}
\caption{Differences between the $H_s$ dependent MSSs-shape and the wind speed only relation from Eq. (\ref{eq:fitmssshape}) and Fig. \ref{fig:fitmssshape}. The thin, middle-thick and thick lines refer to the crosswind, upwind and total MSS-shape, respectively.} 
  \label{fig:dmssshape}
\end{figure}
%/home/guerin/Dropbox/COXMUNK_HARTMANN/IASI_HS_2025/PRG_HS/PRG_FIGURES_PAPIER/plot_mssshape.m

\begin{figure}[H]\centering
  \includegraphics[scale=0.25,angle=-0]{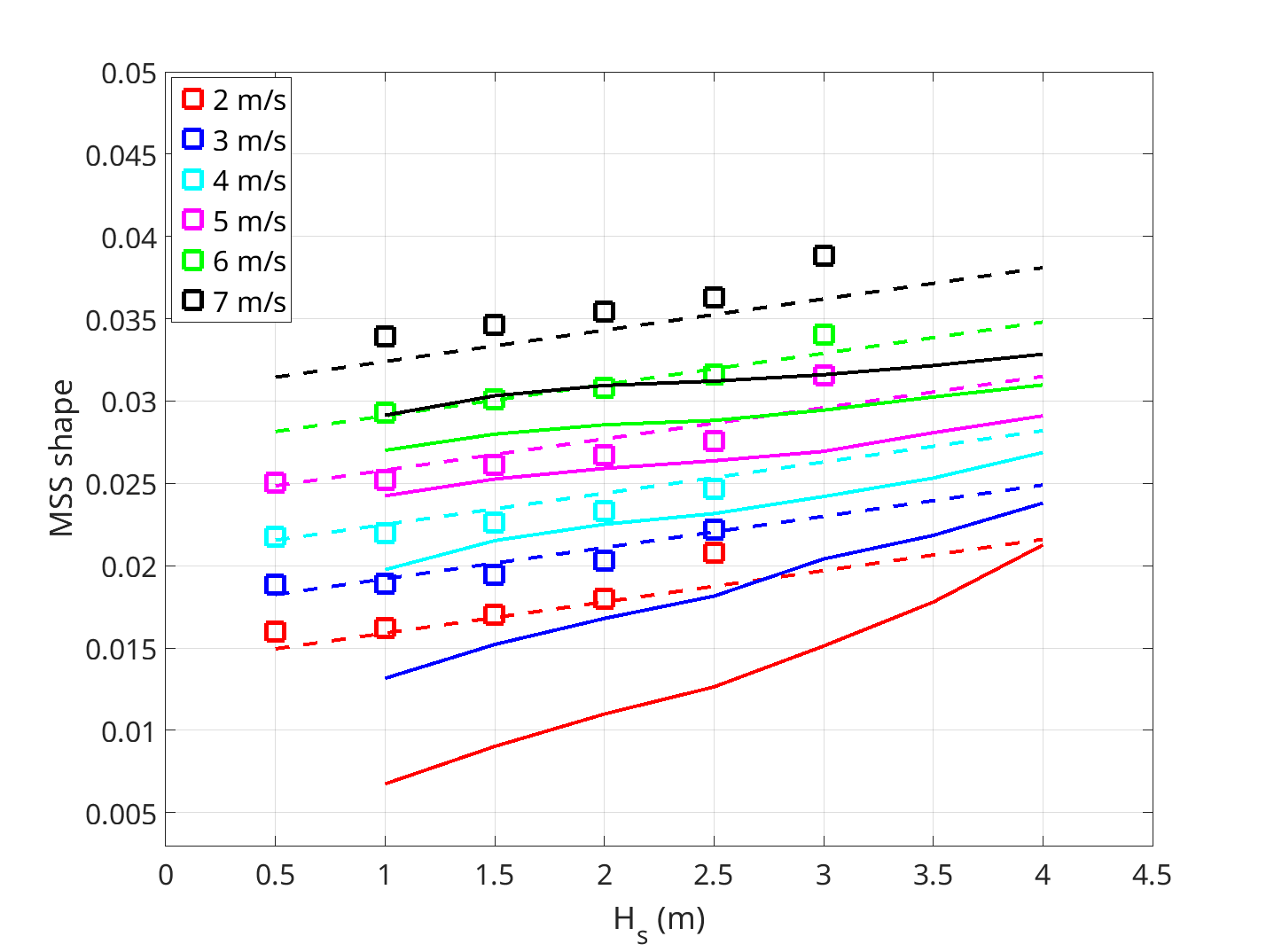}
  \caption{Total MSS-shape $m_{ts}$ as a function of $H_s$ for various wind speed, as estimated from IASI data (squares), and from Ka-band scatterometer observations \cite{Nouguier_GRSL16} (solid lines). The IASI data are limited to $H_s\leq 2.5$ m for slow winds due to lack of data for higher values. The dashed lines are values obtained from Eq. (\ref{eq:MSS_shape})}
    \label{fig:companouguier}
\end{figure}

Then, we considered the MSSs-shape when binning the probabilities in both  $U$ and $H_s$, for all cases where the number of data was sufficient to obtain an accurate azimuthal expansion through Eq. (\ref{eq:azimfit}) (the required minimal number of data was empirically set to 2 10$^{5}$). Figure \ref{fig:dmssshape} displays the results  for  $H_s$ between $0.5$ m and $3$ m, after subtraction of the "all" $H_s$ values in Fig. \ref{fig:fitmssshape}.  It shows that  the MSSs-shape increases with $H_s$ and, for $U$<7 m/s, the variations with $U$ and $H_s$ of the total MSS-shape can be approximated by:
 \be
 m_{ts}=0.0019 H_s +0.0033 U+0.0074,
 \label{eq:MSS_shape}
 \ee
while no such simple relation could be found for the directional MSSs-shape and the larger wind speeds. This trend with both $U$ and $H_s$ is consistent with the results obtained by \cite{Nouguier_GRSL16} using a near-nadir Ka-band scatterometer, as can be seen in Fig. \ref{fig:companouguier}, except for small values of both $U$ and $H_s$. The differences in these cases are likely due to the fact that the backscattering process in the microwave domain filters out the smallest roughness scales. This is not the case for observations in the optical domain, which thus lead to larger MSSs, particularly for slow winds where the sea surface roughness is small. Note that the observed dependences on $H_s$ and $U$, as parameterized by the preceding equation, are weaker than those ($m_{ts}=0.003H_s+0.0052U$) retrieved from altimetric Ka/Ku band data \cite{Guerin_GRS17}. The reason for this discrepancy might be the crude assumption that the long-wave MSS estimated in this last study is simply proportional to $H_s$.

%/home/guerin/Dropbox/COXMUNK_HARTMANN/IASI_HS_2025/PRG_HS/PRG_FIGURES_PAPIER/compaNouguier.m

\section{The MSSs}\label{sec:mss}

\subsection{Determination of the MSSs}

While the MSSs-shape are the primary quantities that can be estimated from radar or optical remote sensing, it is only the MSSs that are intrinsic quantities of the sea-surface slope distribution. Assuming the validity of Eq. (\ref{eq:azimfit}), the directional and total  MSSs  can be expressed in terms of the third moments of the even azimuthal coefficients:
\be\label{eq:defmss2}
\begin{split}
  m_u&=\pi \int_{0}^{+\infty}s^3[a_0(s)+ a_2(s)/2]ds,\\
  m_c&=\pi \int_{0}^{+\infty}s^3[a_0(s)- a_2(s)/2]ds,\\
  m_t&=2\pi \int_{0}^{+\infty}s^3 a_0(s)ds.
\end{split}
\ee
As discussed in Sec. 2.2, the IASI-retrieved probabilities for large $s$ values are strongly contaminated by scatter and potentially by parasitic reflections of solar photons (possibly by foam, breaking waves and aerosols).  As the third order moments of the distribution are significantly sensitive to the tail of the distribution, this  a priori hinders the calculation of the MSSs from Eq. (\ref{eq:defmss2}). To circumvent this problem, \cite{Cox54} used an extrapolation based on the values of the PDF for wave slopes smaller than some limiting threshold. A priori, the most straightforward way to estimate the MSSs and the GC coefficients  is then to find the best fit of the distribution of Eq. (\ref{eq:GC}) to the IASI-derived probabilities for small and moderate values of the wave slope. However,  this turns out to be a highly unstable and inaccurate procedure due to ill-defined minima in the cost function. In our previous study \cite{Guerin_RSE23}, which disregarded the influence of $H_s$, we thus used another method based on a coarse determination of the mean noise level and an approximate expression of the MSSs based on lower moments of the distribution. Here, we propose a new technique which is less sensitive to the  the scatter and eventual bias of the distribution in the tail, and does not require any floating procedure.
Following the idea of a Gram-Charlier cumulant expansion or Edgeworth series, we assume that the omnidirectional coefficient $a_{0}(s)$ can be well represented with a fourth-order polynomial correction to the Gaussian fit that was employed above to determine the MSS-shape and thus write:
\be\label{eq:fit_exppol}
a_{0}(s) =\frac{1}{2\pi m_{0s}}e^{-\demi s^2/m_{0s}^2}\times \left(P_0+\frac{(1-P_0)}{8}\frac{s^4}{m_{0s}^4}\right), 
\ee
where $m_{0s}$ is the omnidirectional MSS-shape, see Eq. (\ref{eq:defmssshape}),  and $P_0$ is some coefficient of value close to unity that accounts for the deviation from the Gaussian distribution. Note that there are no quadratic terms $s^2$ in this polynomial correction, for otherwise this would change the MSS-shape. The factor $(1-P_0)/8$ is imposed by the normalization condition of the PDF, see Eq.  (\ref{eq:normalization}). Such a representation in terms of a ``Gaussian times Polynomial'' (henceforth referred to as ``G$\times$P'') is in the spirit of a cumulant expansion but it is more adapted to describe the distribution when only the MSS-shape is known. We found that an excellent G$\times$P fit generally holds for wave slopes up to at least twice the standard deviation associated to the isotropic MSS-shape, that is for $s\leq 2\sqrt{m_{0s}}$. This representation of $a_0$ enables to correct for a potential overestimation in the tail of the distribution. Note that the polynomial correction is restricted to fourth-order to limit the number of floated parameters and avoid an unstable estimation.

\begin{figure}[H]\centering
  \includegraphics[scale=0.3,angle=-0]{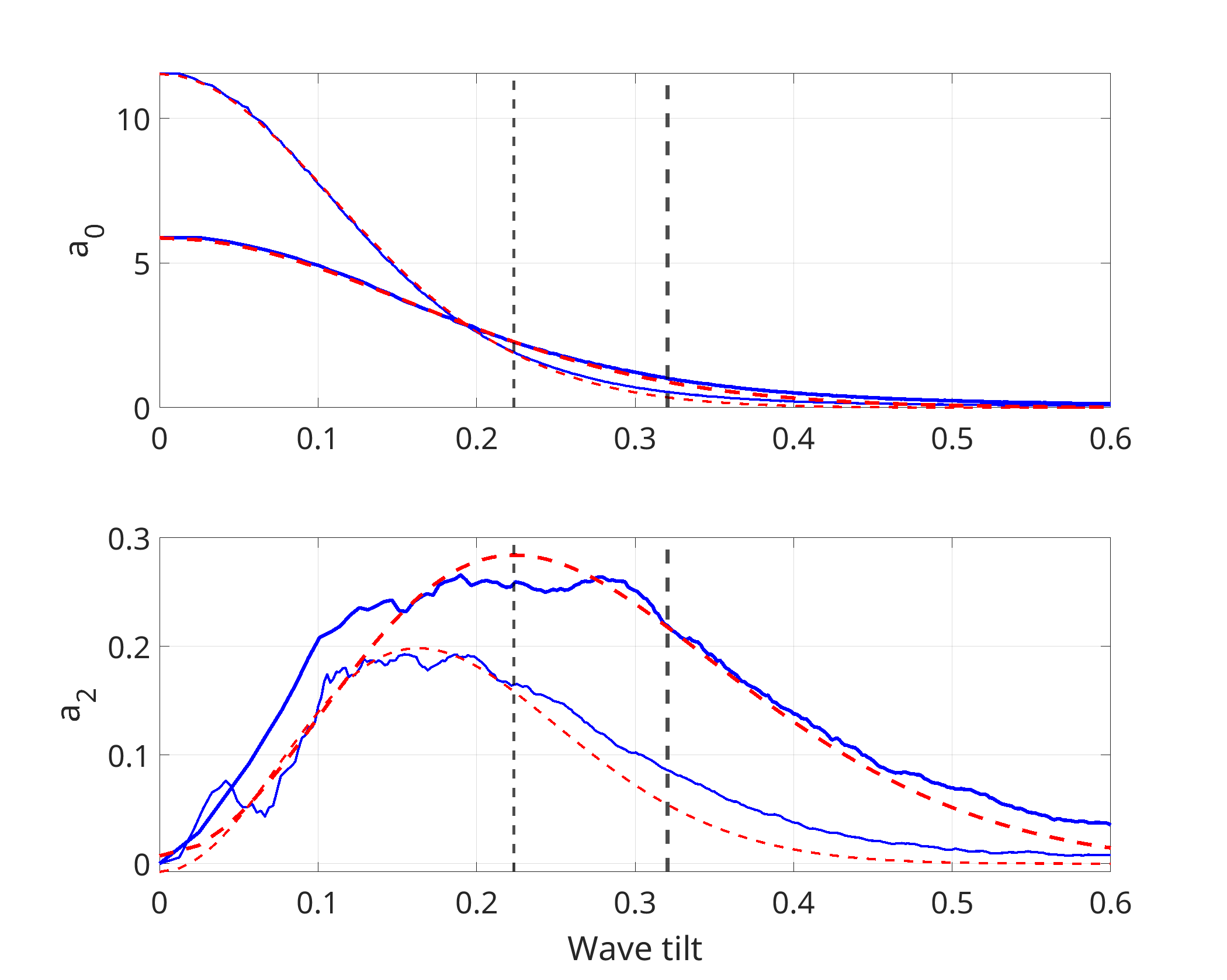}
  \caption{Azimuthal coefficients $a_0$ (top) and $a_2$ (bottom) as  functions of the wave tilt $s$ (solid blue lines) and  their G$\times$P approximations through Eqs. (\ref{eq:fit_exppol}) and (\ref{eq:a0+a2}) (dashed red lines). The results for ($U$=5 m/s ; $H_s$=1 m) and ($U$=10 m/s ; $H_s$=3 m) are shown by thin and thick lines, respectively. The  vertical dashed black lines show the values of $2\sqrt{m_{0s}(U)}.$}
  \label{fig:exampleGPfit}
\end{figure}

We assumed a similar G$\times$P form for the azimuthal coefficient $a_2$ starting from the directional MSSs-shape, and thus represented the upwind and crosswind distributions using the functions:
\be
\begin{split}\label{eq:a0+a2}
 a_{0}(s)+a_2(s)&=\frac{P_0}{2\pi m_{0s}}e^{-\demi s^2/m_{us}^2}\times \left(1+\gamma\frac{s^4}{m_{us}^2}\right) ,\\
 a_{0}(s)-a_2(s)&=\frac{P_0}{2\pi m_{0s}}e^{-\demi s^2/m_{cs}^2}\times \left(1+\gamma'\frac{s^4}{m_{cs}^2}\right),
\end{split}
\ee
which introduce the  coefficients $\gamma,\gamma'$ which must be determined. Once the latter are known, the corrected coefficient $a_2$ is obtained by substraction of the two equations in Eq. (\ref{eq:a0+a2}). Figure \ref{fig:exampleGPfit} shows examples of the  G$\times$P fits of $a_0(s)$ and $a_2(s)$ for $U$=5 m/s with $H_s$=1 m and $U$=10 m/s with $H_s$=3 m. As can be seen the agreement is excellent for $a_0(s)$ for slopes $s$ smaller than $2\sqrt{m_{0s}}$ but less satisfactory for  $a_2(s)$. Note that the latter, which is one order of magnitude smaller than $a_0(s)$, is more difficult to estimate accurately, as seen through the more irregular aspect of the plot. A better fit would require larger probability datasets than those currently available. Replacing $a_{0}$ and $a_2$ by their respective G$\times$P approximations then enables to determine the MSSs using Eq. (\ref{eq:defmss2}).

%/home/guerin/Dropbox/COXMUNK_HARTMANN/IASI_HS_2025/PRG_HS/PRG_FIGURES_PAPIER/plotfit_a0a2.m

\subsection{The dependences of the MSSs on wind speed}

\begin{figure}[H]\centering
\includegraphics[scale=0.28,angle=-0]{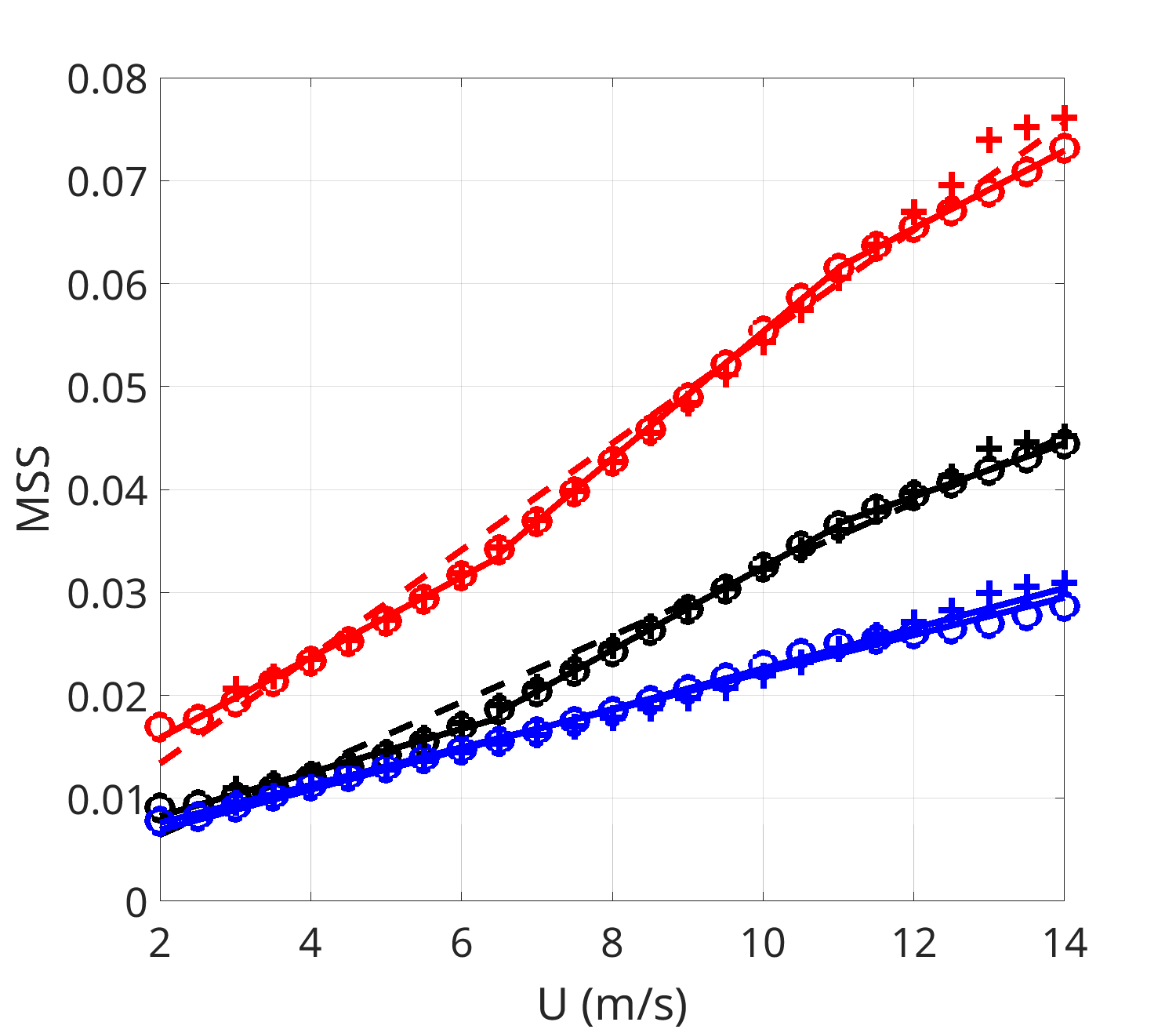}
\caption{MSSs ($m_{t}$ in red, $m_{u}$ in black, and $m_{c}$ in blue) as functions of wind speed only (circles), thus for "all $H_s$", and the corresponding linear fits using Eqs.  (\ref{eq:fitmss} (solid lines). The parametrizations of \cite{Cox54}, in Eq. (\ref{loiCM}), are given for references with dashed lines.\label{fig:msstot}}
\end{figure}

Figure \ref{fig:msstot} shows the obtained total and directional MSSs as functions of $U$ only, regardless of $H_s$, henceforth referred to as the ``All Hs'' MSSs. The presently derived values are in very good agreement with those obtained, using a different method, in our previous study \cite{Guerin_RSE23}, with differences smaller than about $2\%$ in the upwind direction and $4\%$ in the crosswind direction. They confirm a small negative deviation of the upwind values, $m_u$, from the linear relationship in Eq. (\ref{loiCM}) at intermediate wind speeds (4-9 m/s). Since, on the other hand, the crosswind MSS, $m_c$, remains very close to the linear relationship, the difference between the directional MSSs, $m_{u}-m_{c}$, which is a proxy for the sea-surface directivity, grows at a slightly slower rate than predicted by  Eq. (\ref{loiCM}). An accurate representation of the upwind and total MSS can be obtained with a 3 part piecewise linear fit for $U$ between 2 and 14 m/s:
﻿
\be\label{eq:fitmss}
\begin{split}
  m_{u}&=\alpha_{u}U+\beta_{u} ,\\
   m_{c}&=\alpha_{c}U+\beta_{c} ,\\
   m_{t}&=\alpha_{t}U+\beta_{t} ,
\end{split}
\ee

with the coefficients given in Table \ref{eq:tablemss}.
  
% \begin{table}
\begin{center}
  \begin{tabular}{| l | l | l | l | l |}
    \hline
$U_{10}$& 2-6.5 m/s & 6.5-11 m/s & 11-14 m/s \\
    \hline
    $1000\alpha_u$ &2.12 & 4.02 & 2.59 \\
    $1000\beta_u$ &4.03 &-7.76 &8.24 \\
  \hline
        $1000\alpha_c$ &1.82 & 1.82 & 1.82 \\
    $1000\beta_c$ & 3.97 & 3.97 & 3.97 \\
        \hline
      $1000\alpha_t$ &3.91 & 6.15 & 3.78 \\
      $1000\beta_t$ &8.03 & -6.21 & 20.00 \\
 
\hline
  \end{tabular}
 \captionof{table}{Values of the coefficients of the partial linear laws, in Eq. (\ref{eq:fitmss}) describing the evolutions of the MSSs with wind speed. $\beta$  is given in units of s/m.   \label{eq:tablemss}}
%  \end{table}
\end{center}

%/home/guerin/Dropbox/COXMUNK_HARTMANN/IASI_HS_2025/PRG_HS/PRG_FIGURES_PAPIER/plot_fitmss.m

Analysis shows that the differences between our MSSs values and Eq. (\ref{loiCM}) do not indicate discrepancies betwwen our results and those obtained by Cox and Munk. Indeed, a direct comparison between our data and the original individual values provided in Table 1 of \cite{Cox56}, from which Eq. (\ref{loiCM}) was determined, shows, thanks to the scatter of the latter, that all results are fully compatible, as also shown by Fig. 15 of \cite{Guerin_RSE23}.

\subsection{The dependences of the MSSs on wave height}
\begin{figure}[H]\centering
\includegraphics[scale=0.22,angle=-0]{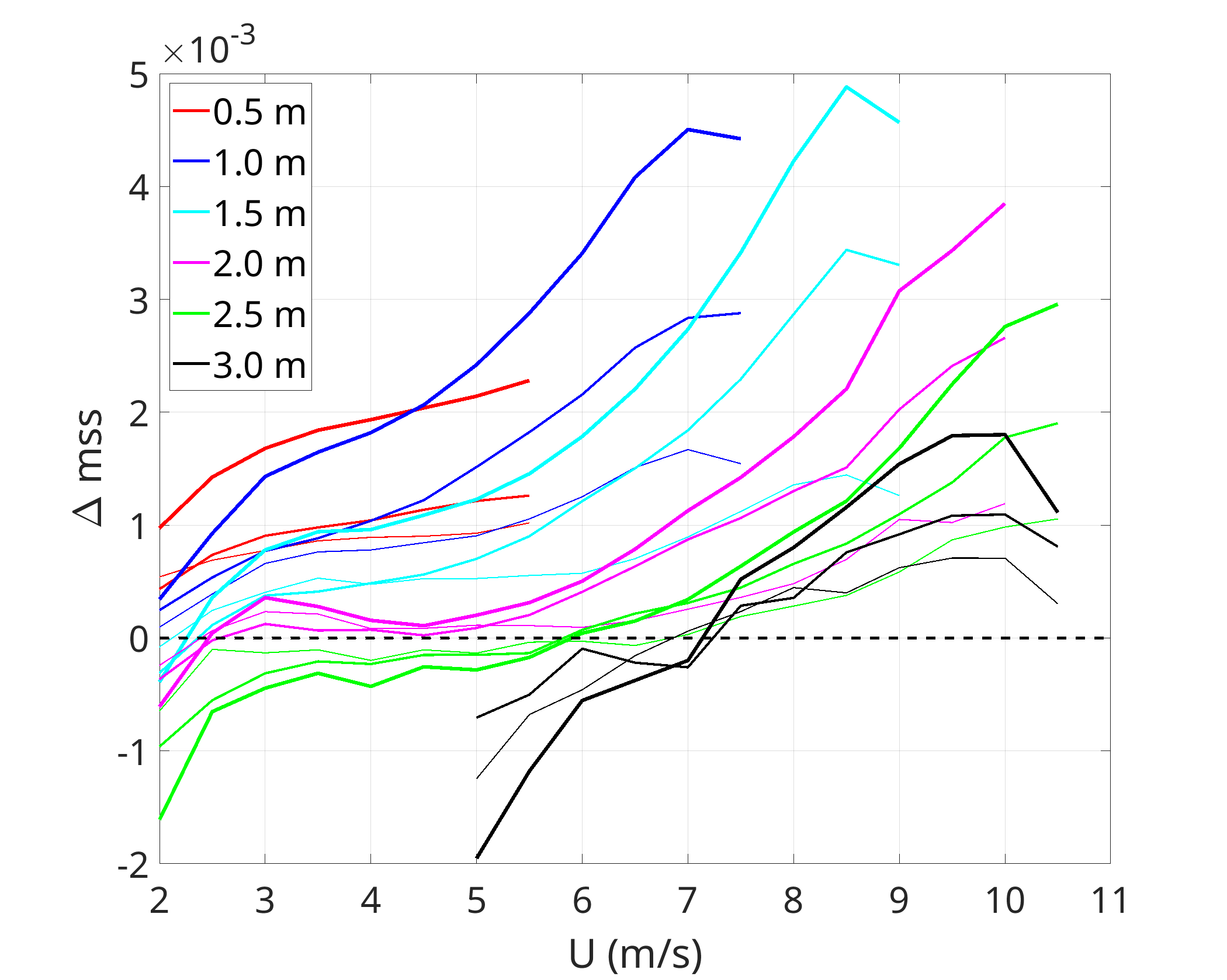}
\caption{Differences between the $H_s$ dependent MSSs and the wind speed only relation from Eq. (\ref{eq:fitmss}) and Figure \ref{fig:msstot}. The thin, middle-thick and thick lines refer to the crosswind, upwind and total MSS, respectively.}
  \label{fig:dmss}
\end{figure}

We now focus on the dependences of the MSSs on both $U$ and $H_s$, for which a smaller range of wind speed could be investigated (2-10.5 m/s), when compared to the previous analysis, due to the reduction of the number of data points after sorting in $H_s$. To highlight the dependences of the MSSs on $H_s$, we substracted the results obtained for the ``All Hs'' MSSs, plotted in Fig.  \ref{fig:msstot}, leading to the results in  Figs. \ref{fig:dmss}. The latter show that  $H_s$ has a small impact with the outcome that the MSSs decrease slightly as $H_s$ increases, a trend which is opposite to that of the MSS-shape (see Fig. \ref{fig:companouguier}). This result is rather counter-intuitive  as one would expect that increasing long-wave components, i.e. the significant wave height, would also increase the MSSs. However, there is no evidence in the literature to support this a priori assumption.  Indeed \cite{Hwang_JGR88} have observed, from a platform-mounted optical device, that the presence of a swell system can either increase or reduce the MSSs, while \cite{hauser_JGR09reply} have, in their airborne radar measurement, found only a small (if any) impact of swell on the crosswind MSS. On the other hand, it has been suggested for a while, on the basis of wind-wave tank measurements \cite{phillips_JFM74,donelan_APL87,chen_JPO00}, that the presence of long mechanically-generated waves has a damping effect on small-scale surface roughness, which is the major contributor to the MSSs. Various mechanisms have been invoked to explain this observation, such as detuning of resonant wave interactions \cite{donelan_APL87}, the reduction of turbulent wind stress by a sheltering mechanism \cite{chen_JPO00}, or the turbulence generated by breaking \cite{ermakov_RemSen20}. Another possible cause is the role played by atmospheric conditions \cite{Hwang_JGR88,shaw}, discussed in Appendix A.5. A last plausible factor is the increased subgrid spatial variability of the local MSSs at the early stage of sea-state development, for which $H_s$ is in general smaller. In that case, the compounding of the random process underlying the wave-slope fields results in an increased equivalent MSS and kurtosis, as was first suggested by \cite{chapron_JGR00}. At this stage, we do not have a definitive answer to this point and leave it to further analysis and confirmation by other field or remote-sensing studies.

%/home/guerin/Dropbox/COXMUNK_HARTMANN/IASI_HS_2025/PRG_HS/PRG_FIGURES_PAPIER/plot_allmss.m + plot_fitmss.m

\begin{figure}[H]\centering
\includegraphics[scale=0.22,angle=-0]{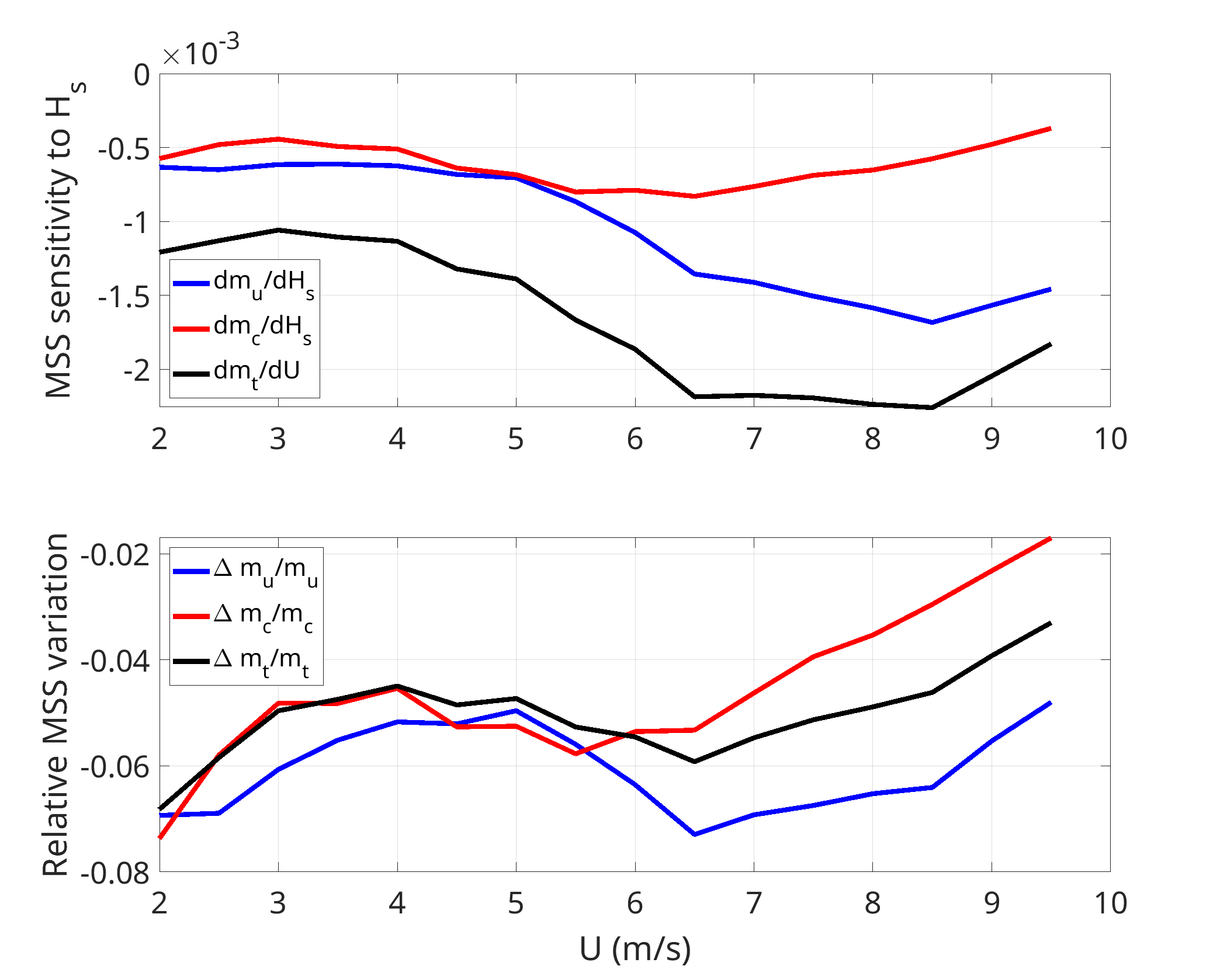}
\caption{Top: Sensitivities (see text) of the directional and total MSS to $H_s$ as a function of wind speed. Bottom: relative variation of MSS induced by a $1$ m change of $H_s$.\label{fig:fig_dmssdHs}}
\end{figure}

    The sensitivity of the total MSS to $H_s$ at a given wind speed, $dm_t/dH_s$, was estimated by averaging the MSS increments $[m_t(U,(n+1)\Delta H_s)- m_t(U,n\Delta H_s)]/\Delta H_s$ for the available values (that is, $\Delta H_s=0.5$ and $n=1,..,6$), with similar equations for the sensitivity of the directional MSSs,  $dm_u/dH_s$ and $dm_c/dH_s$. The results in Fig. \ref{fig:fig_dmssdHs} show that these quantities slightly decrease with wind speed, with average values of:
    \be
    \begin{split}
      \frac{dm_u}{dH_s}&\simeq -0.001\ m^{-1},\\
      \frac{dm_c}{dH_s}&\simeq -0.0006,\ m^{-1},\\
      \frac{dm_t}{dH_s}&\simeq -0.0016\ m^{-1},
      \end{split}
      \ee
 for $U$ between 2 and 9 m/s. To give a better idea of the relative importances of $U$ and $H_s$ in the determination of the MSSs, let us mention that a $1$ m change of  $H_s$ at small wind speed ($2\leq U\leq 6$ m/s) induces a relative variation of the MSS of about $5\%$,  while a 1 m/s change of $U$ induces a change of about $15\%$.  Given the small range of variation of $H_s$ (about 3 m in a $\pm 2$std interval, see Fig. \ref{fig:statHs})), this shows that this parameter can account for up to $15\%$ of the MSS variation for slow winds.

    %/home/guerin/Dropbox/COXMUNK_HARTMANN/IASI_HS_2025/PRG_HS/PRG_FIGURES_PAPIER/plot_allmss.m

\subsection{The directionality of the MSSs}    

\begin{figure}[H]\centering
	  \includegraphics[scale=0.22,angle=-0]{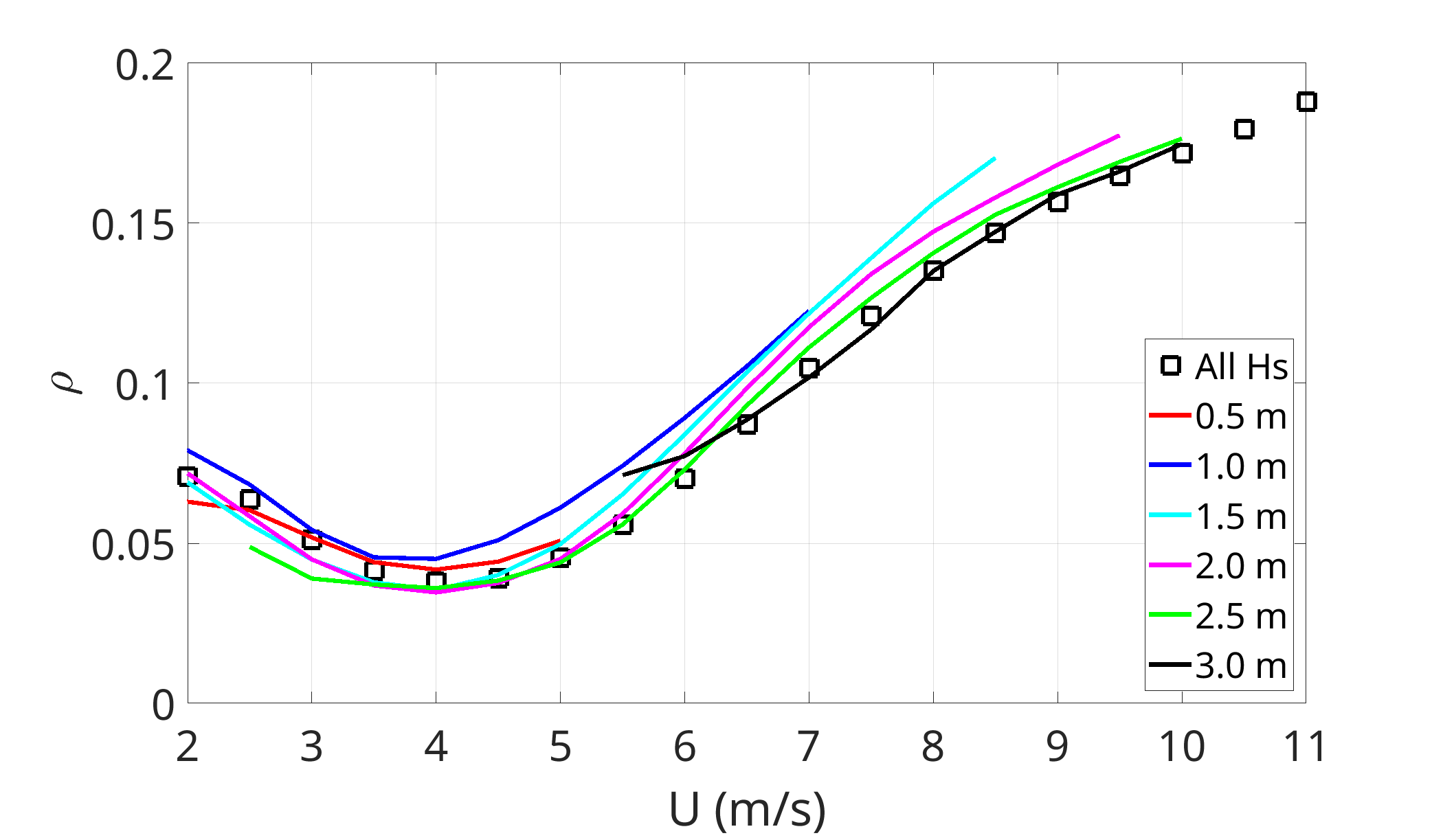}
\caption{Directivity constrast $\rho$ as a function of wind speed for various significant wave heights. \label{fig:constrastmssdir}}
\end{figure}

The relative amplitudes of the upwind and crosswind MSSs can be quantified by their contrast coefficient:
\be
\rho=\frac{m_u-m_c}{m_u+m_c} ,
\ee
whose variations with $U$ and $H_s$ are displayed in Fig. \ref{fig:constrastmssdir}. As shown, $\rho$ varies from about 0.05 to 0.15, with a minimum of directivity obtained at 4-5 m/s and a maximum at the largest wind speed. Increasing $H_s$ reduces the directivity, with variations from 0.1 (small wind speed) to 0.4 (large wind speed).This confirms the trend pointed out using the UCA coefficient (see discussion of Fig.  \ref{fig:UDA_UCA}) namely an increased directivity of wave slopes as $H_s$  decreases.

%/home/guerin/Dropbox/COXMUNK_HARTMANN/IASI_HS_2025/PRG_HS/PRG_FIGURES_PAPIER/plot_allmss.m

\subsection{Skewness and kurtosis}\label{subsec:skewness}

\begin{figure}[H]\centering
\includegraphics[scale=0.3,angle=-0]{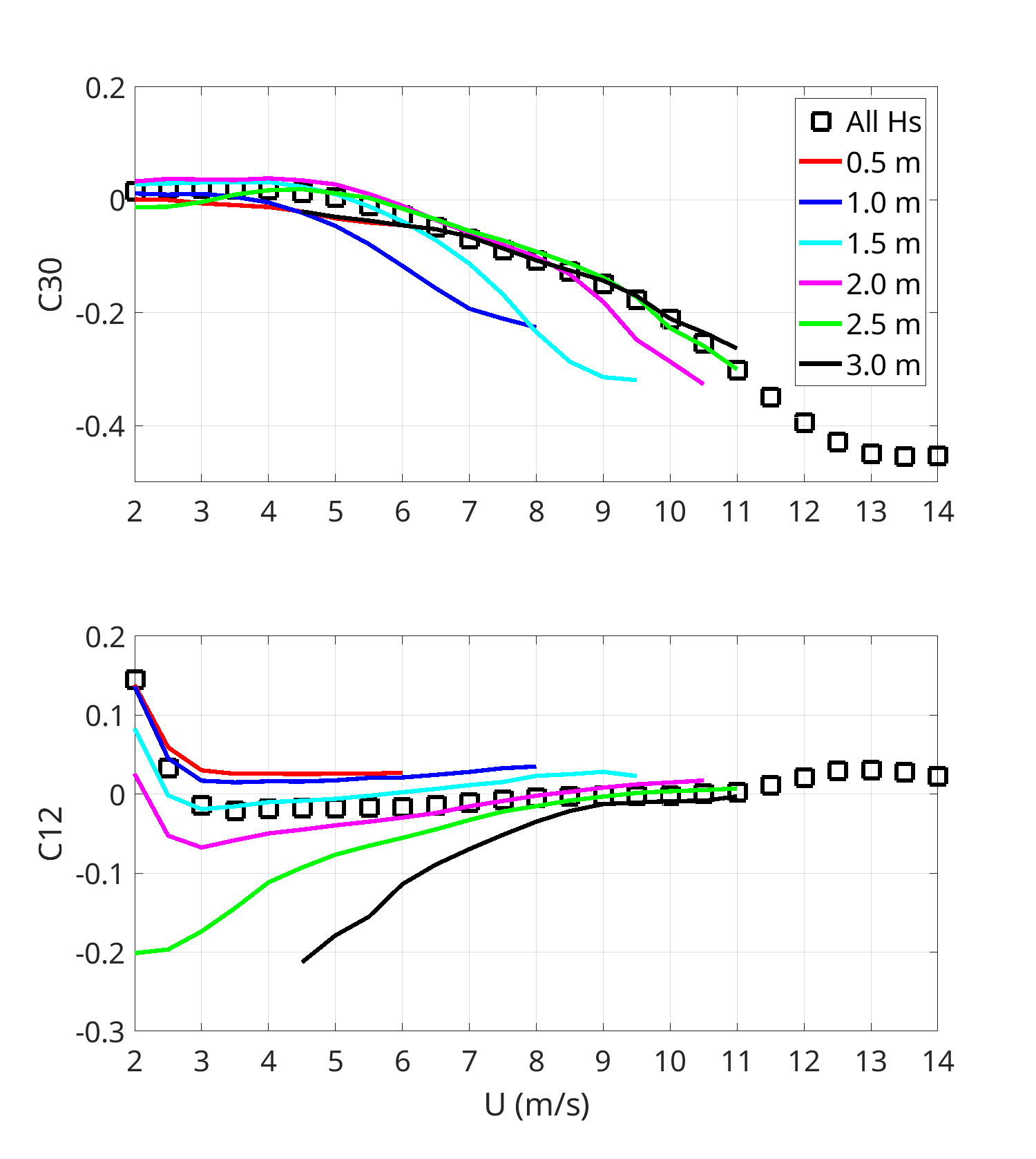}
\caption{Skewness coefficients versus wind speed for various significant wave heights. \label{fig:GCa}}
\end{figure}
%/home/guerin/Dropbox/COXMUNK_HARTMANN/IASI_HS_2025/PRG_HS/PRG_FIGURES_PAPIER/plot_allmss.m

\begin{figure}[H]\centering
\includegraphics[scale=0.3,angle=-0]{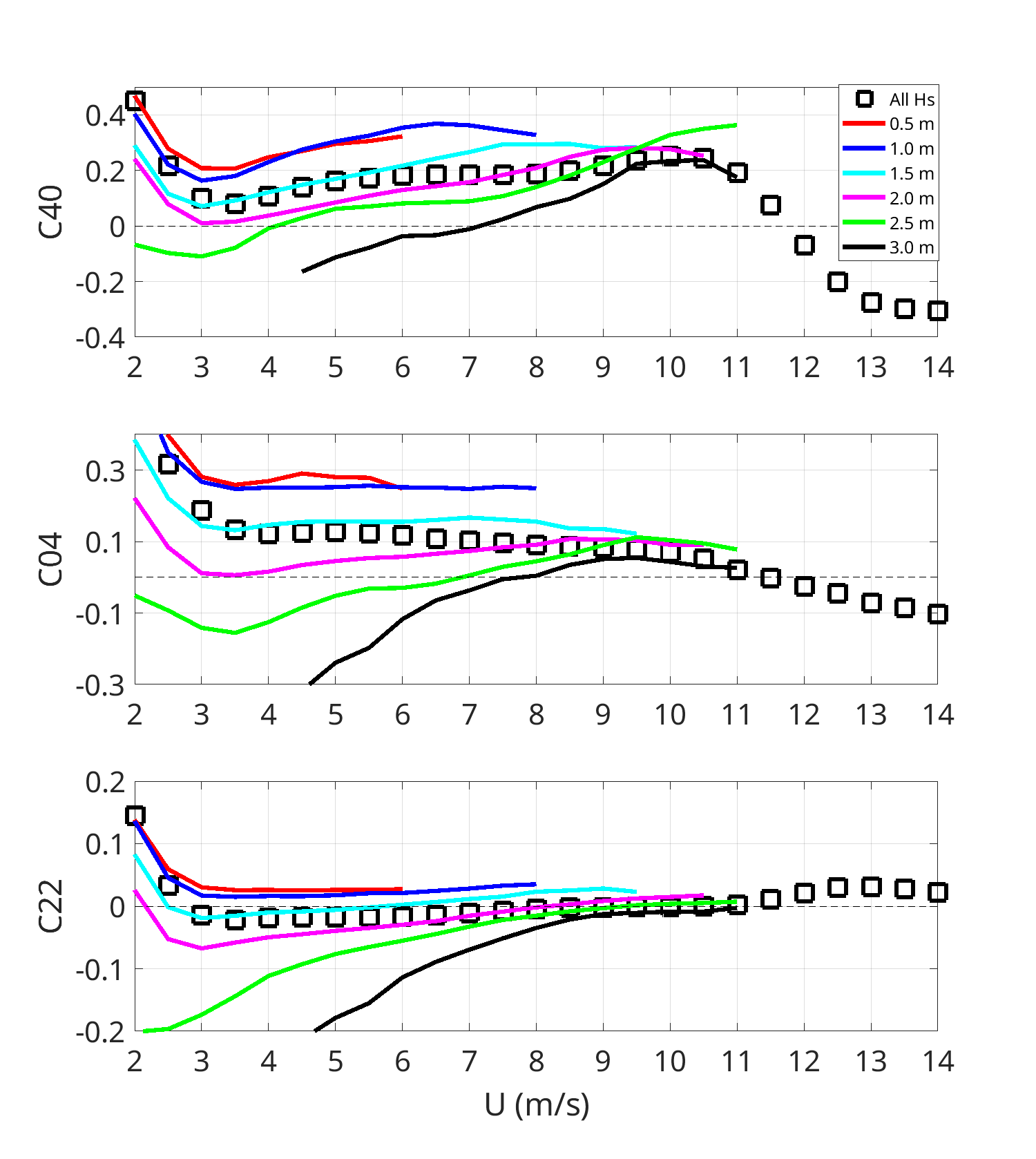}
\caption{Kurtosis coefficients as a function of wind speed for various significant wave height \label{fig:GC}}
\end{figure}
%/home/guerin/Dropbox/COXMUNK_HARTMANN/IASI_HS_2025/PRG_HS/PRG_FIGURES_PAPIER/plot_allmss.m

Now that the directional MSSs have been determined, the skewness ($C_{30},C_{12}$) and kurtosis ($C_{40},C_{04},C_{22}$) coefficients of the Gram-Charlier expansion can be estimated by optimizing the match between Eq. (\ref{eq:GC}) and the symmetric, Eq. (\ref{eq:p_sym}), and antisymmetric, $p_{as}(s,\theta)=a_1(s)\cos(\theta)$), parts of the IASI-derived wave-slope probabilities. Due to the aforementioned issues affecting the probabilities close to $s$=0 at small wind speed, we selected by careful inspection of the PDFs a minimum wind speed of 2 m/s above which the estimation of the GC coefficients is reliable. Figures \ref{fig:GCa} and \ref{fig:GC} show the obtained results. $C_{30}$ is negative and monotonically decreases as $U$ increases, which indicates a growing asymmetry in the upwind plane, a fact  noted in earlier studies \cite{Breon,Guerin_RSE23}. It is also seen that its magnitude is enhanced by smaller $H_s$ at moderate wind speed (6-9 m/s), which is a fully original result. The kurtosis coefficients also show a sensitivity to both $U$ and $H_s$. When considered as functions of wind speed only ('All $H_s$' results), the upwind ($C_{40}$) and crosswind  ($C_{04}$) kurtosis coefficients evolve slowly from positive values for $U=4-11$ m/s, to negative ones for $U\geq 12$ m/s; the cross-kurtosis coefficient $C_{22}$ is close to zero. The upwind coefficient is $C_{40}\simeq 0.1-0.25$ at moderate wind speed, thus consistent with the mean value $C_{40}\simeq 0.23$ first estimated by \cite{Cox54} but smaller than that $C_{40}\simeq 0.4$ found by \cite{Breon}. Note that it is also slightly smaller than the values $\simeq 0.16-0.32$ found in our former study \cite{Guerin_RSE23}, a discrepancy likely due to the small differences in the estimation of the MSSs. Indeed, the calculation of the kurtosis coefficients is extremely sensitive to the input values of the MSSs, which makes their estimation very challenging. It can be shown that the variation of the various kurtosis coefficients induced by a variation of $\pm 1\%$ of the MSSs is of the order of $20\%$. The skewness coefficients, in contrast, are very robust to the estimation of the MSSs and thus probably more accurate that the kurtosis coefficients. The crosswind kurtosis coefficient is $C_{04}\simeq 0.05-0.125$, which is smaller than the mean value  $C_{04}\simeq 0.4$ found by both \cite{Cox54} and \cite{Breon}. While the above mentioned findings are essentially consistent with those of previous studies, much more original is the fact that, when considered as functions of $H_s$, the kurtosis coefficients undergo strong variations, especially at small wind speeds. They are seen to be decreasing functions of $H_s$, following the same trend observed for the MSSs (see Fig. \ref{fig:dmss}). This is consistent with the ratio of the MSS to MSS-shape being a growing function of the kurtosis coefficients, see Eq. (27) of \cite{Guerin_RSE23}). It is interesting to note that the variation of the kurtosis coefficients is essentially sensitive to sea-surface roughness  at large spatial scales (through $H_s$) and not small scales. This statement is supported by the fact that it was shown by \cite{Cox54} that spreading oil significantly damps the small structures and MSSs but does not change the kurtosis.

\section{Large wave tilts and large wind speeds}
\begin{figure}[H]\centering
  \includegraphics[scale=0.25,angle=-0]{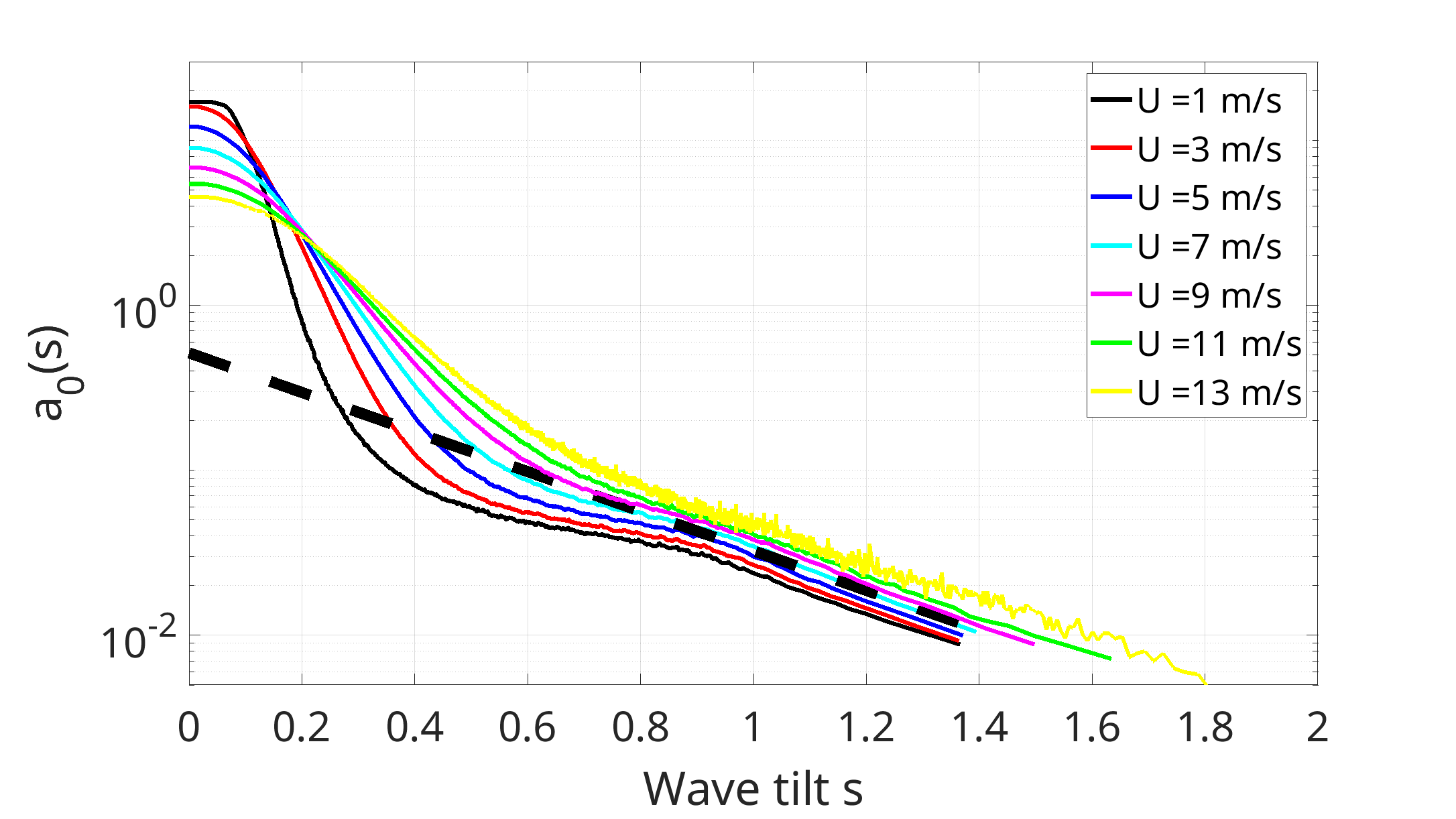}
  \caption{Omnidirectional probabilities (averaged over all azimuth angles of the wind direction and $H_s$ values), $a_0(s,U)$, versus the wave slope $s$ for various wind speed. The thick  dashed black line corresponds to $0.39\times 10^{-1.2s}$. \label{fig:tail}}   \end{figure}

The signal-to-noise ratio and quality of the IASI data enables to investigate the probabilities of wave-slope with  tilts larger than those investigated in previous studies.  However, due to a reduced number of data and a larger scatter of the IASI-retrieved data in the tail (for large wave-slopes, see Fig. \ref{fig:examplespdf_vent2_9}), we restricted the analysis to the sole effect of wind speed. Figure \ref{fig:tail} shows the omnidirectional PDF $a_0(s)$ (that is, averaged over all azimuth angles of the wind direction within various bins of wave slope) for various wind speeds.  While it is nearly Gaussian for moderate wave tilts $s$, this well known behavior breaks down in the tail, for $s$ greater than about 0.4 and 0.6 for $U$=2 and 9 m/s, respectively. Here, there seems to be a universal exponential decrease with $s$, with a quasi wind-independent decay rate, which can be represented by:
\begin{equation}
a_0(s,U)\simeq (0.021\ U+0.244)\ 10^{-1.2s}
\label{exptail}
\end{equation}

This breakdown of the Gaussian behavior for large slopes is qualitatively consistent with those displayed in Fig. 4 of \cite{Breon}, also obtained from space-borne observations in the optical domain,  with air-borne Ka and Ku band radar samplings of the ocean surface \cite{Vandemark_JPO04}, and with the analysis of high resolution camera snapshots in a  wave tank \cite{caulliez_JGR12}. The occurence of large wave slopes, which is much more frequent  than that predicted by a near-Gaussian distribution, is generally attributed to unstable steep waves at the onset of breaking, or to breaking waves themselves as shown in \cite{Voronovich}. Note that a test, described in Appendix A.2, in which nighttime IASI spectra were treated as if it were daytime, showed that the bias on the retrieved wave-slope probabilities (which should be zero) is typically of -0.005 with a RMS of 0.035. This implies that the values for large tilts in Fig. \ref{fig:tail}, which are all greater than 10$^{-2}$,  are meaningful, not due to a retrieval bias, and do result from solar photons reaching the instrument. Let us however mention that we cannot exclude that processes other than reflection on clean sea surfaces may redirect sun radiation toward the IASI instrument. These, which may include light scatterings by white caps, foam, and marine aerosols, would result in a signal wrongly attributed to a sea-surface slope and, because they are much less directional than specular reflection, they would, in particular, artificially enhance the tail of the PDF. This issue is discussed in more detail in Appendix A.3.

%/home/guerin/Dropbox/COXMUNK_HARTMANN/IASI_HS_2025/PRG_HS/PRG_FIGURES_PAPIER/plot_tail_a0.m

\begin{figure}[H]\centering
	  \includegraphics[scale=0.32,angle=-0]{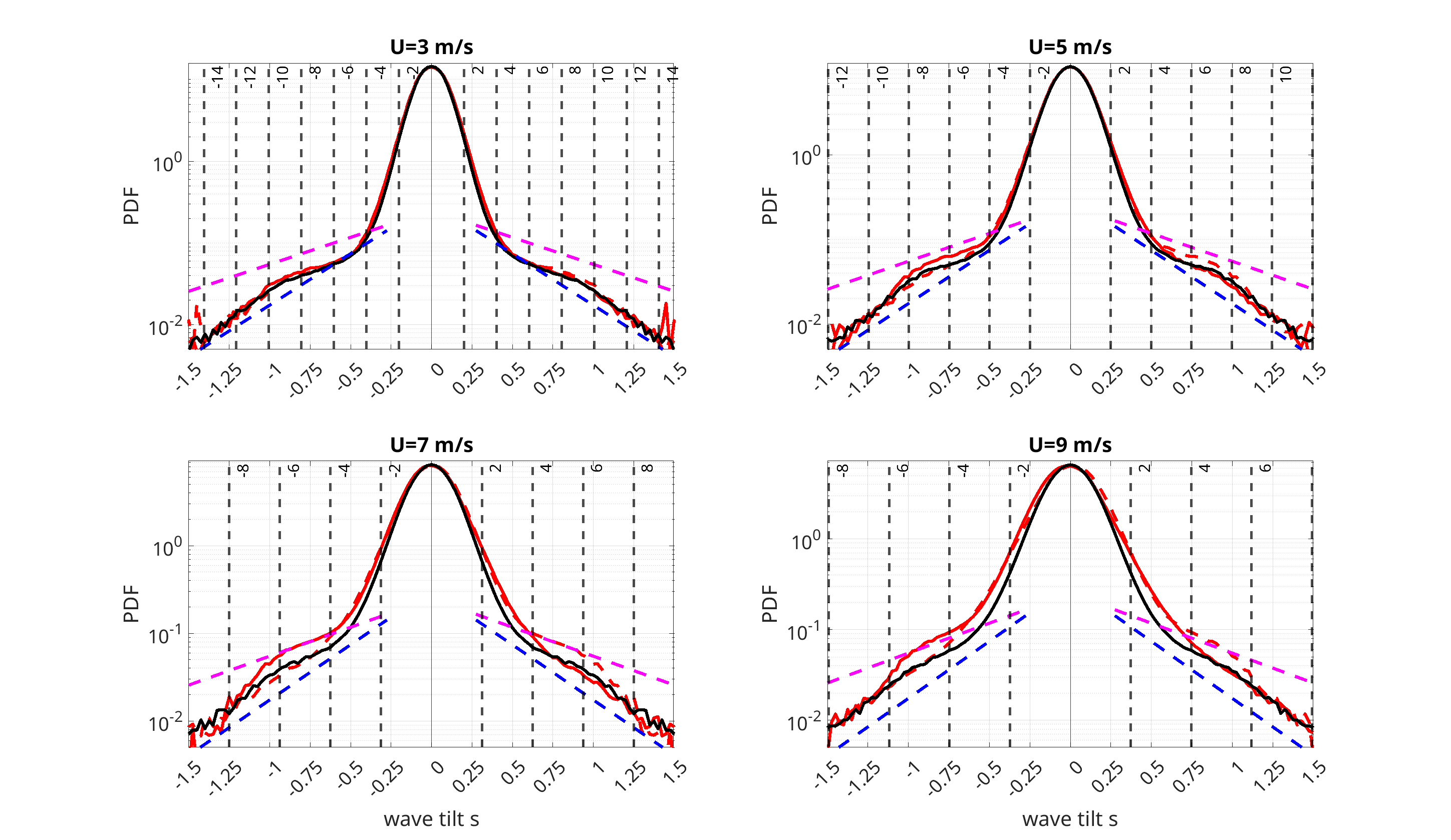}
          \caption{\label{fig:coupe_ventsmoyens}Wave-slope PDFs in the upwind (solid red lines) and crosswind (solid black lines) planes obtained for various wind speeds by averaging the IASI data inside regular bins of slopes, using the median value. To highlight the asymmetry of the upwind distribution, its flipped version ($s_u\leftrightarrow -s_u$) is shown in dashed red lines. The vertical black dashed lines indicate the multiples of the RMS upwind slope.  The dashed blue and magenta lines show, for reference, the functions $10^{-1.26\abs s-0.5}$ and $10^{-0.66\abs s-0.6}$.}
\end{figure}

To see whether this universal exponential decay of the tail depends on the wind direction, we evaluated the median values of the PDFs by only retaining the wave-slope vectors within 5$^{\circ}$ degree from the up- and cross-wind planes, within bins of wave-slope $s$ of width $0.025$. The resulting probabilities for four wind speeds are shown in Fig. \ref{fig:coupe_ventsmoyens} where multiples of the standard deviation $\sigma$ of the upwind slope (i.e. $n\sqrt{m_u}$ for $n=2,3...$) are indicated by vertical dashed lines. As widely pointed out previously, the central parts of the distributions follow a quasi-Gaussian trend for $|s|$ smaller than about $2\sigma$. This behavior is followed by a quasi-exponential regime $~10^{-\alpha \abs s}$ with an exponent $\alpha$ between $0.66$ and $1.26$ (see the dashed lines in Fig. \ref{fig:coupe_ventsmoyens}). Even though the tail of the distribution might be contaminated by scattering by aerosols or foam, as discussed in the Appendix, it remains largely driven by the sea-surface wave facets, since the former processes are likely independent of the wind direction. This as demonstrated by its sensitivity to wind direction shown by the differences between the up- and cross-wind values as well as those, in the upwind case, between the values for $\pm s$. Indeed, in the exponential tail, the downwind probabilities ($s$<0) are much larger than both the upwind ones ($s$>0) and the crosswind values for the same slope, differences which increase with wind speed. The upwind probabilities are smaller than (or equal to) the crosswind ones, an effect that explains the vanishing of the upwind maximum at large wave slopes, as  pointed out above (see  Fig. \ref{fig:examplefita0_azimuth} and Sec. \ref{subsec:firstfindings}). 

%\end{multicols}

%/home/guerin/Dropbox/COXMUNK_HARTMANN/IASI_HS_2025/PRG_HS/PRG_FIGURES_PAPIER/plot_coupe_ventsmoyens.m

%\begin{multicols}{2}
  
\begin{figure}[H]\centering
	  \includegraphics[scale=0.2,angle=-0]{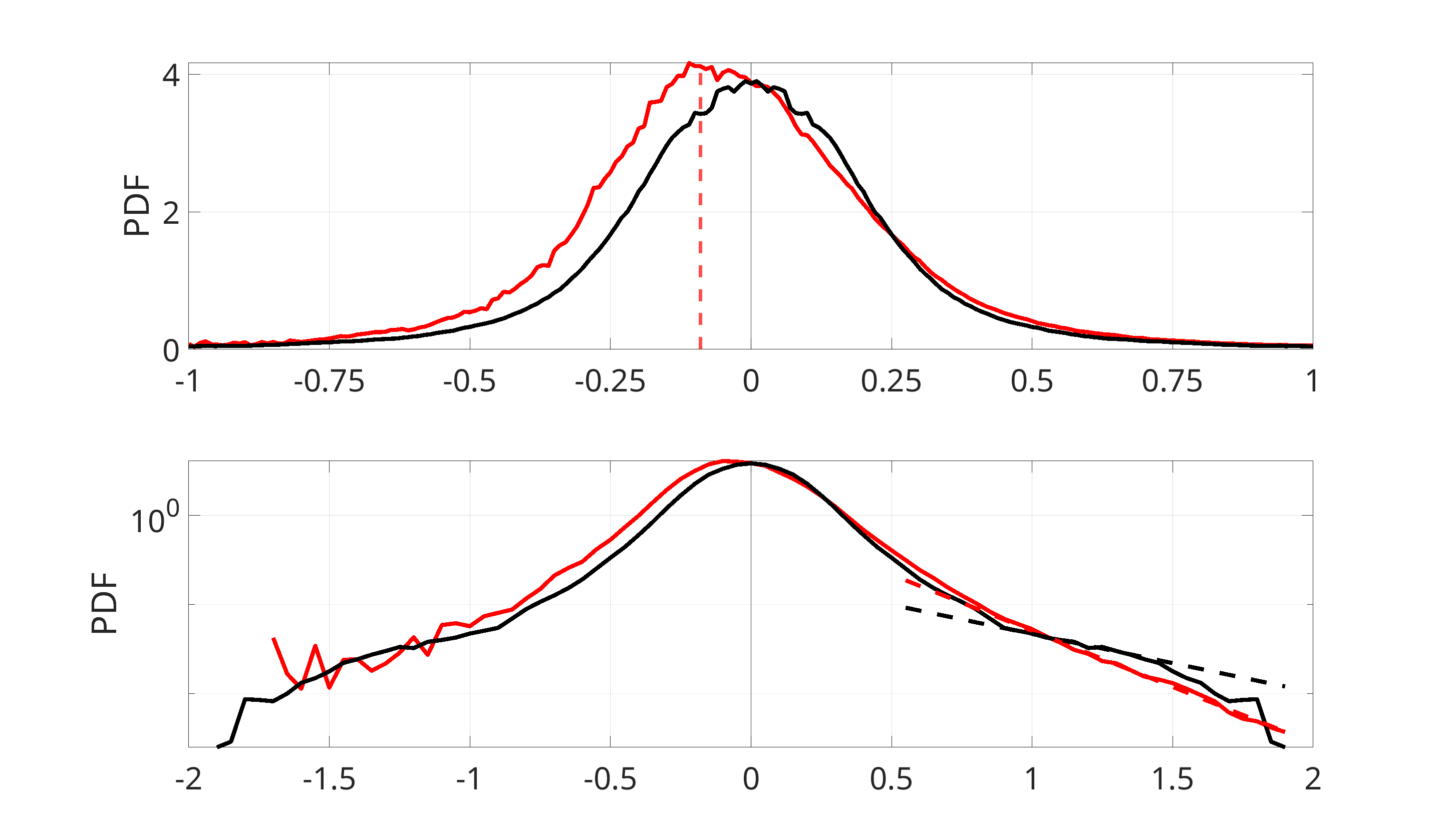}
          \caption{Wave-slope PDFs in the principal planes versus the wave tilt $s$, obtained for wind speeds larger than 15 m/s by averaging the IASI data inside regular bins of slopes, using the median value. The dashed red and black lines show for reference the exponential functions $10^{-1.26\abs s-0.03}$ and $10^{-0.66\abs s-0.67}$.\label{fig:coupe_grandsvents}}
\end{figure}

For strong winds (> 15 m/s), the scarcity of data does not allow regrouping the probabilities within  wind-speed bins. We thus concatenated all the IASI-retrieved probabilities for $U$ larger than 15 m/s, regardless of $H_s$, and again considered the values in the principal planes, (median) averaged within regular wave-slope bins of $0.025$. The resulting PDFs, shown in Fig. \ref{fig:coupe_grandsvents}, confirm the strong upwind-downwind asymmetry, with a most probable slope of $-0.1$ ($\sim -6^{\circ}$ tilt) and negative slopes being significantly more probable than their positive counterpart. In this example, a quasi-exponential decay is confirmed for the probabilities of steep positive slopes in both the up- and cross-wind planes, but with different exponents, i.e.:
\be
\begin{split}
  p(s_u,0)&= 10^{-1.25 s_u-0.5} , \\
  p(0,s_c)&= 10^{-0.66 s_c-0.65} ,
  \end{split}
    \ee
    while the lack of data does not enable to establish a parameterization for largely negative slopes. A last remarkable feature is that the crosswind probabilities become larger than the upwind ones for $|s|$ greater than about 1.1 (48$^{\circ}$), a fact consistent with the observed vanishing of the secondary upwind maximum in azimuth as the wave tilt increases (see Sec. 4 \ref{subsec:firstfindings} and Figs \ref{fig:examplefita0_azimuth}).
    
%/home/guerin/Dropbox/COXMUNK_HARTMANN/IASI_HS_2025/PRG_HS/PRG_FIGURES_PAPIER/plot_coupe_grandsvents.m

Large slopes are often attributed to the steep facets produced by near-breaking waves, and  there is a vast literature discussing the influence of this process on  radar  backscattering cross-sections and Doppler spectra (e.g. \cite{wetzel_chapter86,hwang_JGR08b,jessup_JGR91,yurovsky_RemSens21}). However, despite an extensive bibliographic search we have found only one study discussing the contribution of breaking waves to the tail of the slopes PDF \cite{Voronovich}. In this paper, the latter is determined from the differences between measured K$_u$- and C-bands Radar returns, for the HH-polarization in the upwind direction, and values calculated using the small-slope approximation. The results displayed in Fig. 5 of \cite{Voronovich} show  that the upwind probabilities exponentially decay for $s$ greater than about 0.6, with a rate  that is independent of $U$,  characteristics qualitatively consistent with our observations displayed in Fig. \ref{fig:tail}. However, there are large quantitative differences with the behavior of the IASI-retrieved tail. The first is the dependence on  $U$ of the probabilities for a given tilt, which is much more pronounced than in our data. The second one lies in the magnitude of the tail, which is several orders of magnitude smaller than the IASI-retrieved values. Lastly, according to \cite{Voronovich}, the tail is only present in the upwind direction since the authors a priori assume that the downwind  probabilities are unaffected by wave-breaking and follow a quasi Gaussian law. 
%  Indeed, according to Eq. (19) of \cite{Voronovich} an increase of $U$ by +10 m/s induces a change of the probability by a factor of about 10, much larger than what is seen in Fig. \ref{fig:tail} where the factor is smaller than 2. In addition, for $s$=0.8, \cite{Voronovich} predict probabilities of about 2 10$^{-4}$, 6 10$^{-4}$ and 18 10$^{-4}$ for $U$=5, 10, and 15 m/s, respectively, far below the IASI-retrieved values. Finally, according to \cite{Voronovich}, the tail is only present in the up-wind direction, the authors assuming that the probabilities in the down-wing direction are unaffected by wave-breaking and follow a quasi Gaussian law as proposed by \cite{Cox56}.
 This postulated large asymmetry is not observed in our data. These elements make us believe that the tail in the IASI-retrieved probabilities cannot be fully explained by wave breaking.

\section{Summary of conclusion}
A new technique has been elaborated to analyse the wave-slope probabilities, which mitigates possible defaults in the absolute calibration as well as uncertainties and scatter in the tail of the distribution. Thanks to this refined technique, we have pointed out some new features of the ocean wave-slopes statistics from analysis of the IASI-retrieved probabilities: The influence of the significant wave height, $H_s$, has been analyzed, in addition to that of the wind speed $U$ which is the dominant parameter. The decay of the wave-slope PDF at large wave tilts has also been analyzed and shown to be relevant to the sea surface in spite of a possible contamination by other sources of solar-radiation redirection. Our main findings are the following:

\begin{enumerate}
\item   When only the influence of the wind speed $U$ is considered,  the deviations from the linear relationships of \cite{Cox54}, of up to $10\%$ in the upwind direction, that were pointed out in our previous study \cite{Guerin_RSE23}, have been confirmed. A new and more accurate parameterization based on peace-wise linear functions has been proposed.
\item  When the data are sorted by both $U$ and  $H_s$, it is found that, for the same wind speed, the MSSs decrease slightly when $H_s$ increases, at an average rate of about $-0.0016\ m^{-1}$ for the total MSS. This behavior, here pointed out and quantified for the first time, has several possible physical explanations which remain to be confirmed (short-scale damping by the presence of long waves, atmospheric instability, subgrid variability, and compounding effect on the MSSs).
\item  The kurtosis coefficients $C_{40},C_{04}, C_{22}$ are weakly sensitive to $U$, as previously shown, but strongly depend on $H_s$, in particular at low wind speeds. They are decreasing functions of $H_s$, a finding for which we have no obvious physical explanation.
\item  We recover the well-known facts that: a) The upwind-downwind asymmetry increases with wind speed, as can be measured by the evolution of the most probable wave slope, the main skewness coefficients $C_{30}$, and the tail of the wave-slope PDF in the upwind plane; b) The crosswind-downwind asymmetry increases with wind speed, as can be measured by the evolution of the maximal UCA and the contrast of directional MSSs.  We also unveil the original result that, at equal wind speed, the upwind-downwind and  upwind-crosswind asymmetries decrease with increasing wave height.
\item  For given values of the slope, $U$ and $H_s$, the angular dependence of the wave-slope probability  can be well represented with the first 3 azimuthal harmonics ($a_0,a_1,a_2$). A global maximum is found in the downwind direction with a secondary maximum in the upwind one, while a minimum is found in the crosswind direction. For large wind speeds and wave tilts, the upwind secondary maximum and the crosswind minimum flatten out, and there remains a unique maximum in the downwind direction. This new finding is  opposite to what is observed for the azimuthal variation of the cross-section in radar ocean remote sensing, where the maximum is found in the upwind direction and the secondary downwind maximum eventually vanishes at large incidences. This difference results from the fact that the involved scattering mechanisms (Bragg scattering for radio waves and specular reflection in the optical regime) are not the same.
  
\item  With increasing wave tilt, the omnidirectional probabilities evolve from a near-Gaussian distribution to a quasi-exponential tail, with an exponent $~10^{-1.2 s}$ that remains universal across the various wind speed. The restrictions of the PDF to the (principal) up- and cross-wind planes show the same exponential behavior but with different exponents at large wind speed ($10^{-1.26 s}$  and $10^{-0.66 s}$ in the up- and cross-wind directions, respectively.). 
\end{enumerate}

\section*{Acknowledgment}
This work has been supported in part by CNRS, CNES and Ecole polytechnique. VC and JMH  acknowledge the IPSL mesocenter ESPRI facility for computer simulations, as well as EUMETSAT and the Aeris infrastructure (https://www.aeris-data.fr/) for access to the IASI Level 1c data. CAG acknowledges the CNRS and the University of Toulon for supporting a research leave for the full academic year 2025-2026 (``Délégation'' and ``CRCT''). The authors also warmly thank F. Nouguier
for providing the data plotted in Fig. 3 of \cite{Nouguier_GRSL16}. 

\appendix
\section{The sources of error and scatter}
This Appendix is an attempt to discuss the various potential sources of systematic errors and scatter in the determination of the wave-slope PDF and its parameterization. For this we first recall the retrieval approach applied to the IASI-collected radiances and the procedure used for the analysis of the obtained probabilites. This enables to point out the various assumptions made, which are then discussed one by one in the following sub-sections.
\subsection{The retrieval and its underlying assumptions}
The approach used to determine a wave-slope probability from each retained IASI daytime observation has been detailed in \cite{Capelledaytime,Guerin_RSE23}. Briefly, the probability is obtained, together with the sea-surface temperature (SST), from a fit of each IASI radiance spectrum in two spectral windows centered at  3.7 $\mu$m (2594-2760 cm$^{-1}$) and 4.0 $\mu$m (2480-2528 cm$^{-1}$) containing 185 and 107 individual radiances, respectively. This implicitly assumes that the forward model and its input data are "exact" (\textit{Hypothesis 1}) and that the solar photons collected by the instrument have all been (specularly) reflected by a clean water surface (\textit{Hypothesis 2}). Thanks to this last assumption, each retrieved probability can be ascribed to given wave-slopes $s_x$ and $s_y$ in the East-West and North-South planes, accurately determined from the (precise) knowledge of the positions of the sun, instrument, and observed sea surface pixel center in the considered measurement. This leads to a set of probabilities $p[s_x(k),s_y(k)]$, where $k$, which is a short-cut for the spectrum number, is fully characterized by the observed pixel spatial position and the time at which it has been looked at. Note that the latter only depends on the date since the plateforms (Metop-A, -B and -C) carrying IASI operate in a sun-synchronous polar orbit with a constant 9:30 AM (and 9.30 PM for nighttime) local time at the Equator.

In a second step, $s_x(k)$ and $s_y(k)$ are converted into slopes in the upwind, $s_u(k)$, and crosswind, $s_c(k)$, directions from knowledge of the wind direction (from ERA5, see Sec. \ref{Dataused}.1). At the same time, values of the wind-speed (at 10 m), $U(k)$, and significant wave height, $H_s(k)$, also provided by ERA5, are associated to each observation. This implicitly assumes that $\overrightarrow{U(k)}$ (module and direction) is exact and fully characterizes the wind in the observed pixel (\textit{Hypothesis 3}), with a similar assumption for $H_s(k)$ (\textit{Hypothesis 4}). Further assuming that  $U(k)$ and $H_s(k)$ are the only oceano-atmospheric parameters driving the wave-slope probabilities (\textit{Hypothesis 5}) enables to convert the original set of $p[s_x(k),s_y(k)]$ into a set of $p[s_u(k),s_c(k),U(k),H_s(k)]$.

In the following, we step-by-step discuss how the assumptions
made (\textit{Hypotheses 1-6}) may affect the IASI-retrieved probabilities as well as their analysis and parameterization.

\subsection{Retrieval errors (Hypothesis 1)}
As shown by Eq. (1) of \cite{Capelledaytime} the forward calculation of  daytime IASI-collected  radiance spectra is based on the modeling of four terms: $I_{Surf}$, $I_{Atm-up}$,  $I_{Atm-down}$, and  $I_{Sun}$, which respectively are the contributions of the sea-surface emission transmitted up to the satellite, of the upward emission of the atmosphere in the observed field of view, of the downward atmospheric emission toward the surface then reflected to the instrument, and of the sun transmitted down to the observed pixel and then reflected up within the field of view. Their calculations are based on knowledge of the sea-surface spectral emission and reflection coefficients, of the atmospheric  state (vertical profiles of temperature, pressure, and absorbing species fractions), of the spectroscopic parameters of the atmospheric gaseous species which absorb in the considered spectral range, of the extra-terrestrial solar radiance, and of the sea-surface temperature (SST) and wave-slope probability $p[s_x,s_y]$, these last two quantities being floated in the retrieval procedure. For nightttime observations \cite{Capellenighttime},  $I_{Sun}$ is  omitted which, together with the fact that only the SST is floated, is the only difference between the night- and day-time models. In \cite{Capellenighttime}, a very large number of nighttime IASI-retrieved SSTs were compared with collocated in situ temperature values measured by buoys. This showed an excellent agreement with a mean difference <0.02 K on average, with a similar agreement (<0.05 K) obtained for daytime \cite{Capelledaytime}. The associated standard deviations are also small, typically 0.3 K, a large part of them being attributable to the buoy measurements uncertainties. These error statistics, together with the sensitivity analyses  made in these two papers show that the systematic and random errors on the retrieved values of $p[s_x,s_y]$ are likely small, significantly below the scatter of the points in Fig. \ref{fig:examplespdf_vent2_9}. In order to further verify this statement, a "trick" was used, which consists in treating nighttime IASI spectra as if they had been recorded during the day.  We carried this exercise using spectra collected under clear sky conditions during the entire year 2020. The associated radiances, recorded at 9:30 PM LT, were treated as explained in  \cite{Capelledaytime} with the presence of an "artificial" sun located at its position 12 hours later. In such a case, the retrieved probabilities $p[s_x,s_y]$ should, in the absence of any error in the spectrum fit procedure, be equal to zero. The exercise leads to a mean obtained probability of -0.005 with a RMS of 0.035 (the median being -0.0001 with a RMS of 0.0093) with no noticeable dependence of $p[s_x,s_y]$ on the geometry of the observation (IASI line of sight and sun position). The RMS, being significantly smaller than the scatter of the probabilities shown in Fig. \ref{fig:examplespdf_vent2_9}, confirms that the latter are not (or only marginally) due to retrieval errors. Furthermore, the mean error is significantly smaller than the values for large wave slopes, which are of a few 10$^{-2}$ (see Figs. \ref{fig:examplespdf_vent2_9} and \ref{fig:tail}), implying that the exponential tail of the PDF is "real". In other words, it is not due a systematic bias of the retrieval and indeed results from solar photons received by the IASI detector. Some processes, other than reflection on clean water, potentially participating to this solar contribution are discussed in the next section.

\subsection{Clean water surface specular reflection (Hypothesis 2)}
As recalled above, the determination of wave-slope probabilities from IASI spectra assumes (\textit{Hypothesis 2}) that: (i) The solar photons collected by the instrument have all reached the sea surface, and that (ii)  the latter involves clean water. Assumption (i) thus disregards scattering by aerosols, while (ii) does not consider the role played by the eventual "pollution" of the sea surface by white caps, foam, or other natural or artificial materials. A break down of these hypotheses, which are discussed below, would lead to the erroneous attribution of all the collected solar signal to a wave-slope probability, thus leading to a positive or negative bias on the latter.

\subsubsection{Clouds/Aerosols}
\begin{figure}[H]\centering
 \includegraphics[scale=0.18,angle=-0]{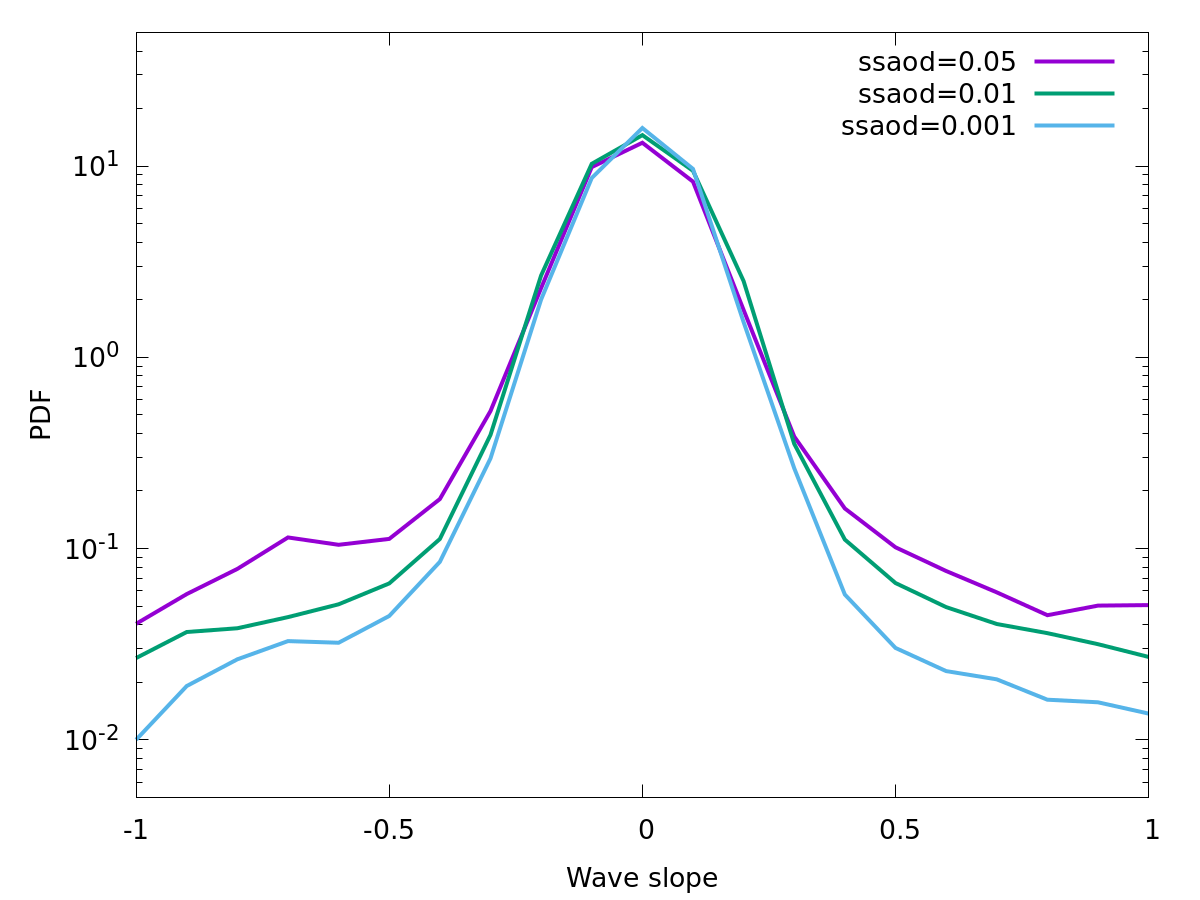}
 \caption{Wave-slope PDFs in the upwind-cross wind principal plane ($s_c$=0, $s=s_u$) obtained for  $U$=2 $\pm$ 0.5 m/s ,  $H_s$=1$\pm$ 0.5 m, and 3 maximum values of sea-salt 550nm AOD, by averaging the IASI data inside regular and 0.1-broad bins of slope $s$, using  the median value. } 
   \label{fig:Effet sea-salt}
\end{figure}

Obviously, aerosols (and clouds) of different types, if present in the atmospheric paths followed by solar radiation, will scatter a fraction of the latter toward the instrument. In this study, a large fraction of clouds and coarse aerosol particles, primarily composed of mineral dust, were filtered out beforehand. This is achieved using criteria on brightness temperatures from IASI and the collocated AVHRR instrument, as explained in \cite{Capellenighttime}, together with a dedicated retrieval of the dust particles optical depth at 10 $\mu$m using the method described in \cite{Capelledust}, that is used here for aerosol screening. It should be noted that sea-salt aerosols, which contain a substantial fraction of coarse particles, may also have been partly filtered out by this algorithm. Nevertheless, owing to their generally low aerosol optical depth (AOD) and their confinement to the planetary boundary layer --where satellite infrared radiances have limited sensitivity-- a substantial fraction of these aerosols is likely to have remained in the dataset. Moreover, since the retrieval in \cite{Capelledust} is based on channels that are insensitive to solar radiation, it is expected to be more sensitive to  absorption by coarse particles than to their scattering effect. Consequently, it may not have screened out particles that are purely scattering.
Nevertheless, even if this screening is not perfect, desert dust particles are strongly localized in both time and space and are therefore unlikely to have a statistically significant effect. The same applies to pollution and biomass-burning aerosols, which are also less likely to affect the infrared domain because of their small particle size. In contrast, sea-salt aerosols are frequently present nearby all ocean surfaces (with amounts depending on the local conditions). Unfortunately, quantitatively analyzing the contribution of solar photons that they scatter to the signals collected by IASI around 3.9  $\mu$m is practically impossible. Indeed, these aerosols involve a great variety of chemical compositions and sizes, with radii (between typically 0.01 to 100 $\mu$m) ranging from much smaller to much larger than the considered wavelengths. They are generated by a number of different processes and their amounts above the oceans significantly vary with the wind speed and relative humidity which also significantly influence their size distribution, as discussed in \cite{AODSize,Revueaerosols1,MAprod1,MAprod2,Revueaerosols3,Revueaerosols2} and references therein. As a result, the light absorption and scattering that they induce also vary considerably according to the oceano-atmospheric conditions, as exemplified for instance in \cite{MAAOD3,AODSize, MAAOD2,MAAOD1,MAAOD4}. Modeling the contribution of marine aerosols to the probabilities derived in the present study is thus not only far beyond the scope of the present paper, but also likely hopeless. However note that, since aerosol scattering is much less directive than specular reflection at the air-sea interface, one can predict that their presence will increase the amplitude of the tail of the distribution, creating a bias in the probabilities for large slopes.  This is confirmed by  Fig. \ref{fig:Effet sea-salt} for which the IASI spectra used in the present work have been collocated with CAMS global reanalysis (EAC4)  sea-salt 550nm AOD  \cite{Inness2019,CAMS_EAC4_2020} and further filtered, together with the associated wave-slope probabilities, according to the letter AOD. IN this plot, the remaining  median PDFs averaged inside regular bins of slopes are plotted as a function of wave slope for different sea-salt AOD limits, revealing a clear dependence of the tail of the distribution on aerosol loading. Note that the moderate dependence of the wave-slope PDF tail on $U$ shown in Fig. \ref{fig:tail} is a priori not fully compatible with the (exponential) rapid increase of the marine aerosols production and of their resulting concentration shown in \cite{Revueaerosols1,Revueaerosols2}. This may indicate that the scatter of solar photons by the latter plays a relatively minor role.  Moreover Fig. \ref{fig:Effet sea-salt} shows that the exponential behavior of the tail persists even for very low AOD limit values, indicating the presence of other processes redirecting solar photons toward the instrument. The same analysis was extended (results not shown) using the CAMS total AOD at 550 nm, which accounts not only for sea salt but also for dust, black carbon, organic matter, and sulfate aerosols. A similar behavior was observed: filtering aerosol-loaded scenes reduces the amplitude of the tail, but does not eliminate it.

%%%%%%%%%%%%%%%%%%%%%%%%%%%%%%%

\subsubsection{Sea surface "pollution"}
  First note that, due to the very strong absorption coefficient ($\simeq$10 mm$^{-1}$) of liquid water at the IASI wavelengths used (between 3.62 and 4.03 $\mu$m), any water "pollutant" (air bubbles, biogenic or artificial materials) located deeper than typically 0.3 mm play a negligible role in the collected radiances.  Above this depth, the clean-water "pollution" can result from white caps, foam or other materials (plastics, oil slicks, algae, phytoplankton, etc). These may not only bias the wave-slope probability retrieval  but also change the wave-slope PDF itself.  Indeed, it is  well known that natural or artificial slicks (surfactants) at the air-sea interface damp the waves and reduce the MSS, as shown and discussed in \cite{Cox54,Alpers,Behroozi,Hu,Kudryavtsev,Laxague,Wind_effect} and references therein. However, this seldom occurs, both temporally and spatially, so that a very small fraction of the IASI spectra are affected, and thus only the much more frequent white caps and foam patches must be considered here. Unfortunately, for the latter we have found no data providing the needed bidirectional reflectivity distribution function (BRDF). However, due to the (very) complex geometric structure of the foam layer, e.g.  \cite{ZHANG2025}, the dependence of its BRDF on the incident and outgoing angles is much smoother when compared to a clean see surface for which reflection is specular, as shown in \cite{BRDFFoam1,BRDFFoam2}. A part of the incoming sun radiation can thus be scattered toward the instrument, even in angular configurations corresponding to large wave tilts. As for the marine aerosols, the presence of white caps and foam thus likely enhances the probabilities in the tail of the wave-slope PDF. For any modeling, another difficulty comes from the needed knowledge of the white-cap fraction $W$ (percentage of a sea surface covered by foam). Since the pioneering work of \cite{Monahan}, large progress has been made, as discussed in \cite{Albert_foam, Brumer, Anguelova19} and references therein. The dependence of $W$ on $U$ has been reduced with respect to the original law according to which $W$ is proportional to $U^q$ with $q \simeq 3.5$ \cite{Monahan}, but 
uncertainties and large differences between the various $W$ determinations remain. This is exemplified by Fig. 1 of \cite{Brumer} which compares various models, and it was stated that the scatter displayed by wind-speed only
parameterizations may be largely due to varying
sea states. Furthermore, as shown by Figs. 8c,d of \cite{Anguelova19}, simultaneous observations for different microwave frequencies and polarizations lead to inconsistent white-cap coverage values. Large uncertainties thus remain but the numerous studies all indicate that changing $U$ from a few m/s to 15 m/s results in an increase of $W$ by, at least, one order of magnitude, in contrast with the values in  Fig. \ref{fig:tail}. However, it is worth mentioning that the  $W$ does not fully characterize the properties of white caps, since wind also influences the foam layer thickness and its structure including void fraction and bubble size distribution, which vary greatly in space and time \cite{Reul}. Due to this and to the large uncertainties on the measurements and parameterizations of $W$, we believe that the above mentioned differences in the wind dependences is, at this step, insufficient to  fully rule out that white caps contribute the IASI spectra and, in particular, positively bias the probabilities of large slopes. 

\subsubsection{Conclusion}
As discussed above, the disregarded eventual presence of marine aerosols and/or white caps likely enhances the small wave-slope probabilities and reduces the large ones through scattering and absorption which has the effect of increasing the apparent MSSs.
Note, however, that this possible effect does not compromise our main finding that the MSSs decrease with $H_s$. Indeed, because the amounts of aerosols and white caps are positively correlated to $H_s$, due to the enhancement of wave-breaking and sea sprays with increasing $H_s$, the MSSs of sea states with a larger significant wave height (at equal wind speed) are expected to suffer from a larger positive bias. Removing this possible bias would thus reinforce our conclusion.

\subsection{Wind and significant wave height errors (Hypotheses 3-4)}
As recalled in Appendix A.1, each IASI-retrieved probability $p(s_u,s_c,U,H_s)$ is ascribed to a quadruplet of variables: The up- and cross-wind slopes, $s_u$ and $s_c$, determined from knowledge of the wind direction and wind speed $U$, and  significant wave height $H_s$. Random errors on these quantities, which we all took in ERA5 (see Sec. \ref{Dataused}.1), would result in a scatter of the the obtained values of $p(s_u,s_c,U,H_s)$ for a given assumed quadruplet.  

\subsubsection{Wind-speed errors}
\begin{figure}[H]\centering
 \includegraphics[scale=0.27,angle=-0]{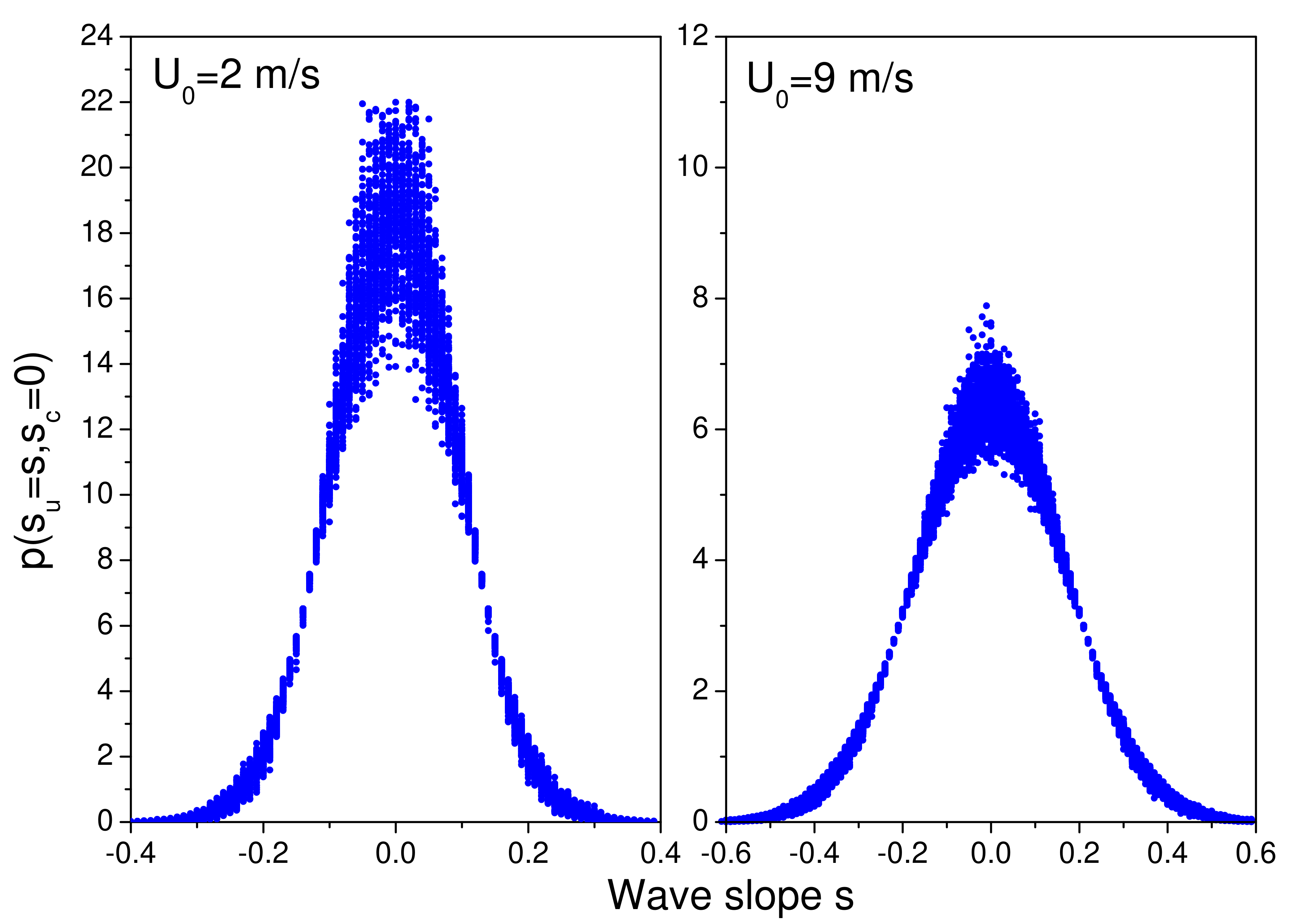}
 \caption{Computed influences of errors on the assumed wind speed on the PDF for a RMSE on $u$ of $\delta U$=0.5 m/s.} 
   \label{fig:Effet erreur vent}
\end{figure}

As mentioned in Sec. \ref{Dataused}.1, the evaluations of the ERA5 wind-speed accuracy made in \cite{Belmonte,Guerin_RSE23} using ASCAT \cite{ASCAT} collocated measurements show that the associated RMSE is of about 1 m/s, a value which can be reduced to  $\delta U\simeq$0.5 m/s when the uncertainty of the ASCAT data \cite{precisionASCAT} is taken into account. In order to evaluate the influence of such a random error on the IASI-retrieved wave-slope PDF, we computed a set of probabilities $p_i$ for numerous wind speeds $U_i=U_0+\delta u_i$, assuming that $\delta u_i$ follows a normal distribution, centered on zero and of RMS equal to  $\delta U$. For this, we neglected, for simplicity, the GC corrections in Eq. (\ref{eq:GC}), and used:

\begin{equation}
p(s_u,s_c,U_i)\frac{e^{-\demi s_u^2/m_u(U_i)}e^{-\demi s_c^2/m_c(U_i)}}{2\pi\sqrt{m_u(U_i)m_c(U_i)}}
\label{eq:effetvent}
\end{equation}
where $m_u(U)$ and $m_c(U)$ were computed using Eq. (\ref{eq:fitmss}) and the data in Table \ref{eq:tablemss}. Calculations were then made  for $U_0$=2 and 9 m/s, $\delta U$=0.5 m/s and $s_c$=0 for a  comparison with the  values plotted in Fig. \ref{fig:examplespdf_vent2_9}, leading to the results  displayed in Fig. \ref{fig:Effet erreur vent}. In order to mimic the clipping of high probabilities due to the rejection of the associated IASI spectra (see Sec. 2.2), all computed probabilities greater than 22 were disregarded in this exercise.  Comparing the results obatained, displayed in Fig. \ref{fig:Effet erreur vent}, with those in Fig. \ref{fig:examplespdf_vent2_9}  shows that a large part of the scatter of the IASI-retrieved data can be explained by a RMS error $\delta U \simeq$0.5 m/s, consistent with the above discussed uncertainty of the ERA5 values of $U$. Indeed, there is a good agreement, around the maximum, between the observed and calculated dispersion-averaged probabilities and scatter amplitudes. However, this is not the case around $s_0(U_0)=\pm$0.13 and $\pm$0.22 for $U_0$=2 and 9 m/s, respectively, where, because Eq. (\ref{eq:effetvent}) here leads to $dp(s_c=0,s_u=s_0,U_0)/dU=0$, the scatter in the computed PDF is significantly smaller than the observed one. This indicates that some other factors are involved.  

\subsubsection{Wind-direction errors}
As recalled in Sec. \ref{Dataused}.1, the assessments of the ERA5 wind-direction accuracy made in \cite{Belmonte,Guerin_RSE23}  show that the associated RMSE varies (with wind speed) in the $[6^{\circ}-20^{\circ}]$ range, thus with a mean value of about $15^{\circ}$.  In order to evaluate the influence of such a random error on the IASI-retrieved probabilities, we computed a set of probabilities $p_i$ for numerous wind directions $\phi_i$, assuming that $\phi_i$ follows a normal distribution, centered on zero and of RMS equal to  $\delta \phi=15^{\circ}$. For a comparison with Fig. \ref{fig:examplespdf_vent2_9}, wind speeds of 2 and 9 m/s were retained and Eq. (\ref{eq:effetvent}) was used with $s_{u,i}=scos(\phi_i)$ and $s_{c,i}=ssin(\phi_i)$. As in Sec. A.4.1, $m_u(U)$ and $m_c(U)$ were computed using Eq. (\ref{eq:fitmss}) and the data in Table \ref{eq:tablemss}. The results of this exercise show that uncertainties on the wind direction lead to a scatter of the probabilities that is very small and negligible when compared with those in Figs. \ref{fig:examplespdf_vent2_9} and \ref{fig:Effet erreur vent}, as could be expected from the small value of $\delta \phi$
and limited difference between $m_u(U)$ and $m_c(U)$.

\subsubsection{Significant wave height errors}
The results presented in Sec. 6.3 show that the significant wave height has a limited influence on the MSSs, much smaller than that of the wind speed. Due to this, and the fact that the errors in the ERA5 values for $H_s$ (see Sec. \ref{Dataused}.1) are also small, the influence of the latter on the scatter of the probabilities can confidently be expected to be much smaller that that of wind-speed errors discussed in Sec. A.4.1. For this reason it was not investigated.

\subsubsection{Variability with the IASI-observed pixel}
The wind and significant wave height information provided by ERA5 are on an hourly basis over a $0.25^\circ\times0.25^\circ$ spatial-grid. This implies that even if they are exact and perfectly coincide in space and time with the IASI observations, they would only characterize the averaged values over the spatial extent of the observed pixel (of radius 6 km at nadir, i.e. $\simeq$110 km$^2$). As is well known, variabilities at much smaller spatial scales and over time intervals of much less than one hour are present. This is particularly the case for the wind speed and direction which are affected by gustiness for periods from seconds to minutes over areas as small as a few hundred m$^2$. Obviously, the associated limited representativity of the EAR5 data leads to a scatter of the probabilities, for reasons similar to those discussed in Appendix A.4. Also recall that subpixel inhomogeneities of the surface roughness may conduct to an increase of the apparent MSS and kurtosis when modeling the surface slopes as a spatially homogeneous random process \cite{chapron_JGR00}.

\subsection{Other driving processes (Hypothesis 5)}
\begin{figure}[H]\centering
 \includegraphics[scale=0.2,angle=-0]{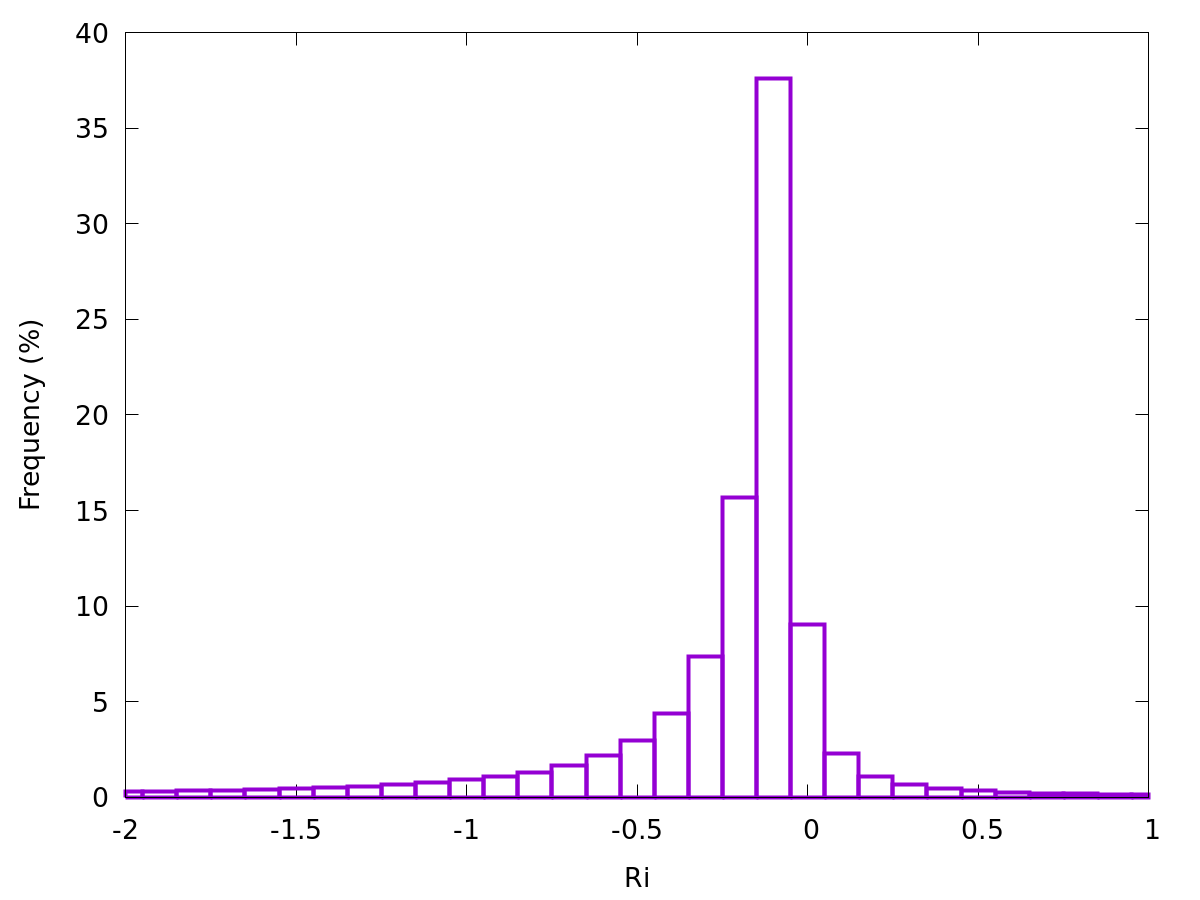}
 \caption{Histogram of the Richardson number for the IASI observations used in the present study.} 
   \label{fig:DistribRi}
\end{figure}

Besides the other assumptions discussed in this Appendix, a key hypothesis a priori made in the present study is that the wave-slope PDF only depends on the local and instantaneous values of the wind vector and significant wave height. This not only disregards the influence of the past, but also assumes that no other oceano-atmospheric process participates to the driving the sea-surface roughness. In this study, we chose, for practical reasons, to classify the data according to the simple $H_s$ parameter, regardless of the origin (swell or wind-wave sea) and direction of long waves (with respect to the wind). However note that a few studies \cite{hwang_JGR08,hauser_JGR09reply} have shown some influence, on the MSSs, of the nature of the sea state (wind-wave only or mixed with swell) and of the swell direction. 
The influence of another variable, the degree of atmospheric stability, was investigated a while ago by \cite{Hwang_JGR88} from an analysis the diffraction of the beam emitted by an underwater. They found that, with respect to those of \cite{Cox54}, there is a sharp increase of the MSSs for unstable conditions and a significant reduction for stable ones (conclusions  based on less than 10 data points), findings confirmed later \cite{wu_JGR91,shaw}. In these studies the degree of atmospheric stability was characterized using the bulk Richardson number $R_i$. Modifications of the MSSs of \cite{Cox54} were proposed, which change the original ones by up to +100\%  and  -35\% in  unstable ($R_i\approx{-0.2}$)   and stable ($R_i\geq{+0.5}$) cases, respectively, with no change for $R_i\simeq$0.15. The agreement in this case was attributed  to the fact that the observations of \cite{Cox54} were made under near neutral or slightly stable conditions. Later on, \cite{Ross} used observations of the sun glint to deduce MSSs which again show a reduction for very stable atmospheres, a result however supported by 8 values only, while the previously observed increase for unstable ones was not confirmed. Finally, the most recent study \cite{Lenain_JPO19}, based on the analysis of the glint of an airborne laser, makes the situation further unclear since no meaningful trend versus $R_i$ was detected in the [-0.3,+0.1] range. 
In order to see which atmospheric conditions are involved by the IASI observations used in the present work, we computed the  bulk Richardson number from  \cite{Louis}: 
\begin{equation}
R_i=\frac{g(T_A-T_W)z_0}{T_WU^2(z_0)} ,
 \label{eqRichardson}
\end{equation}
where $g$=9.81 m/s$^2$ and $z_0$=10 m, while $T_A$ and $T_W$ denote the air and water temperatures (both in $^\circ$C), respectively. As for $U$ and $H_s$ (see Sec. 2.1), the values of $T_A$ and $T_W$, were taken from ERA5 reanalyses. The obtained statistical distribution, displayed in Fig. \ref{fig:DistribRi}, shows that most IASI-retrieved probabilities are for (near) neutral conditions (small $|R_i|$) but that a number of unstable cases (significantly negative values of $R_i$) are also involved, while few very stable cases (large positive $R_i$) are probed. This last finding is likely explained by the observations time (9:30 AM LT) which is generally not late enough in the day for the solar insulation to have raised the temperature of the air significantly above that of the water. Although, as discussed above, the influence of atmospheric stability remains unclear, we cannot fully rule out that its variation from one IASI observation to another contributes to the scatter of the probabilities shown, for instance, in Fig. \ref{fig:examplespdf_vent2_9}. 

%\end{multicols}

\begin{thebibliography}{10}

\bibitem{Albert_foam}
M.~F. M.~A. Albert, M.~D. Anguelova, A.~M.~M. Manders, M.~Schaap, and
  G.~de~Leeuw.
\newblock Parameterization of oceanic whitecap fraction based on satellite
  observations.
\newblock {\em Atm. Chem. Phys.}, 16:13725--13751, 2016.

\bibitem{Alpers}
Werner Alpers and Heinrich Hühnerfuss.
\newblock The damping of ocean waves by surface films: \uppercase{A} new look
  at an old problem.
\newblock {\em J. Geophys. Res.: Oceans}, 94(C5):6251--6265, 1989.

\bibitem{Hs-vs-U}
Jose Henrique G.~M. Alves, Michael~L. Banner, and Ian~R. Young.
\newblock Revisiting the \uppercase{P}ierson-\uppercase{M}oskowitz asymptotic
  limits for fully developed wind waves.
\newblock {\em J. Phys. Oceanogr.}, 33(7):1301--1323, 2003.

\bibitem{Anguelova19}
Magdalena~D. Anguelova and Michael~H. Bettenhausen.
\newblock Whitecap fraction from satellite measurements: \uppercase{A}lgorithm
  description.
\newblock {\em J. Geophys. Res.: Oceans}, 124(3):1827--1857, 2019.

\bibitem{ANUSREE}
A.~Anusree and V.Sanil Kumar.
\newblock Mean wave direction and wave height in the \uppercase{ERA}5
  reanalysis dataset: \uppercase{C}omparison with measured data in the coastal
  waters of \uppercase{I}ndia.
\newblock {\em Dyn. Atmos. Oceans}, 107:101478, 2024.

\bibitem{MAAOD2}
Marcel Babin, André Morel, Vincent Fournier-Sicre, Frank Fell, and Dariusz
  Stramski.
\newblock Light scattering properties of marine particles in coastal and open
  ocean waters as related to the particle mass concentration.
\newblock {\em L\&O}, 48:843--859, 2003.

\bibitem{Behroozi}
Peter Behroozi, Kimberly Cordray, William Griffin, and Feredoon Behroozi.
\newblock The calming effect of oil on water.
\newblock {\em Am. J. Phys.}, 75(5):407--414, 05 2007.

\bibitem{Belmonte}
M.~Belmonte~Rivas and A.~Stoffelen.
\newblock Characterizing \uppercase{ERA}-\uppercase{I}nterim and
  \uppercase{ERA5} surface wind biases using \uppercase{ASCAT}.
\newblock {\em Ocean Sci.}, 15:831--852, 2019.

\bibitem{Breon}
F.~M. Br{\'e}on and N.~Henriot.
\newblock Spaceborne observations of ocean glint reflectance and modeling of
  wave slope distributions.
\newblock {\em J. Geophys. Res.: Oceans}, 111:C06005, 2006.

\bibitem{Brumer}
Sophia~E. Brumer, Christopher~J. Zappa, Ian~M. Brooks, Hitoshi Tamura, Scott~M.
  Brown, Byron~W. Blomquist, Christopher~W. Fairall, and Alejandro
  Cifuentes-Lorenzen.
\newblock Whitecap coverage dependence on wind and wave statistics as observed
  during \uppercase{SO} \uppercase{G}as\uppercase{E}x and
  \uppercase{H}i\uppercase{W}in\uppercase{GS}.
\newblock {\em J. Phys. Oceanogr.}, 47:2211--2235, 2017.

\bibitem{Capelledust}
V.~Capelle, A.~Ch\'edin, M.~Pondrom, C.~Crevoisier, R.~Armante, L.~Cr\'epeau,
  and N.A. Scott.
\newblock Infrared dust aerosol optical depth retrieved daily from
  \uppercase{IASI} and comparison with \uppercase{AERONET} over the period
  2007-2016.
\newblock {\em Remote Sens. Environ.}, 206:15--32, 2018.

\bibitem{Capelledaytime}
Virginie Capelle and Jean-Michel Hartmann.
\newblock Use of hyperspectral sounder to retrieve daytime sea-surface
  temperature from mid-infrared radiances: \uppercase{A}pplication to
  \uppercase{IASI}.
\newblock {\em Remote Sens. Environ.}, 280:113171, 2022.

\bibitem{Capellenighttime}
Virginie Capelle, Jean-Michel Hartmann, and Cyril Crevoisier.
\newblock A full physics algorithm to retrieve nighttime sea surface
  temperature with \uppercase{IASI}: \uppercase{T}oward an independent
  homogeneous long time-series for climate studies.
\newblock {\em Remote Sens. Environ.}, 269:112838, 2022.

\bibitem{caulliez_JGR13}
Guillemette Caulliez.
\newblock Dissipation regimes for short wind waves.
\newblock {\em J. Geophys. Res.: Oceans}, 118(2):672--684, 2013.

\bibitem{caulliez_JGR12}
Guillemette Caulliez and Charles-Antoine Gu\'erin.
\newblock Higher-order statistical analysis of short wind wave fields.
\newblock {\em J. Geophy. Res.: Oceans}, 117(C6), 2012.

\bibitem{chapron_JGR00}
Bertrand Chapron, V~Kerbaol, D~Vandemark, and T~Elfouhaily.
\newblock Importance of peakedness in sea surface slope measurements and
  applications.
\newblock {\em J. Geophys. Res.: Oceans}, 105(C7):17195--17202, 2000.

\bibitem{chen_JPO00}
Gang Chen and Stephen~E Belcher.
\newblock Effects of long waves on wind-generated waves.
\newblock {\em J. Phys. Oceanogr.}, 30(9):2246--2256, 2000.

\bibitem{revue_vagues}
C.~O. et~al Collins.
\newblock Measuring ocean surface waves.
\newblock {\em Reviews of Geophysics (submitted)}, (preprint available at
  https://hal.science/hal-05401283v1)(?), 2026.

\bibitem{CAMS_EAC4_2020}
{Copernicus Atmosphere Monitoring Service}.
\newblock \uppercase{CAMS} global reanalysis (\uppercase{EAC}4).
\newblock Atmosphere Data Store, 2020.
\newblock doi = 10.24381/d58bbf47.

\bibitem{Cox54}
Charles Cox and Walter Munk.
\newblock Measurement of the roughness of the sea surface from photographs of
  the sun's glitter.
\newblock {\em J. Opt. Soc. Am.}, 44:838--850, 1954.

\bibitem{Cox54b}
Charles Cox and Walter Munk.
\newblock Statistics of the sea surface derived from sun glitter.
\newblock {\em J. Mar. Res.}, 13:198--227 (available at
  https://images.peabody.yale.edu/publications/jmr/jmr13--02--04.pdf), 1954.

\bibitem{Cox56}
Charles Cox and Walter Munk.
\newblock Slopes of the sea surface deduced from photographs of sun glitter.
\newblock {\em Bull. Scripps Inst. Oceanogr.}, 6:401--488 (available at
  https://escholarship.org/uc/item/1p202179), 1956.

\bibitem{donelan_APL87}
Mark.~A Donelan.
\newblock The effect of swell on the growth of wind waves.
\newblock {\em John Hopkins APL Technical Digest}, 8(1), 1987.

\bibitem{Wind_effect}
A.~Engel, G.~Friedrichs, K.~E. Krall, and B.~J\"ahne.
\newblock Wind-induced collapse of the biopolymeric surface microlayer induces
  sudden changes in sea surface roughness.
\newblock {\em Biogeosciences}, 23(6):2101--2117, 2026.

\bibitem{ermakov_RemSen20}
Stanislav~A Ermakov, Vladimir~A Dobrokhotov, Irina~A Sergievskaya, and Ivan~A
  Kapustin.
\newblock Suppression of wind ripples and microwave backscattering due to
  turbulence generated by breaking surface waves.
\newblock {\em Remote Sens.}, 12(21):3618, 2020.

\bibitem{IASI}
Fiona~Hilton et~al.
\newblock Hyperspectral earth observation from \uppercase{IASI}:
  \uppercase{F}ive years of accomplishments.
\newblock {\em Bull. Am. Meteorol. Soc.}, 93:347--370, 2012.

\bibitem{ASCAT}
J.~Figa-Saldana, J~J.W. Wilson, E.~Attema, R.~Gelsthorpe, M~R Drinkwater, and
  A.~Stoffelen.
\newblock The advanced scatterometer (\uppercase{ASCAT}) on the meteorological
  operational (\uppercase{M}et\uppercase{O}p) platform: \uppercase{A} follow on
  for european wind scatterometers.
\newblock {\em Can. J. Remote Sens.}, 28:404--412, 2002.

\bibitem{Revueaerosols1}
James~W. Fitzgerald.
\newblock Marine aerosols: \uppercase{A} review.
\newblock {\em Atmos. Environ. Part A. General Topics}, 25:533--545, 1991.

\bibitem{MAAOD3}
S.~G. Gathman.
\newblock Optical properties of the marine aerosol as predicted by the navy
  aerosol model.
\newblock {\em Optical Engineering}, 22:057--62, 1983.

\bibitem{Guerin_GRS17}
Charles-Antoine Gu\'erin, Jean-Christophe Poisson, Fanny Piras, Laiba
  Amarouche, and Jean-Claude Lalaurie.
\newblock \uppercase{K}u-/\uppercase{K}a-band extrapolation of the altimeter
  cross section and assessment with
  \uppercase{J}ason2/\uppercase{A}lti\uppercase{K}a data.
\newblock {\em IEEE Trans. Geosci. Remote Sens.}, 55(10):5679--5686, 2017.

\bibitem{guerraou_igarss18}
Zaynab Guerraou, S{\'e}bastien Angelliaume, and Charles-Antoine Gu{\'e}rin.
\newblock Physical modeling of the upwind-downwind asymmetry in microwave
  return from the sea surface.
\newblock In {\em IGARSS 2018-2018 IEEE International Geoscience and Remote
  Sensing Symposium}, pages 228--231. IEEE, 2018.

\bibitem{guerraou_GRS16}
Zaynab Guerraou, S{\'e}bastien Angelliaume, Luke Rosenberg, and Charles-Antoine
  Gu{\'e}rin.
\newblock Investigation of azimuthal variations from \uppercase{X}-band
  medium-grazing-angle sea clutter.
\newblock {\em IEEE Trans. Geosci. Remote Sens.}, 54(10):6110--6118, 2016.

\bibitem{Guerin_RSE23}
Charles-Antoine Guérin, Virginie Capelle, and Jean-Michel Hartmann.
\newblock Revisiting the cox and munk wave-slope statistics using
  \uppercase{IASI} observations of the sea surface.
\newblock {\em Remote Sens. Environ.}, 288:113508, 2023.

\bibitem{Hauser_JGR08}
D.~Hauser, G.~Caudal, S.~Guimbard, and A.~A. Mouche.
\newblock A study of the slope probability density function of the ocean waves
  from radar observations.
\newblock {\em J. Geophys. Res.: Oceans}, 113:C02006, 2008.

\bibitem{hauser_JGR09reply}
Dani{\`e}le Hauser, G{\'e}rard Caudal, S{\'e}bastien Guimbard, and Alexis
  Mouche.
\newblock Reply to comment by \uppercase{P}aul \uppercase{A}. \uppercase{H}wang
  on '\uppercase{A} study of the slope probablity density function of the ocean
  waves from radar observations' by \uppercase{D}. \uppercase{H}auser,
  \uppercase{G}. \uppercase{C}audal, \uppercase{S}. \uppercase{G}uimbard, and
  \uppercase{A}. \uppercase{M}ouche [2007jc004264rr].
\newblock {\em J. Geophys. Res.: Oceans}, 114(C2):C02009, 2009.

\bibitem{Revueaerosols3}
J.~Heintzenberg, D.~C. Covert, and R.~Van Dingenen.
\newblock Size distribution and chemical composition of marine aerosols:
  \uppercase{A} compilation and review.
\newblock {\em Tellus B: Chem. Phy. Meteorol.}, 52(4):1104--1122, 2000.

\bibitem{ERA5data}
Hans et~al Hersbach.
\newblock \uppercase{ERA5} hourly data on single levels from 1959 to present.
  \uppercase{C}opernicus \uppercase{C}limate \uppercase{C}hange
  \uppercase{S}ervice (\uppercase{C}3\uppercase{S}) \uppercase{C}limate
  \uppercase{D}ata \uppercase{S}tore (\uppercase{CDS}).
\newblock 2018.

\bibitem{ERA5}
Hans et~al Hersbach.
\newblock The \uppercase{ERA5} global reanalysis.
\newblock {\em Q. J. R. Meteorol. Soc.}, 146:1999--2049, 2020.

\bibitem{holthuisen_book10}
Leo Holthuijsen.
\newblock {\em Waves in Oceanic and Coastal Waters}.
\newblock International series of monographs on physics. Cambridge University
  Press, 2010.

\bibitem{AODSize}
W.~A. Hoppel, J.~W. Fitzgerald, G.~M. Frick, R.~E. Larson, and E.~J. Mack.
\newblock Aerosol size distributions and optical properties found in the marine
  boundary layer over the \uppercase{A}tlantic ocean.
\newblock {\em J. Geophys. Res.: Atmospheres}, 95(D4):3659--3686, 1990.

\bibitem{Hu}
Chuanmin Hu, Xiaofeng Li, William~G. Pichel, and Frank~E. Muller-Karger.
\newblock Detection of natural oil slicks in the \uppercase{NW} gulf of
  \uppercase{M}exico using \uppercase{MODIS} imagery.
\newblock {\em Geophys. Res. Lett.}, 36(1), 2009.

\bibitem{hu_atmosChem08}
Yongxiang et~al Hu.
\newblock Sea surface wind speed estimation from space-based lidar
  measurements.
\newblock {\em Atm. Chem. Phys.}, 8:3593--3601, 2008.

\bibitem{hwang_JGR05}
Paul~A Hwang.
\newblock Wave number spectrum and mean square slope of intermediate-scale
  ocean surface waves.
\newblock {\em J. Geophys. Res.: Oceans}, 110(C10), 2005.

\bibitem{hwang_JGR08}
Paul~A Hwang.
\newblock Observations of swell influence on ocean surface roughness.
\newblock {\em J. Geophys. Res.: Oceans}, 113:C12024, 2008.

\bibitem{hwang_JGR09comment}
Paul~A Hwang.
\newblock Comment on “\uppercase{A} study of the slope probability density
  function of the ocean waves from radar observations” by \uppercase{D}.
  \uppercase{H}auser et al.
\newblock {\em J. Geopyhs. Res.: Oceans}, 114(C2), 2009.

\bibitem{Hwang_JGR88}
Paul~A. Hwang and Omar~H. Shemdin.
\newblock The dependence of sea surface slope on atmospheric stability and
  swell conditions.
\newblock {\em J. Geophys. Res.: Oceans}, 93:13903--13912, 1988.

\bibitem{hwang_JGR08b}
Paul~A Hwang, Mark~A Sletten, and Jakov~V Toporkov.
\newblock Analysis of radar sea return for breaking wave investigation.
\newblock {\em J. Geophys. Res.: Oceans}, 113(C2), 2008.

\bibitem{Inness2019}
Antje Inness, Melanie Ades, Anna Agust{\'\i}-Panareda, J{\'e}r{\^o}me
  Barr{\'e}, Anna Benedictow, Anne-Marlene Blechschmidt, Juan~Jose Dominguez,
  Richard Engelen, Henk Eskes, and Johannes et~al Flemming.
\newblock The \uppercase{CAMS} reanalysis of atmospheric composition.
\newblock {\em Atmos. Chem. Phys.}, 19(6):3515--3556, 2019.

\bibitem{jackson_JGR92}
FC~Jackson, WT~Walton, DE~Hines, BA~Walter, and CY~Peng.
\newblock Sea surface mean square slope from \uppercase{K}u-band backscatter
  data.
\newblock {\em J. Geophys. Res.: Oceans}, 97(C7), 1992.

\bibitem{jessup_JGR91}
AT~Jessup, WK~Melville, and WC~Keller.
\newblock Breaking waves affecting microwave backscatter: 1.
  \uppercase{D}etection and verification.
\newblock {\em J. Geophys. Res.: Oceans}, 96(C11):20547--20559, 1991.

\bibitem{MAAOD4}
Gennady~A. Kaloshin.
\newblock Visible and infrared extinction of atmospheric aerosol in the marine
  and coastal environment.
\newblock {\em Appl. Opt.}, 50:2124--2133, 2011.

\bibitem{Kudryavtsev}
Vladimir Kudryavtsev, Alexander Myasoedov, Bertrand Chapron, Johnny~A.
  Johannessen, and Fabrice Collard.
\newblock Joint sun-glitter and radar imagery of surface slicks.
\newblock {\em Remote Sens. Environ.}, 120:123--132, 2012.

\bibitem{Laxague}
Nathan J.~M. Laxague, Christopher~J. Zappa, Shantanu Soumya, and Oliver Wurl.
\newblock The suppression of ocean waves by biogenic slicks.
\newblock {\em J. Roy. Soc. Interface}, 21:20240385, 2024.

\bibitem{Lefevre}
J.~M. Lefevre, J.~Barckicke, and Y.~Ménard.
\newblock A significant wave height dependent function for
  \uppercase{TOPEX}/\uppercase{POSEIDON} wind speed retrieval.
\newblock {\em J. Geophys. Res.: Oceans}, 99(C12):25035--25049, 1994.

\bibitem{Lenain_JPO19}
Luc Lenain, Nicholas~M. Statom, and W.~Kendall Melville.
\newblock Airborne measurements of surface wind and slope statistics over the
  ocean.
\newblock {\em J. Phys. Oceanogr.}, 49:2799--2814, 2019.

\bibitem{Li_IntJournRemSen13}
Shuiqing Li, Dongliang Zhao, Liangming Zhou, and Bin Liu.
\newblock Dependence of mean square slope on wave state and its application in
  altimeter wind speed retrieval.
\newblock {\em Int. J. Remote Sens.}, 34(1):264--275, 2013.

\bibitem{Liu_Atmos22}
J.~Liu, B.~Li, W.~Chen, J.~Li, and J.~Yan.
\newblock Evaluation of \uppercase{ERA}5 wave parameters with in situ data in
  the south \uppercase{C}hina sea.
\newblock {\em Atmosphere}, 13:935, 2022.

\bibitem{LH63}
MS~Longuet-Higgins.
\newblock The generation of capillary waves by steep gravity waves.
\newblock {\em J. Fluid Mech.}, 16(1):138--159, 1963.

\bibitem{Louis}
J.-F. Louis.
\newblock A parametric model of vertical eddy fluxes in the atmosphere.
\newblock {\em Boundary Layer Meteorology}, 17:187–202, 1979.

\bibitem{BRDFFoam2}
L.~X. Ma, F.~Q. Wang, C.~A. Wang, C.~C. Wang, and J.~Y. Tan.
\newblock Investigation of the spectral reflectance and bidirectional
  reflectance distribution function of sea foam layer by the monte carlo
  method.
\newblock {\em Appl. Opt.}, 54:9863--9874, 2015.

\bibitem{IASIoverflow}
C.~Maraldi and E.~Jacquette.
\newblock \uppercase{IASI} quarterly performance report from 2014/03/01 to
  2014/05/31.
\newblock {\em Document number: IA-RP-2000-4162-CNE}, pages available at
  https://cnes.fr/sites/default/files/migration/smsc/iasi/QPR/IASI$\_$M01$\_$quarterly$\_$2014--03$\_$2014--05.pdf,
  2014.

\bibitem{masuko_JGR86}
Harunobu Masuko, Ken'ichi Okamoto, Masanobu Shimada, and Shuntaro Niwa.
\newblock Measurement of microwave backscattering signatures of the ocean
  surface using \uppercase{X} band and \uppercase{K}a band airborne
  scatterometers.
\newblock {\em J. Geophys. Res.: Oceans}, 91(C11):13065--13083, 1986.

\bibitem{Monahan}
E.~C. Monahan and I.~Muircheartaigh.
\newblock Optimal power-law description of oceanic whitecap coverage dependence
  on wind speed.
\newblock {\em J. Phys. Oceanogr.}, 10:2094–2099, 1980.

\bibitem{mouche_JGR07}
Alexis~A Mouche, Bertrand Chapron, Nicolas Reul, Dani{\`e}le Hauser, and Yves
  Quilfen.
\newblock Importance of the sea surface curvature to interpret the normalized
  radar cross section.
\newblock {\em J. Geophys. Res.: Oceans}, 112(C10), 2007.

\bibitem{MAAOD1}
J.~P. Mulcahy, C.~D. O'Dowd, S.~G. Jennings, and D.~Ceburnis.
\newblock Significant enhancement of aerosol optical depth in marine air under
  high wind conditions.
\newblock {\em Geophys. Res. Lett.}, 35(16), 2008.

\bibitem{Nouguier_GRSL16}
Frederic Nouguier, Alexis Mouche, Nicolas Rascle, Bertrand Chapron, and Douglas
  Vandemark.
\newblock Analysis of dual-frequency ocean backscatter measurements at
  \uppercase{K}u- and \uppercase{K}a-bands using near-nadir incidence
  \uppercase{GPM} radar data.
\newblock {\em IEEE Trans. Geosci. Remote Sens. Lett.}, 13:1310--1314, 2016.

\bibitem{Revueaerosols2}
Colin~D O'Dowd and Gerrit de~Leeuw.
\newblock Marine aerosol production: \uppercase{A} review of the current
  knowledge.
\newblock {\em Phil. Trans. Roy. Soc. A: Mathematical, Physical and Engineering
  Sciences}, 365(1856):1753--1774, 2007.

\bibitem{MAprod2}
Colin~D. O'Dowd and Michael~H. Smith.
\newblock Physicochemical properties of aerosols over the northeast
  \uppercase{A}tlantic: \uppercase{E}vidence for wind-speed-related submicron
  sea-salt aerosol production.
\newblock {\em J. Geophys. Res.: Atmospheres}, 98(D1):1137--1149, 1993.

\bibitem{phillips_JFM74}
OM~Phillips and ML~Banner.
\newblock Wave breaking in the presence of wind drift and swell.
\newblock {\em J. Fluid Mech.}, 66(4):625--640, 1974.

\bibitem{pierson}
Willard~J. Pierson~Jr.
\newblock The interpretation of wave spectrums in terms of the wind profile
  instead of the wind measured at a constant height.
\newblock {\em J. Geophys. Res.}, 69:5191--5203, 1964.

\bibitem{Reul}
N.~Reul and B.~Chapron.
\newblock A model of sea-foam thickness distribution for passive microwave
  remote sensing applications.
\newblock {\em J. Geophys. Res.: Oceans}, 108(C10), 2003.

\bibitem{Ross}
Vincent Ross and Denis Dion.
\newblock Sea surface slope statistics derived from sun glint radiance
  measurements and their apparent dependence on sensor elevation.
\newblock {\em J. Geophys. Res.: Oceans}, 112:C09015, 2007.

\bibitem{ERA5wind}
I.~Sandu, P.~Bechtold, L.~Nuijens, A.~Beljaars, and A~Brown.
\newblock On the causes of systematic forecast biases in near-surface wind
  direction over the oceans.
\newblock {\em ECMWF technical memo}, available at
  https://www.ecmwf.int/en/elibrary/19545-causes-systematic-forecast-biases-near-surface-wind-direction-over-oceans,
  2020.

\bibitem{shaw}
Joseph~A. Shaw and James~H. Churnside.
\newblock Scanning-laser glint measurements of sea-surface slope statistics.
\newblock {\em Appl. Opt.}, 36:4202--4213, 1997.

\bibitem{Shi}
H.~Shi, X.~Cao, Q.~Li, D.~Li, J.~Sun, Z.~You, and Q.~Sun.
\newblock Evaluating the accuracy of \uppercase{ERA}5 wave reanalysis in the
  water around \uppercase{C}hina.
\newblock {\em J. Ocean. Univ. China}, 28:1--9, 2021.

\bibitem{shrira_JFM96}
Victor~I Shrira, Sergei~I Badulin, and Christian Kharif.
\newblock A model of water wave ‘horse-shoe’ patterns.
\newblock {\em J. Fluid Mech.}, 318:375--405, 1996.

\bibitem{MAprod1}
M.~H. Smith, P.~M. Park, and I.~E. Consterdine.
\newblock Marine aerosol concentrations and estimated fluxes over the sea.
\newblock {\em Quart. J. Roy. Meteo. Soc.}, 119(512):809--824, 1993.

\bibitem{Vandemark_JPO04}
D.~Vandemark, B.~Chapron, J.~Sun, G.~H. Crescenti, and H.~C. Graber.
\newblock Ocean wave slope observations using radar backscatter and laser
  altimeters.
\newblock {\em J. Phys. Oceanogr.}, 34:2825--2842, 2004.

\bibitem{precisionASCAT}
Jur Vogelzang, Ad~Stoffelen, Anton Verhoef, and Julia Figa-SaldaÃ±a.
\newblock On the quality of high-resolution scatterometer winds.
\newblock {\em J. Geophys. Res.: Oceans}, 116(C10), 2011.

\bibitem{Schuckmann20082021}
K.~et~al Von~Schuckmann.
\newblock Copernicus marine service ocean state report, issue 5.
\newblock {\em J. Operational Oceanogr.}, 14:1--185, 2021.

\bibitem{Voronovich}
A.G.Y Voronovich and V.~U. Zavorotny.
\newblock Theoretical model for scattering of radar signals in \uppercase{K}u-
  and \uppercase{C}-bands from a rough sea surface with breaking waves.
\newblock {\em Waves in Random Media}, 11:248--269, 2001.

\bibitem{BRDFFoam1}
Chengchao Wang, Chengwei Jia, Qingzhi Lai, Rifeng Zhou, Yinmo Xie, Linhua Liu,
  and Lanxin Ma.
\newblock Experimental investigation on bidirectional reflection
  characteristics of sea foam in the visible and near-infrared bands.
\newblock {\em Infrared Phys. Technol.}, 151:106065, 2025.

\bibitem{Wang04032022}
Jichao Wang and Yue Wang.
\newblock Evaluation of the \uppercase{ERA}5 \uppercase{S}ignificant
  \uppercase{W}ave \uppercase{H}eight against \uppercase{NDBC} buoy data from
  1979 to 2019.
\newblock {\em Marine Geodesy}, 45:151--165, 2022.

\bibitem{wetzel_chapter86}
Lewis Wetzel.
\newblock On microwave scattering by breaking waves.
\newblock In {\em Wave dynamics and radio probing of the ocean surface}, pages
  273--284. Springer, 1986.

\bibitem{wu_JGR91}
Jin Wu.
\newblock Effects of atmospheric stability on ocean ripples: \uppercase{A}
  comparison between optical and microwave measurements.
\newblock {\em J. Geophys. Res.: Oceans}, 96(C4):7265--7269, 1991.

\bibitem{yurovsky_RemSens21}
Yury~Yu Yurovsky, Vladimir~N Kudryavtsev, Semyon~A Grodsky, and Bertrand
  Chapron.
\newblock Ka-band radar cross-section of breaking wind waves.
\newblock {\em Remote Sens.}, 13(10):1929, 2021.

\bibitem{zhang_JFM95}
Xin Zhang.
\newblock Capillary--gravity and capillary waves generated in a wind wave tank:
  \uppercase{o}bservations and theories.
\newblock {\em J. Fluid Mech.}, 289:51--82, 1995.

\bibitem{ZHANG2025}
Yue Zhang, Xiaoxiao Yu, Peng Gao, Chunlin Huang, Qixiang Chen, Yuan Yuan,
  Shikui Dong, and Kaifeng Lin.
\newblock An effective computational method and analysis of scattering
  characteristics for sea surface foam layer.
\newblock {\em J. Quant. Spectrosc. Radiat. Transf.}, 333:109332, 2025.

\end{thebibliography}
\end{document}